\documentclass[11pt]{article}
\pdfoutput=1
\usepackage{jcapmod}

\usepackage{tikz}
\usetikzlibrary{shapes.misc,positioning,arrows,arrows.meta,bending,matrix,shapes,fit,tikzmark,calc,external,patterns,topaths,decorations.pathmorphing,decorations.markings,decorations.pathreplacing}

\tikzset{>=latex} 

\usepackage[export]{adjustbox}
\usepackage{array}
\usepackage{mathtools}
\usepackage{booktabs}
\usepackage[english]{babel}
\usepackage{amsmath,amssymb,amsbsy,amstext, amsthm, simplewick, amsfonts}
\usepackage{hyperref}
\usepackage{graphicx}
\usepackage[small]{caption}
\usepackage{siunitx}
\usepackage{upgreek}
\usepackage{framed}
\usepackage{wrapfig}
\usepackage{multirow}
\usepackage{bbm}
\usepackage{selinput}
\usepackage{bm}
\usepackage{float}
\usepackage{geometry}
\usepackage{yfonts}
\usepackage{caption}
\usepackage{subcaption}
\usepackage{sidecap}
\usepackage{longtable}
\usepackage{anyfontsize}
\usepackage{dsfont}
\usepackage{tikz}
\usepackage{relsize}
\usepackage{tcolorbox}
\usepackage{xcolor}
\usepackage{xparse}
\usepackage{slashed}
\usepackage{simpler-wick}

\usepackage{shorthand}
\usepackage{TikzMaker}
\usepackage{SpecialColors}

\usepackage{array}
\newcolumntype{L}[1]{>{\raggedright\let\newline\\\arraybackslash\hspace{0pt}}m{#1}}
\newcolumntype{C}[1]{>{\centering\let\newline\\\arraybackslash\hspace{0pt}}m{#1}}
\newcolumntype{R}[1]{>{\raggedleft\let\newline\\\arraybackslash\hspace{0pt}}m{#1}}

\def\half{\frac{1}{2}}

\newcommand{\bee}{\begin{equation*}}
\newcommand{\eee}{\end{equation*}}
\def\ba{\begin{equation}\begin{aligned}}
\def\ea{\end{aligned}\end{equation}}
\def\dif{{\rm{d}}}

\newcommand{\Vvec}[1]{{\bf{#1}}}
\def\dl{\ud \log}
\def\dev{\xi}

\tikzset{cross/.style={cross out, draw=black, minimum size=2*(#1-\pgflinewidth), inner sep=0pt, outer sep=0pt},
cross/.default={3pt}}
\newcommand{\reef}[1]{(\ref{#1})}

\def\hs{\hskip 1pt}

\def\beq{\begin{equation}}
\def\eeq{\end{equation}}
\def\be{\begin{equation}}
\def\ee{\end{equation}}

\def\k{\vec k}

\newcommand{\ud}{{\rm d}}

\makeatletter
\newlength{\apb@width}
\newcommand{\autoparbox}[2][c]{\settowidth{\apb@width}{#2}\parbox[#1]{\apb@width}{#2}}

\makeatother


\allowdisplaybreaks[1]
\begin{document}
\newgeometry{top=2cm, bottom=2cm, left=2cm, right=2cm}

\begin{titlepage}
\setcounter{page}{1} \baselineskip=15.5pt 
\thispagestyle{empty}

\begin{center}
{\fontsize{21}{18} \bf Differential Equations for Massive Correlators}
\end{center}

\vskip 20pt
		\begin{center}
			\noindent
			{\fontsize{14}{18}\selectfont 
				Daniel Baumann\hs$^{1,2,3}$, Austin Joyce\hs$^{4,5}$, 
				Hayden Lee\hs$^{6}$ and Kamran Salehi Vaziri\hs$^{1}$}
		\end{center}

		\begin{center}
			\vskip8pt
			\textit{$^1$ Institute of Physics, University of Amsterdam, Amsterdam, 1098 XH, The Netherlands}
			
			\vskip8pt
			\textit{$^2$  Leung Center for Cosmology and Particle Astrophysics,
				Taipei 10617, Taiwan}
			
			  \vskip8pt
\textit{$^3$  Max Planck--IAS--NTU Center for Particle Physics, Cosmology and Geometry,
Taipei 10617, Taiwan}

			\vskip 8pt
			\textit{$^4$ 
				Department of Astronomy and Astrophysics,
				The University of Chicago, Chicago, IL 60637, USA}
			
			\vskip 8pt
			\textit{$^5$ Kavli Institute for Cosmological Physics, 
				The University of Chicago, Chicago, IL 60637, USA}
			
			\vskip 8pt
			\textit{$^6$ Center for Particle Cosmology, Department of Physics and Astronomy, \\ University of Pennsylvania, Philadelphia, PA 19104, USA}
			
		\end{center}

\vspace{0.4cm}
\begin{center}{\bf Abstract}
\end{center}
\noindent
We uncover a combinatorial structure governing the differential equations satisfied by wavefunction coefficients of scalar fields with generic masses in de Sitter space. Using an integral representation of the massive mode functions, we express the Feynman integrals underlying cosmological correlators as twisted integrals of rational functions. In this formulation, the integrals belong to a finite set of master integrals obeying a first-order system of differential equations, which can be derived efficiently in the time-integral representation. We show that these equations admit a simple graphical description in terms of graph tubings, which encode the couplings among basis functions and the evolution of singularities. This structure provides an efficient algorithm to derive the differential equations, and a boundary-centric perspective on massive cosmological correlators in which their analytic structure emerges from underlying combinatorial data. As an illustration, we solve the system in the limits of small and large masses.

\end{titlepage}
\restoregeometry

\newpage
\setcounter{tocdepth}{3}
\setcounter{page}{2}

\linespread{1.2}
\tableofcontents
\linespread{1.1}

\newpage
\section{Introduction}

An important feature of cosmology is that space itself is dynamical and evolves in time.
This complicates the quantum-mechanical sum over all possible times that processes can occur, because particle interactions typically also change in time.
Consequently, even tree-level computations of cosmological correlators involve nontrivial time integrals over propagators and interaction vertices~\cite{corrlectures,Baumann:2022jpr,Benincasa:2022gtd}. In addition, in the curved spacetimes relevant for cosmology, the free-field mode functions are generically complicated special functions. Combined, these features make it difficult to evaluate the Feynman integrals underlying cosmological correlation functions in closed form.

\vskip4pt
Despite these challenges, there are remarkable structures underlying cosmological observables that make their study tractable.
One way to glimpse these features is to consider how cosmological correlations change as we change the parameters they depend on.
For example, if we smoothly vary the kinematics, 
we expect observables to change continuously and therefore to satisfy differential equations that describe how their strength changes. In~\cite{Arkani-Hamed:2023kig}, these differential equations were studied in a simplified setting involving conformally coupled scalars with polynomial interactions in a power-law cosmology~\cite{Arkani-Hamed:2017fdk}. The integrals of interest are then generalized Euler integrals, which have close connections to the mathematics of twisted cohomology and hyperplane arrangements, as well as a beautiful geometric interpretation related to cosmological polytopes~\cite{Arkani-Hamed:2017fdk}. A consequence of these mathematical connections is that these cosmological integrals belong to a finite-dimensional basis of master integrals $I_a$, which satisfy differential equations of the form
\beq
{\rm d} I_a = A_{ab} \,I_b\,,
\eeq
where ${\rm d}$ is a differential with respect to kinematic variables, and $A_{ab}$ is a matrix of one-forms that encodes the singularities of the differential system.
Remarkably, these differential equations have a simple combinatorial description, called {\it kinematic flow}~\cite{Arkani-Hamed:2023kig,Arkani-Hamed:2023bsv,Baumann:2025qjx}, which generates the connection matrix $A_{ab}$. (Additional features have been studied in~\cite{De:2023xue,De:2024zic,Fan:2024iek,He:2024olr,Glew:2025ypb,Capuano:2025ehm,McLeod:2026jpz,Fu:2026dqb}.) Aside from providing an efficient algorithm to derive these differential equations, this structure reveals an intriguing boundary-centric perspective on cosmological correlations, where they arise  from purely combinatorial origins~\cite{Arkani-Hamed:2023bsv}.

\vskip4pt
It is natural to wonder whether the structures seen in the conformally coupled case are merely artifacts of the toy nature of the setup, or whether they can be abstracted to more general (and phenomenologically relevant) settings involving massive particles~\cite{Chen:2009zp,Baumann:2011nk,Noumi:2012vr,Arkani-Hamed:2015bza,Lee:2016vti,Arkani-Hamed:2018kmz,Baumann:2019oyu,Sleight:2019hfp,Wang:2019gbi,Bodas:2020yho,Pimentel:2022fsc,Jazayeri:2022kjy}. In this paper, we address this question by finding similar combinatorial structures in the differential equations satisfied by correlators and wavefunction coefficients of scalar fields with arbitrary masses in exact de Sitter space.\footnote{Wavefunctions and correlation functions in de Sitter space satisfy the same differential equations; they differ only in the boundary conditions used to solve them. As a result, the methods developed here apply equally to both cases. The same tools can also be used to compute boundary correlators in Euclidean AdS.}
Related aspects of this problem have also been considered recently in~\cite{Benincasa:2019vqr,Chen:2023iix,Chen:2024glu,Liu:2024str,Gasparotto:2024bku,Melville:2024ove,Werth:2024mjg,Raman:2025tsg,Xianyu:2025lbk,Belrhali:2026ktb}.

\vskip4pt
The most obvious complication is that the mode functions of
massive fields in de Sitter space are Hankel functions. This leads to a richer structure of integrands, compared to the conformally coupled case.
One way to deal with this is to express the Hankel functions as integrals over auxiliary parameters, which makes it possible to cast the associated wavefunction coefficients as integrals over rational functions times a twist factor that depends on the mass parameter~\cite{Gasparotto:2024bku}. 
 An interesting feature of this perspective is that the wavefunction for an arbitrary process involving generic masses can be written in terms of a {\it universal integrand}, which itself has a combinatorial definition. This guarantees that the wavefunction of interest still lies in a finite-dimensional space of twisted integrals which satisfies a system of differential equations.

\vskip4pt
While the twisted-integral representation guarantees the existence of a system of differential equations, in order to actually derive these equations it is simplest to go back to the time-integral representation.
This representation assembles the wavefunction from a natural set of time-ordered pieces. 
In addition, it is convenient to enlarge the function basis by essentially including a second copy of all basis functions where we replace the mode functions with their first derivatives~\cite{Chen:2023iix,Chen:2024glu}. We then write all basis elements in terms of the following functions
\be
h_k^\pm(\eta) \propto (-k\eta)^\nu H_\nu^{(2)}(-k\eta) \mp i \frac{\partial}{\partial k\eta}\Big[(-k\eta)^\nu H_\nu^{(2)}(-k\eta)\Big]\,,
\ee
which are linear combinations of the de Sitter mode function and its first derivative. The benefit of these combinations is that they satisfy {\it first-order} differential equations. 
Fundamentally, this underlies the existence of a first-order system in the time-integral representation. As an example, we can write the (time-ordered) wavefunction coefficient for a single massive exchange as a sum of functions of the form 
\be
\label{eq:introfeyn}
\begin{aligned}
\psi^{(\pmr \pmb)}\,&=
\scalebox{1.0}{
\raisebox{-26pt}{
\begin{tikzpicture}[line width=1. pt, scale=2]
\draw[line width=1.pt,lightgray] (0,0) -- (-0.25,0.55);
\draw[line width=1.pt,lightgray] (0,0) -- (0.25,0.55);
\draw[line width=1.pt,lightgray] (1,0) -- (0.75,0.55);
\draw[line width=1.pt,lightgray] (1,0) -- (1.25,0.55);
\draw[lightgray, line width=2.pt] (-0.5,0.55) -- (1.5,0.55);
\draw[line width=2.pt,darkgray] (0,0) -- (1,0);
\draw[fill=black] (0,0) circle (.03cm);
\draw[fill=black] (1,0) circle (.03cm);
\draw[fill=white] (.5,0) circle (.0275cm);
\node[scale=1] at (0,-.15) {$X_1$};
\node[scale=1] at (1,-.15) {$X_2$};
\node[scale=1] at (.5,-.15) {$Y$};
\node[scale=1] at (.25,.12) {$\pmr$};
\node[scale=1] at (.75,.12) {$\pmb$};
\end{tikzpicture}
} }\\[4pt]
&=  \int \ud\eta_1\ud\eta_2 \,\frac{e^{iX_1\eta_1}e^{iX_2\eta_2} }{(-\eta_1)^{1+\alpha_1+\dev}(-\eta_2)^{1+\alpha_2 + \dev}}  \Big[ \bar h^{\pmr}_Y (\eta_1)  h^{\pmb}_Y (\eta_2)\, \theta_{12} +  h^{\pmr}_Y (\eta_1)  \bar h^{\pmb}_Y (\eta_2)\, \theta_{21} \Big]\, ,
\end{aligned}
\ee
where $\theta_{ij} \equiv  \theta(\eta_i-\eta_j)$ is the Heaviside theta function, $X_1$ and $X_2$ are the external energies entering each vertex and $Y$ is the internal energy flowing between the vertices. The parameters $\alpha_1$ and $\alpha_2$ depend on the precise interactions involved in the process, while $\dev \equiv \nu - \tfrac{1}{2} $ characterizes the deviation of the mass from the conformally coupled value. Differentials of these basis functions have an elegant interpretation in terms of collapsing the propagators of the underlying Feynman diagram. For example, the differential of the functions in~\eqref{eq:introfeyn} is
\ba
\label{eq:introdiff}
\ud \psi^{(\pmr\pmb)} &= \Big[\alpha_1\, \dl (X_1\,\pmr\, Y) + \alpha_2\, \dl (X_2\,\pmb\, Y)\Big]\, \psi^{(\pmr\pmb)} - \frac{\pmr 1 \mpb 1}{2} \hs \dl \left(\frac{X_1\,\pmr\, Y}{X_2\,\pmb\, Y}\right)
J^{(c)}\\[1pt]
&\quad + \dev \bigg[ \dl \left(\frac{X_1\,\pmr\, Y}{Y}\right)\,  \psi^{(\mpr\pmb)}+   \dl \left(\frac{X_2\,\pmb\, Y}{Y}\right) \psi^{(\pmr\mpb)}\bigg]  \,,
\ea
where ${\rm d} \equiv \sum_i {\rm d} X_i \partial_{X_i} + {\rm d} Y \partial_Y$. Notice that the source functions appearing on the right-hand side of~\eqref{eq:introdiff} are related to the original function by sign flips of the $\pm$ indices, along with the function
$J^{(c)}$ that arises from collapsing the internal propagator to a contact diagram. The dlog-forms (``letters") multiplying these source functions also display apparent structure.
A similar pattern persists for generic diagrams with arbitrary masses, including loop integrands.

\vskip4pt
The differential equations satisfied by the wavefunction coefficients clearly have some combinatorial underpinnings.
In~\cite{Arkani-Hamed:2023kig, Baumann:2025qjx}, the structures underlying similar differential equations were elucidated by representing the basis functions and possible singularities by graph tubings.
Two simple rules then governed how the functions are coupled and how the singularities ``evolve" when taking derivatives with respect to the kinematic variables.
We will show that a similar graphical representation also reveals the underlying structure of the differential equations in the massive case.
For example, the differential for the function
$\psi^{(-+)} \equiv \psi_{\includegraphics[scale=0.6]{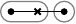}}$
can be written graphically as
\beq
\label{eq:intrographicaleq}
\begin{aligned}
{\rm d}\psi_{\includegraphics[scale=0.6]{Figures/Tubings/Functions/psi-.pdf}} &\, = \, \Big(
\alpha_1   \, \includegraphics[scale=0.9,valign=c]{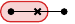}   \, +\, \alpha_2     \includegraphics[scale=0.9,valign=c]{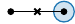}  \Big) \,  \psi_{\includegraphics[scale=0.6]{Figures/Tubings/Functions/psi-.pdf}} 
  \, +\, 
  \Big(
 \includegraphics[scale=0.9,valign=c]{Figures/Tubings/two/Lam.pdf}   \, -\,     \includegraphics[scale=0.9,valign=c]{Figures/Tubings/two/Lbp.pdf}  \Big)  \, J_{\includegraphics[scale=0.6]{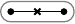}}  \\
 &\qquad + \dev \Big[ \Big(\includegraphics[scale=0.9,valign=c]{Figures/Tubings/two/Lam.pdf}  -  \Lm\Big) \psi_{\includegraphics[scale=0.6]{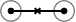}} 
 + \Big(\includegraphics[scale=0.9,valign=c]{Figures/Tubings/two/Lbp.pdf}  -  \Lm\Big) \psi_{\includegraphics[scale=0.6]{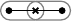}} 
  \Big]\,,
\end{aligned}
\eeq
where each colored tube represents a letter related to the total energy enclosed by the tube.
This pictorial representation makes clear that the differential equations have a dynamical interpretation in the space of tubings. 
The first line in~\eqref{eq:intrographicaleq} is the same as for conformally coupled fields, while the second line (proportional to $\dev$) is a novelty of massive fields.  We see that massive propagators lead to additional basis functions, and we expand the meaning of the graph tubings to capture these functions. In the conformally coupled case, tubes can only ``grow" by the merger of adjacent tubes, corresponding to the collapse of the internal propagator (like in the function $J_{\includegraphics[scale=0.6]{Figures/Tubings/Functions/psic.pdf}}$ in the above example).  Tubes enclosing a massive propagator, on the other hand, can also ``shrink" to create an exposed cross or grow to produce overlapping tubes. In the above examples, this creates the new functions $\psi_{\includegraphics[scale=0.6]{Figures/Tubings/Functions/psipp.pdf}}$ and $\psi_{\includegraphics[scale=0.6]{Figures/Tubings/Functions/psimm.pdf}}$, respectively. We formulate an additional kinematic flow rule that accounts for this new phenomenon.

\vskip4pt
The first-order system produced by this massive flow provides some interesting new perspectives on familiar phenomena. Broadly in this formalism, physical solutions are selected by specifying boundary data for the system, which further reinforces the perspective of cosmological observables as arising from a flow through kinematic space. Beyond this, the organization of the equations makes taking parametric mass limits particularly transparent.
In order to illustrate this, we solve the differential equations for a single massive exchange between conformal scalars both in the limit where the mass is taken to infinity, and where it approaches its conformally coupled value.
The large-mass limit is closely connected to the effective field theory expansion, which arises in an intriguing way from the first-order system. In the infinite-mass limit, the equations become algebraic, and so the perturbative expansion about this point can be solved recursively. This expansion can be formally re-summed to match the EFT expansion of~\cite{Arkani-Hamed:2018kmz}.
In the conformally coupled limit, the observables simplify to polylogarithmic functions, and  the differential equations encode the symbol (sequence of branch points) of these functions~\cite{Arkani-Hamed:2023kig}. As we expand about this limit, the perturbative corrections in mass can also be expressed in terms of polylogarithmic functions~\cite{Gasparotto:2024bku}, though with a more intricate structure.

\vskip4pt
It is rather interesting that the differential equations for massive fields display a similar richness to the conformally coupled case. This suggests that the interesting structures discovered in the conformal case are more far-reaching than has been previously appreciated.
In addition, the fact that all masses can be treated uniformly provides a universal parameterization of processes relevant for phenomenology, but encoded in a language that is more amenable to the formal study of their fundamental properties.

\vspace{-2pt}
\paragraph{Outline} 
The paper is organized as follows. In Section~\ref{sec:background}, we introduce the Feynman integrals that compute wavefunction coefficients involving massive fields, and explain that they are part of a finite-dimensional vector space of master integrals. In Section~\ref{sec:DE}, we choose a convenient basis and derive the associated differential equations for several representative examples, uncovering simple and universal patterns in the connection matrix that governs the couplings among the basis functions. In Section~\ref{sec:Flow}, we show that these patterns are naturally explained by a straightforward generalization of the kinematic flow rules. In Section~\ref{sec:Solutions}, we solve the differential equations in the limits of small and large masses, illustrating the method with the example of the single massive exchange. Finally, we summarize our conclusions in Section~\ref{sec:Conclusions}.

\vskip 4pt
Three appendices contain supplemental material. In Appendix~\ref{app:massless}, we describe a mapping between massive fields in de Sitter space and massless fields in a power-law cosmology, which allows the results for massive fields to be translated directly to the massless setting by appropriately mapping parameters. In Appendix~\ref{app:combin}, we show how the massive integrals can be expressed as twisted integrals of rational functions, with twists given by the deviation from the conformally coupled mass. We further discuss the geometry and combinatorics underlying the universal integrand. Finally, in Appendix~\ref{app:loop}, we
demonstrate that the kinematic flow algorithm also applies to loop integrands, in a way that is more uniform than the conformal case.

\section{Massive Integrals}
\label{sec:background}

We begin by introducing the Feynman integrals that appear in the computation of cosmological correlators. In theories containing massive fields, the integrands involve products of Hankel functions, which makes these integrals difficult to evaluate directly. By employing an integral representation of the Hankel functions, we rewrite these expressions as twisted integrals of rational functions. This reformulation allows us to show that the relevant integrals belong to a finite-dimensional vector space spanned by a set of master integrals. In Section~\ref{sec:DE}, we derive differential equations satisfied by these integrals.

\subsection{Massive Fields in de Sitter}
We will consider minimally coupled massive scalar fields in a rigid $(d+1)$-dimensional de Sitter spacetime, with polynomial interactions
\be
\label{eq: Lagrangian}
S=\int {\rm{d}}^{d+1}x \, \sqrt{-g} \left[-\half (\partial \phi)^2- \half m^2 \phi^2 - \sum_{p=3}^{\infty}\frac{\lambda_p}{p!} \phi^p \right] .
\ee
The spacetime metric is of the Robertson--Walker form ${\dif s}^2 = a^2(\eta)(-\dif \eta^2+\dif {\bf{x}}^2)$, with scale factor
\be\label{eq: metric}
a(\eta)= \frac{1}{H(-\eta)}\,,
\ee
where $\eta \in (-\infty,0)$ is conformal time and $H$ is the Hubble constant. We are interested in computing correlation functions at the future boundary, which we denote by $\eta_*=0$.
The initial conditions are imposed at $\eta\to -\infty$, when all modes are deep inside the horizon.

\vskip4pt
A free massive field satisfies the Euler--Lagrange equation derived from the quadratic part of the action~\reef{eq: Lagrangian}. In momentum space, this is given by
\be\label{eq:FRLW EoM}
\left[\frac{\partial^2}{\partial \eta^2}- \frac{(d-1)}{\eta} \frac{\partial}{\partial \eta}+ \left(k^2 +\frac{m^2}{H^2\eta^2}  \right)\right]  \phi_\Vvec{k}(\eta)=0\,.
\ee
This differential equation can be cast into the form of Bessel's equation, whose solution can be written as
\beq
\label{eq:modefunction}
\phi_\Vvec{k}(\eta) = f_k(\eta) \hs \alpha_{\Vvec{k}} \,,
\eeq
where $\alpha_{\Vvec{k}}$ are constants that depend on initial conditions, and $f_k(\eta)$ is a mode function.\footnote{The mode function is normalized so that its Wronskian is $f^*\partial_\eta f - f\partial_\eta f^* =i (- H\eta)^{d-1}$. This ensures that the canonically quantized field has the correct commutation relations when $\alpha_\textbf{k}$ is promoted to an annihilation operator.} 
Assuming Bunch--Davies initial conditions, the mode function takes the following explicit form
\be\label{eq: modefunction}
f_k(\eta)= \frac{\sqrt{\pi H^{d-1}}}{2} e^{-i\frac{\pi\nu }{2}}  
(-\eta)^{d/2}H^{(2)}_{\nu}(-k\eta) \,,
\ee
where $H^{(2)}_\nu$ is the Hankel function of the second kind, with 
\beq
\nu \equiv \sqrt{\frac{d^2}{4} - \frac{m^2}{H^2}}\,.
\label{eq:defofnu}
\eeq
Depending on the mass value, the parameter $\nu$ is either purely real or purely imaginary. Unitary representations correspond to the range $-d/2<\nu<d/2$ ({\it complementary series}) and $\nu = i\mu$, with $\mu\in {\mathbb R}$ ({\it principal series}).  Since the system is invariant under the shadow symmetry $\nu \leftrightarrow -\nu$, it is convenient to adopt the convention  $\Re \nu\geq 0$ or $\Im \nu\geq 0$.
The conformally coupled mass $m^2/H^2 = (d^2-1)/4$ corresponds to $\nu = \tfrac{1}{2}$, and it will be useful to define the deviation away from this value as $\dev \equiv \nu-\tfrac{1}{2}$.

\subsection{Wavefunction Coefficients}
\label{sec:WFC}

Cosmological data is encoded in correlation functions. We are interested in correlation functions of the scalar field $\phi(\eta, \Vvec{x})$ at the late-time boundary  $\eta_*=0$. 
These correlation functions can be calculated using the Born rule:
\be
\langle \varphi(\Vvec{x}_1)\cdots \varphi(\Vvec{x}_n)\rangle = \int \mathcal{D}  \varphi \;\varphi(\Vvec{x}_1)\cdots \varphi(\Vvec{x}_n)\left|\Psi \left[\varphi\right]\right|^2\,,
\ee
where $\varphi(\Vvec{x})\equiv \phi(\eta_*,\Vvec{x})$ is the boundary value of the field. In other words, by summing over all possible profiles of $n$ insertions of fields with the weight given by the square of the wavefunction, $|\Psi[\varphi]|^2$, one obtains the $n$-point correlation function. For small fluctuations, the wavefunction can be expanded as
\be
    \Psi[\varphi]=\exp\left[-\sum_{n=2}^\infty\frac{1}{n!} \int \frac{\ud^d k_1}{(2\pi)^d}\cdots \frac{\ud^d k_n}{(2\pi)^d}\,(2\pi)^d \delta(\textbf{k}_1+\dots+\textbf{k}_n)\,\psi_n(\underbar{\textbf{k}}) \,\varphi_{\textbf{k}_1}\cdots \varphi_{\textbf{k}_n}\right] ,
\ee
where $\psi_n$ are the {\it wavefunction coefficients}.  These wavefunction coefficients can be expressed as a sum over Feynman diagrams, and it is these diagrams that we aim to compute.

\vskip 4pt
In momentum space, the $n$-point wavefunction coefficients depend on 
$n$ Euclidean momentum vectors, $\underbar{\Vvec{k}} = \{\Vvec{k}_1,\cdots\hskip -1pt,\Vvec{k}_n\}$.  
We denote their magnitudes by 
$k_a =|\Vvec{k}_a|$, and refer to them as ``energies" in a small abuse of terminology.
Spatial translation invariance of the background spacetime implies momentum conservation at each vertex, so that the momenta form a closed polygon. The shape of this polygon is determined by the lengths of its edges and diagonals, so
wavefunction coefficients can be expressed in terms of external and internal energies.

\subsubsection*{Feynman rules}

The contributions to the wavefunction are computed using the following {\it Feynman rules}:
\begin{itemize}
\item \textit{External lines:} Every external line corresponds to a bulk-to-boundary propagator: 
\ba\label{eq: K def}
\raisebox{-10pt}{
\begin{tikzpicture}[line width=1. pt, scale=2]
\draw[line width=1.pt] (0,0) -- (0.0,0.35);
\draw[lightgray, line width=2.pt] (-0.2,0.35) -- (0.2,0.35);
\draw[fill=black] (0,0) circle (.03cm);
\end{tikzpicture}
} 
\  &\equiv& \ 
K(k,\eta)&&=&&\frac{f_k(\eta)}{f_k(\eta_*)}\,,
\ea
where the mode function $f_k(\eta)$ is a solution to the equation of motion for the field $\phi_{\Vvec{k}}(\eta)$. For Bunch--Davies initial conditions, the mode functions at large $|k\eta|$ (i.e.~at early times or on small scales) reduce to their Minkowski space counterparts, $ f_k(\eta)\sim e^{+ik\eta}$. 

\item \textit{Internal lines:} Every internal line corresponds to a bulk-to-bulk propagator: 
\be
\begin{aligned}
\includegraphics[valign=c,scale=0.6]{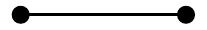}  \ \ \ \equiv \  \ 
G(k,\eta, \eta')  &\ \ = \ \ \ G_\text{F}(k,\eta,\eta')\ \, -\  \ \, G_\text{D}(k,\eta,\eta')  \\ 
&\ \ = \ \ 
\includegraphics[valign=c,scale=0.6]{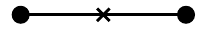} \  - \  
\includegraphics[valign=c,scale=0.6]{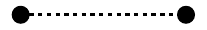} 
\end{aligned}
\ee
which is the difference between the Feynman propagator and a disconnected (non-time-ordered) propagator: 
\ba\label{eq: Fand D}
\includegraphics[valign=c,scale=0.6]{Figures/Equations/CrossedPropagator.pdf}  \  &\equiv \ \ \
G_\text{F}(k,\eta,\eta')&&=&& {f}^*_k(\eta) {f}_k(\eta')\hs \theta(\eta-\eta')+   {f}_k(\eta) {f}^*_k(\eta') \hs \theta(\eta'-\eta)\,,\\
\includegraphics[valign=c,scale=0.6]{Figures/Equations/DashedPropagator.pdf}  \  &\equiv \ \ \
G_\text{D}(k,\eta,\eta')&&=&&\frac{{f}^*_k(\eta_*)}{f_k(\eta_*)}f_k(\eta)f_k(\eta')\,.
\ea
Note that the relative normalization is chosen so that the bulk-to-bulk propagator vanishes on the late-time boundary, $G(k,\eta,\eta_*)  
=0$\,. 
\item  \textit{Vertices:} To each vertex $j$, we assign a (time-dependent) coupling  $i \lambda_{j}\, a(\eta_j)^{d+1}$ and  integrate over the interaction time $\eta_j$.
\item \textit{Loop integrals:} For diagrams involving loops, we integrate over the loop momenta.
\end{itemize}
As we will see, the mode functions of massive fields are Hankel functions so the wavefunction coefficients become nested integrals over products of these Hankel functions. Performing these integrals explicitly is challenging, so we will instead derive differential equations for an associated basis of master integrals.

\subsubsection*{Simplifying the wavefunction}

It will be useful to extract some overall factors and write the mode function  in \eqref{eq: modefunction} as\footnote{Here, we have used that $\nu$ is either purely real or purely imaginary, so that
\beq
\left(e^{-i\frac{\pi\nu}{2}} H_\nu^{(2)}(-k\eta)\right)^* = e^{i\frac{\pi\nu}{2}} H_\nu^{(1)}(-k\eta)\,.
\eeq
Moreover, the prefactor and $k$-scaling of the functions $g_k(\eta)$ and $\bar g_k(\eta)$ are chosen for later convenience.} 
\beq
\begin{aligned}
f_k(\eta) &= i {H^\frac{d-1}{2}} \hs e^{-i\frac{\pi\nu}{2}} (-\eta)^{d/2-\nu} \, \frac{g_k(\eta)}{\sqrt{2k}}\,,\\[4pt]
f_k^*(\eta) &= -i {H^\frac{d-1}{2}} \hs e^{+i\frac{\pi\nu}{2}} (-\eta)^{d/2-\nu} \, \frac{\bar g_k(\eta)}{\sqrt{2k}} \,,
\end{aligned}
\label{equ:f+g}
\eeq
where 
we defined the new functions 
\beq
\begin{aligned}
g_k(\eta) &\equiv -i \sqrt{\frac{\pi}{2}} \,k^{-\dev}  (-k\eta)^\nu H_\nu^{(2)}(-k\eta) \,,\\
\bar g_k(\eta) &\equiv i \sqrt{\frac{\pi}{2}} \, k^{-\dev} (-k\eta)^\nu H_\nu^{(1)}(-k\eta) \,.
\end{aligned}
\label{equ:g}
\eeq
Note that $\bar g_k(\eta) \neq g_k^*(\eta)$. Substituting (\ref{equ:f+g}) into (\ref{eq: K def}) and (\ref{eq: Fand D}), we find that the propagators are given by 
\beq
\label{eq: F C N}
\begin{aligned}
K(k,\eta) &= N_k\hs (-\eta)^{d/2-\nu} \,\hat K(k,\eta)\,, \qquad\qquad \  N_k \equiv   \frac{i}{(-\eta_*)^{d/2}  H_\nu^{(2)}(-k\eta_*)}\sqrt{\frac{2}{\pi k}}\,,\\[4pt]
G_{\rm F}(k,\eta,\eta') &= C_k^{({\rm F})}\hs (\eta \eta')^{d/2-\nu} \,\hat G_{\rm F}(k,\eta,\eta') \,, \qquad\,  C_k^{({\rm F})} \equiv \frac{H^{d-1}}{2k} \,,\\[4pt]
G_{\rm D}(k,\eta,\eta') &=  C_k^{({\rm D})}\hs  (\eta \eta')^{d/2-\nu} \,\hat G_{\rm D}(k,\eta,\eta')\,, \qquad   C_k^{({\rm D})} \equiv -\frac{H^{d-1}}{2k}  \hs \frac{H_\nu^{(1)}(-k\eta_*)}{H_\nu^{(2)}(-k\eta_*)}\,,
\end{aligned}
\eeq
where we have defined the simplified propagators
\beq
\begin{aligned}
\hat K(k,\eta) &\equiv g_k(\eta)\,,\\
\hat G_{\rm F}(k,\eta,\eta') &\equiv \bar{g}_k(\eta) {g}_k(\eta')\hs \theta(\eta-\eta')+   {g}_k(\eta) \bar{g}_k(\eta') \hs \theta(\eta'-\eta) \,,\\
\hat G_{\rm D}(k,\eta,\eta') &\equiv g_k(\eta) g_k(\eta')\,. 
\end{aligned}
\label{equ:scale-prop}
\eeq
We will compute a rescaled wavefunction $\hat \psi$ using these rescaled propagators $\hat K$, $\hat G_{\rm F}$ and $\hat G_{\rm D}$, because they will be somewhat simpler than the original objects. 
To do this, we first absorb the additional factors of conformal time in $K, G$ relative to $\hat K, \hat G$ into
 modified vertex factors: 
\beq
i \lambda_{i}\,a(\eta_i)^{d+1} \prod_{j=1}^{p_i} (-\eta_i)^{d/2-\nu_j}  = i \lambda_i H^{-(d+1)} (-\eta_i)^{-(1+\alpha_i + \dev_i)}\,,
\label{equ:scale-vertex}
\eeq
where we have defined the parameters
\beq
\label{eq:paramsaxi}
\alpha_i \equiv d - \frac{p_i}{2}(d-1)  \quad {\rm and} \quad \dev_i  \equiv \sum_{j=1}^{p_i} \left(\nu_j - \frac{1}{2} \right)  ,
\eeq
with $p_i$ being the number of lines emanating from vertex $i$.
To derive the wavefunction coefficients coming from arbitrary graphs, we simply apply the Feynman rules of Section~\ref{sec:WFC}, but with the rescaled propagators~\eqref{equ:scale-prop} and the modified vertex factors~\eqref{equ:scale-vertex}. 
It is convenient to decompose the full answer into pieces and grade the various contributions by how many disconnected propagators appear in a given piece.
We denote these modified wavefunction coefficients by $\hat \psi^{(r)}$, where $r$ counts the number of disconnected propagators.
The original wavefunction of interest can then be reassembled as
\beq\label{eq:psi to psihat}
\psi \,=\,\sum_{r=0}^{n_I} \left( \prod_{e=1}^{n_E} N_{k_e}\prod_{i=1}^{n_I-r} C_{k_{I_i}}^{({\rm F})}\prod_{j=1}^{r} C_{k_{I_j}}^{({\rm D})}\,\hat \psi^{(r)} \right) ,
\eeq
where $n_E$ and $n_I$ are the number of external and internal lines, respectively, and $k_{I_i}$ are the internal momenta corresponding to either the Feynman (F) or disconnected (D) propagators. 
In the rest of the paper, we omit the hat on $\hat\psi$, with the implicit understanding that we are always discussing the rescaled wavefunction.

\subsection{Twisted Integrals}
\label{sec:twistedint}

Applying the Feynman rules described in the previous section leads to a representation of the wavefunction coefficients as nested time integrals over products of Hankel functions. It is not obvious what the general features of such integrals are, or what kind of special functions they should evaluate to. 
To gain further insight, it is therefore useful to recast the integrals of interest in a form where a broader range of mathematical tools can be applied.

\vskip4pt
Concretely, we would like to cast the time integrals of interest in a rational form, so that some results from geometry and combinatorics can be applied to them. To do so, we use the fact that Hankel functions admit an integral representation of the form
\be
H_\nu^{(2)}(x) = \frac{i}{\sqrt{\pi}}\frac{2^{1+\nu}x^{-\nu}}{\Gamma[\tfrac{1}{2}-\nu]}\int_1^{\infty}\ud t \,e^{-ixt}(t^2-1)^{-\nu-\frac{1}{2}}\, ,
\ee
which expresses them as a twisted exponential integral. 
The benefit of this representation is that it takes the same form regardless of the order $\nu$ of the Hankel function, which only enters via the twist factor $(t^2-1)^{-\nu-\frac{1}{2}}$.
By substituting this expression (and its analogue for $H^{(1)}_\nu(x)$) into~\eqref{equ:scale-prop}, and representing the temporal vertex factors in Fourier space, the time integrals only involve products of plane waves, and can be done explicitly. 

\vskip4pt
After performing the time integrals, the wavefunction is expressed as an integral of a {\it universal integrand} times twist factors. As a concrete example, we consider the four-point function of conformally coupled scalar fields exchanging a field of general mass:
\be
\begin{aligned}
\psi&=
\scalebox{1.0}{
\raisebox{-24pt}{
\begin{tikzpicture}[line width=1. pt, scale=2]
\draw[line width=1.pt,lightgray] (0,0) -- (-0.25,0.55);
\draw[line width=1.pt,lightgray] (0,0) -- (0.25,0.55);
\draw[line width=1.pt,lightgray] (1,0) -- (0.75,0.55);
\draw[line width=1.pt,lightgray] (1,0) -- (1.25,0.55);
\draw[lightgray, line width=2.pt] (-0.5,0.55) -- (1.5,0.55);
\draw[line width=2.pt,darkgray] (0,0) -- (1,0);
\draw[fill=black] (0,0) circle (.03cm);
\draw[fill=black] (1,0) circle (.03cm);
\draw[fill=white] (.5,0) circle (.0275cm);
\node[scale=1] at (0,-.15) {$X_1$};
\node[scale=1] at (1,-.15) {$X_2$};
\node[scale=1] at (.5,-.15) {$Y$};
\node[scale=.85] at (.25,.12) {$t_{L}$};
\node[scale=.85] at (.75,.12) {$t_{R}$};
\end{tikzpicture}
} }\, .
\end{aligned}
\ee
In Appendix~\ref{app:combin}, we show that this
has the integral representation
\beq
\psi = \int \bigg(\prod_{a=1,2} \ud x_a \,x_a^\xi \prod_{b=L,R} \ud t_b \,(t_b^2-1)^{-\nu-\frac{1}{2}}  \bigg)\,\uppsi(X_a+x_a,t_b Y)\,,
\label{eq:massiveintegralmain}
\eeq
where the rational function appearing inside the integral is $\uppsi \equiv \uppsi_{\rm F} - A \,\uppsi_{\rm D}$, with
\beq
\begin{aligned}
\uppsi_{\rm F} &\equiv  \frac{1}{2Y} \bigg[\frac{1}{X_1+X_2+(t_L-t_R)Y}\frac{1}{X_1+t_LY} +\frac{1}{X_1+X_2+(t_R-t_L)Y}\frac{1}{X_2+t_RY}\bigg]\,,\\
\uppsi_{\rm D} &\equiv   \frac{1}{2Y}\, \frac{1}{X_1+t_LY}\frac{1}{X_2+t_RY}\,.
\end{aligned}
\eeq
Note that, for $t_L = t_R \equiv 1$, this function reduces to the flat-space wavefunction. The relative normalization of the time-ordered (F) and disconnected (D) contributions is given by the parameter $A$. For heavy fields, this is a complex number, while, for light fields, it becomes $A=1$.
The universal integrand can also be written as
\be
\uppsi \equiv \frac{1}{2Y}\frac{(1-A)(X_1+X_2)^2+2(t_L X_1+t_R X_2)Y+(1+A)(t_L-t_R)^2Y^2}{(X_1+t_LY)(X_2+t_RY)(X_1+X_2+(t_L-t_R)Y)(X_1+X_2+(t_R-t_L)Y)}\,,
\label{eq:mainpsiintegrand}
\ee
which makes its singularity structure manifest.
For more general external states (e.g.~scalar fields of different masses), the integrand in~\eqref{eq:massiveintegralmain} would remain the same; the only difference would be the way that we would integrate it.

\vskip4pt
The integrand~\eqref{eq:mainpsiintegrand}, together with its generalizations to arbitrary bulk processes, admits an elegant definition in terms of combinatorial tubings of a decorated graph, with suggestive connections to underlying geometric structures. These features are briefly described in Appendix~\ref{app:combin}. However, for our present purposes, the mere existence of a twisted integral representation like~\eqref{eq:massiveintegralmain} is the most relevant fact. This structure of the integral means that we can apply the machinery of twisted cohomology~\cite{Aomoto:2011ggg,Gasparotto:2024bku}.

\subsubsection*{Integral basis}
The geometric viewpoint implies
in particular that the wavefunction $\psi$ is a member of a finite-dimensional vector space of integrals. This fact guarantees that it will satisfy a first-order matrix differential equation as a function of the external kinematics. The natural basis integrals have the same singularities as~\eqref{eq:massiveintegralmain}, but generalized to allow the twisted lines to have different parameters. We therefore consider integrals of the form\footnote{To write compact expressions like this, we have color-coded the signs in $\psi^{(\pmr\pmb)}$. This will be even more relevant below. We hope that this does not cause any confusion in the black-and-white printing of this paper.}
\beq
\psi^{(\pmr\pmb)} \equiv  \frac{1}{4}\int \bigg(\prod_{a=1,2} \ud x_a \,x_a^\xi \prod_{b=L,R} \ud t_b \,(t_b^2-1)^{-\nu-\frac{1}{2}}  \bigg) \, (1\,\mpr\, t_L)(1\,\pmb\, t_R)\,\uppsi^{(2)} (X_a+x_a,t_b Y)\,.
\label{eq:FintegralsX}
\eeq
These integrands shift the twists of the $(1\pm t_{L,R})$ singularities by one unit, and so these integrals serve as a natural basis for this integral family.
The integrals~\eqref{eq:FintegralsX} arise from replacing $g_k(\eta)$ appearing in the propagator \eqref{equ:scale-prop} with
\ba \label{eq:hrational}
g_k(\eta) \equiv h^+_k(\eta)+h^-_k(\eta)\,, \quad {\rm where} \quad
h_k^{\pm}(\eta)&\equiv  \frac{2^{\dev}}{\Gamma[-\dev]\,k^\dev} \int_1^\infty \ud t \, e^{i k\eta t} \frac{1\,\pm\, t}{(t^2-1)^{1+\dev}}\, , 
\ea
and similarly for $\bar g_k(\eta) \equiv \bar h^+_k(\eta)+ \bar h^-_k(\eta)$. The benefit of this decomposition is that the functions $h_k^{\pm}(\eta)$ satisfy a {\it first-order} differential equation in $\eta$, which leads to a simple first-order system for the basis functions $\psi^{(\pm \pm)}$. One recovers the original wavefunction by summing over all these basis elements as $\psi =\psi^{(++)}+\psi^{(--)}+\psi^{(+-)}+\psi^{(-+)}$.

\section{Time for Differential Equations}
\label{sec:DE}

Although the twisted integral representation is useful for establishing the {\it existence} of a first-order system of differential equations, the explicit derivation of these equations is most transparent in the time-integral representation.
In this section, we show how bulk perturbation theory naturally organizes the basis of master integrals and  derive the differential equations that these integrals satisfy.
In doing so, we will find combinatorial patterns in the equations that we explore more fully in Section~\ref{sec:Flow}.

\subsection{Master Integrals}
\label{sec:masterI}

Perturbation theory in terms of {\it time} provides a natural organization of the basis of possible functions that arise in correlators. In~\cite{Baumann:2025qjx,Glew:2025ypb}, this organization was used to derive a first-order system of differential equations satisfied by the wavefunction of conformally coupled scalars in power-law cosmologies. (See also~\cite{De:2023xue,He:2024olr,De:2024zic,Glew:2025ypb,Capuano:2025ehm}.)
In the present context, the existence of a first-order system is somewhat more surprising. In contrast to the conformal case
(where the mode functions are plane waves), the propagators~\eqref{equ:scale-prop} are assembled from mode functions that satisfy the {\it second-order} Klein--Gordon equation. 

\vskip4pt
We can connect the twisted-integral and time-integral perspectives 
by adding the time derivative of the mode function as a separate basis element. The time derivative of this function will then reduce back to the original mode function via the equation of motion. In particular, it is natural to consider the functions~\cite{Chen:2023iix,Chen:2024glu}
\be
h^\pm(x) = x^\nu H_\nu^{(2)}(x) \pm i\partial_x[x^\nu H_\nu^{(2)}(x)]= \frac{i}{\sqrt{\pi}}\frac{2^{1+\nu}}{\Gamma[\tfrac{1}{2}-\nu]}\int_1^{\infty}\ud t \,e^{-ixt}\frac{1\pm t}{(t^2-1)^{\nu+\frac{1}{2}}} \,,
\ee
where in the last equality we have written this combination as a twisted integral, which is precisely~\eqref{eq:hrational}. The functions $h^\pm$ satisfy a
system of first-order equations
\be
\partial_x h^\pm(x)  \pm i h^\pm(x)\pm\frac{\nu-\tfrac{1}{2}}{x}\Big[h^-(x)-h^+(x)\Big] = 0\,.
\label{eq:hankel1storder}
\ee
This suggests that we should try to  build the basis of bulk integrals in terms of these combinations. The time-ordered structure of perturbation theory will then organize the differential system.

\vskip4pt
Concretely, 
we first split each mode function $g_k(\eta)$ appearing in~\eqref{equ:g} 
into two auxiliary functions 
\ba\label{eq: h def}
g_k(\eta) \equiv h^+_k(\eta)+h^-_k(\eta)\,, \quad {\rm where} \quad
h_k^{\pm}(\eta)&\equiv  \frac{1}{2} \left(g_k(\eta)\mp \frac{i}{k} \partial_\eta g_{k}(\eta)\right) , 
\ea
and similarly for $\bar g_k(\eta) \equiv \bar h^+_k(\eta)+ \bar h^-_k(\eta)$.
Up to a $k$-dependent factor, the functions $h_k^\pm$ are equal to the functions $h^\pm$ in~\eqref{eq:hankel1storder}, so that they satisfy the following first-order equations\hs\footnote{Note that in~\eqref{eq: h dif k} the term proportional to $\dev$ is purely off-diagonal, which is a consequence of the $k^{-\dev}$ factor in \eqref{equ:g}. This will simplify the resulting differential equations.}
\begin{align}
\label{eq: h dif}
\partial_\eta h_k^\pm(\eta)&= \pm i k  h_k^\pm (\eta) \pm \frac{\dev}{\eta}\left[h_k^+(\eta)-h_k^-(\eta) \right] , \\
\label{eq: h dif k}
\partial_k h_k^\pm(\eta) &= \pm i \eta  h_k^\pm (\eta) - \frac{\dev}{k}h_k^\mp(\eta)\, ,
\end{align}
where again $\dev =  \nu - \frac{1}{2}$. Identical equations hold for the functions $\bar h_k^\pm(\eta)$. We can then replace $g_k(\eta)$  in the propagators~\eqref{equ:scale-prop} with $h^\pm_k(\eta)$, which effectively defines new propagators $\hat K^{\pm}$, $\hat G_{\rm F}^{\pm\pm}$ and $\hat G_{\rm D}^{\pm\pm}$. We then use these propagators in the Feynman rules.
This separates the wavefunction $\psi$ into a set of functions 
$\psi^{(\pm \pm \cdots \pm)}$ assembled from strings of the functions $h_k^{\pm}(\eta)$ and $\bar h_k^{\pm}(\eta)$.  Since $h^+_k+h^-_k = g_k$, adding up all possible assignments of $\pm$ in $\psi^{(\pm \pm \cdots \pm)}$ reproduces the original function of interest $\psi$.

\vskip4pt
By taking time derivatives of the integrand and using~\eqref{eq: h dif} we will derive first-order 
differential equations for the functions $\psi^{(\pm \pm \cdots \pm)}$, along with functions obtained by collapsing internal lines.
As in~\cite{Baumann:2025qjx}, the equations separate into independent sectors graded by the number of disconnected propagators.  Expressed in terms of the functions $h^\pm_k(\eta)$, each internal massive propagator leads to a four-fold multiplicity of basis functions, while each external line leads to an additional two-fold multiplicity. As an example, a 
Feynman graph with a single massive exchange and conformally coupled external fields (with $\dev = 0$) leads to four functions $\{\psi^{(++)}\,, \psi^{(+-)}\,, \psi^{(-+)}\,, \psi^{(--)}\}$ in each of the time-ordered sector and the disconnected sector. In the time-ordered sector, the differential equations for $\psi^{(+-)}$ and $\psi^{(-+)}$ contain an additional source function $J^{(c)}$ corresponding to a contact solution arising from the collapse of the propagator, so that there are $9$ functions total.\footnote{Note that this is a substantial expansion of the size of the basis relative to~\cite{Baumann:2025qjx}. We can understand this by noting that in the conformally coupled limit, half of the $h^\pm$ functions vanish, and the other ones simplify
\be
\label{eq: h cc}
h^-_k(\eta) =\bar{h}^+_k(\eta)= 0\,,\hspace{1cm}\,h^+_k(\eta) =\left(\bar{h}^-_k(\eta)\right)^*=  e^{i k \eta}\,.
\ee
This reduces the size of the basis in the conformally coupled limit, and the corresponding differential equations and diagrammatic rules simplify to those of~\cite{Baumann:2025qjx}.}

\vskip 4pt
In summary, to deal with time integrals over mode functions, we will increase the number of integrals by replacing each mode function with two auxiliary functions $h^\pm$. The benefit is that these satisfy a system of first-order differential equations. The desired original wavefunction coefficients can be recovered by adding up all the pieces of the new basis. In what follows, we illustrate this method by working out a few concrete examples.

\subsection{Contact Diagram}
\label{ssec:contact}

We first study
 a three-point contact diagram involving one generic massive leg and two conformally coupled legs. This is perhaps the simplest process involving a field of general mass. The wavefunction coefficient of interest is given by the integral
\beq
\psi_c \ \equiv \  \raisebox{-16pt}{
\begin{tikzpicture}[line width=1. pt, scale=2]
\draw[lightgray, line width=1.pt] (0,0) -- (0.0,0.55);
\draw[lightgray, line width=1.pt] (0,0) -- (0.25,0.55);
\draw[darkgray, line width=2.pt] (0,0) -- (-0.25,0.55);
\draw[lightgray, line width=2.pt] (-0.5,0.55) -- (0.5,0.55);
\draw[fill=black] (0,0) circle (.03cm);
\end{tikzpicture}
} 
=  \int \frac{\ud\eta}{ (-\eta)^{1+\alpha + \dev}}\, e^{iX\eta} \,  g_k (\eta)\,,
\label{eq: psi def contact}
\eeq
where $k \equiv k_1$ and $X\equiv k_2+k_3$. To avoid clutter, we have dropped the overall normalization $N_c = i \lambda$. In this particular case, the time integral can be computed directly, and evaluates to a hypergeometric function. However, it is convenient to use this simplified setting to illustrate how integrals like this can be computed using differential equations.  

\vskip 4pt
Recall that we can decompose $g_k(\eta)= h^+_k(\eta)+h^-_k(\eta)$, which means that~\eqref{eq: psi def contact} can be split into two
basis functions: 
\ba
\label{eq:contact3pts}
\psi^+_c&= \int \frac{\ud\eta}{ (-\eta)^{1+\alpha+\dev}}\, e^{iX\eta} \,  h^+_k (\eta)\,,\\
\psi^-_c&=   \int \frac{\ud\eta}{ (-\eta)^{1+\alpha+\dev}}\, e^{iX\eta} \,  h^-_k (\eta)\,,
\ea
where $\psi_c= \psi^+_c+\psi^-_c$.
In the inset below, we will show that these basis integrals satisfy the following differential equations: 
\be\label{eq: contact final}
\ud \psi^\pm_c =\alpha\, \dl (X\pm k)\,\psi^\pm_c \, + \, \dev \Big[\dl (X\pm k)-\dl k  \Big] \psi^\mp_c\,,
\ee
where ${\rm d} = {\rm d} X \hs \partial_X + {\rm d}k \hs \partial_k$ is the differential with respect to kinematic variables. Notice that only dlog-forms of certain combinations of kinematic variables (called ``letters") appear in the differential. 
It is also illuminating to write the differential equation as a matrix equation
\be\label{eq: dI AI}
\ud \begin{bmatrix}
\psi^+_c\\
\psi^-_c
\end{bmatrix} = A \cdot \begin{bmatrix}
\psi^+_c\\
\psi^-_c
\end{bmatrix} \,,
\ee
where the ``connection matrix" is
\be\label{eq:amatrix contact} 
A = \alpha \begin{bmatrix}
 \dl (X+ k)& 0 \\
0 &  \dl (X- k)  
\end{bmatrix}
+ \dev \begin{bmatrix}
0  &\  \dl \left(\frac{X+k}{k}\right) \\
\ \dl \left(\frac{X- k}{k}\right) &0  
\end{bmatrix} .
\ee
We see that $\dev \ne 0$ induces off-diagonal mixings between the basis functions. 

\vskip 6pt
\begin{eBox3}
{\bf Derivation}: \ The derivation of equation~\reef{eq: contact final} proceeds in two steps.
\begin{itemize}
\item \textit{Shifted vertex factor}:  First, we differentiate $\psi_c^\pm$ with respect to $X$ and $k$.
These derivatives can be brought inside the integrals and act on the mode functions. In cases where the momentum factors multiply $\eta$, these derivatives effectively raise the power of conformal time in the integral. This can also be thought of as shifting the vertex factor~$\alpha$. For this reason, we make the dependence on $\alpha$ explicit by writing $ \psi^\pm_c(\alpha)$. First, we consider the derivative with respect to~$X$:
\begin{align}\label{eq: x der to alpha-1}
\partial_X \psi^\pm_c(\alpha) &=  \int \frac{\ud\eta}{ (-\eta)^{1+\alpha+\dev}}\,  \partial_X\big[e^{iX\eta} \big]\,  h^\pm_k (\eta) \nonumber\\
&= -i  \int \frac{\ud\eta}{ (-\eta)^{\alpha+\dev}}\, e^{iX\eta} \,  h^\pm_k (\eta) \nonumber \\
&= -i \psi^\pm_c(\alpha-1)\,.
\end{align}
To compute $\partial_k \psi^\pm_c(\alpha)$, we use \eqref{eq: h dif k}, so that
\begin{align}
\partial_k \psi^\pm_c(\alpha) &=  \int \frac{\ud\eta}{ (-\eta)^{1+\alpha+\dev}}\, e^{iX\eta} \,  \partial_k\big[h^\pm_k (\eta)\big] \nonumber \\
&=\mp i \psi^\pm_c(\alpha-1)-\frac{\dev}{k}\psi^\mp_c(\alpha)\, . \label{eq: k der contact  to alpha-1}
\end{align}
To find a closed system of equations, we need to find relations between the basis elements $\psi^\pm_c(\alpha)$ and their counterparts $\psi^\pm_c(\alpha-1)$ with a shifted vertex factor. Such relations follow from integration-by-parts identities.

\item \textit{Integration-by-parts}: Since the integral of a total derivative vanishes, we have
\begin{align}
0&=\int \ud\eta\;\partial_{\eta}\big[ { (-\eta)^{-\alpha-\dev}}  e^{iX\eta}   \,h^\pm_k (\eta)\big] \nonumber \\
&=(\alpha+\dev) \, \psi^\pm_c(\alpha) + \int\frac{ \ud\eta}{(-\eta)^{\alpha+\dev}} \partial_{\eta}\big[e^{iX\eta}\big]\, h^\pm_k(\eta) +  \int\frac{ \ud\eta}{(-\eta)^{\alpha+\dev}}e^{iX\eta} \partial_{\eta} \big[ h^\pm_k (\eta)\big] \nonumber \\
&=(\alpha+\dev) \, \psi^\pm_c(\alpha)+ iX \psi^\pm_c(\alpha-1) \,\pm ik \psi^\pm_c(\alpha-1) - {\dev}\,\Big[\psi^\pm_c(\alpha) - \psi^\mp_c (\alpha)\Big]\,,
\label{eq: IBP contact}
\end{align}
where we used~\reef{eq: h dif}  in the last equality. 
We then use the identity~\eqref{eq: IBP contact} to replace $\psi^\pm_c(\alpha-1)$ in~\reef{eq: x der to alpha-1} and \reef{eq: k der contact  to alpha-1}, so that
\ba\label{eq: contact der}
\partial_X \psi^\pm_c& =\frac{1}{X\pm k} \Big[\alpha\,\psi^\pm_c+ \dev\,\psi^\mp_c \Big]\,,\\
\partial_k \psi^\pm_c& =\pm\frac{1}{X\pm k} \Big[\alpha\,\psi^\pm_c +\dev\,\psi^\mp_c \Big] -\frac{\dev}{k}\,\psi^\mp_c\, ,
\ea
which leads to \reef{eq: contact final} in terms of differential forms.
\end{itemize}
\end{eBox3}

\vskip 6pt
The system of equations~\eqref{eq: dI AI} is two-dimensional, which implies that each of the basis elements (and their sum) satisfy a second-order differential equation. Indeed, combining the equations for $\psi_c^\pm$, it is easy to show that $\psi_c = \psi_c^+ + \psi_c^-$ satisfies 
\be\label{eq: contact diff eq}
\Big[\left(X^2- k^2\right)\partial_{X}^2 +2 X(1-\alpha)\partial_{X}- (\alpha+\dev)(1-\alpha+\dev) \Big] \psi_{c} = 0\,,
\ee
which has the following solution 
\be
\label{eq:contact3pt}
\psi_c= A(\dev,\alpha)\,(2k)^{\alpha}\, {}_2F_1\bigg[\begin{array}{c}
-\alpha-\dev\ ,\  1-\alpha+\dev\\[-1pt]
1-\alpha
\end{array}\bigg\rvert \, \frac{k-X}{2k}\,\bigg] \,,
\ee
where we imposed regularity of the wavefunction at $k=X$, and the $k^\alpha$ prefactor follows from dilation invariance.\footnote{In checking this, recall that $N_k\sim k^\dev$, and the physical wavefunction has dimension $\nu+1-d/2$.}  The overall  normalization 
$A(\dev,\alpha)$ 
can be fixed by directly evaluating the integral in~\reef{eq: psi def contact} for a specific value of $k$.

\subsection{Single Exchange}
\label{sec:singleexc}

As a more complex example, we next consider 
the tree-level exchange of a massive particle ($\dev\ne 0$) with an arbitrary number of  conformally coupled external fields ($\dev=0$). The wavefunction coefficient for this process is computed from the Feynman diagram
\begin{equation} 
\psi_{e}^{(2)}  \ \equiv \ 
\scalebox{1.0}{
 \raisebox{-26pt}{
\begin{tikzpicture}[line width=1. pt, scale=2]
\draw[line width=2.pt,darkgray] (0,0) -- (1,0);
\draw[line width=1.pt,lightgray] (0,0) -- (-0.25,0.55);
\draw[line width=1.pt,lightgray] (0,0) -- (0,0.55);
\draw[line width=1.pt,lightgray] (0,0) -- (0.25,0.55);
\draw[line width=1.pt,lightgray] (1,0) -- (0.75,0.55);
\draw[line width=1.pt,lightgray] (1,0) -- (1,0.55);
\draw[line width=1.pt,lightgray] (1,0) -- (1.25,0.55);
\draw[lightgray, line width=2.pt] (-0.5,0.55) -- (1.5,0.55);
\draw[fill=Red,Red] (0,0) circle (.03cm);
\draw[fill=Blue,Blue] (1,0) circle (.03cm);
\node[scale=1] at (0,-.15) {$X_1$};
\node[scale=1] at (1,-.15) {$X_2$};
\node[scale=1] at (0.5,-.12) {$Y$};
\end{tikzpicture}
} 
} \, .
\end{equation}
Since the bulk-to-bulk propagator has two terms, the wavefunction coefficient naturally splits into two pieces: 
\be
\psi_{e}^{(2)} \,=\, \psi_{\BtoB}  - \psi_{\BtoBDis}\, ,
\ee
where each individual piece has the time integral representation
\ba\label{eq: psi plus psi}
\psi_{\BtoB} &\equiv  \int \frac{\ud\eta_1\ud\eta_2}{(-\eta_1)^{1+\alpha_1+\dev}(-\eta_2)^{1+\alpha_2+\dev}}\,e^{iX_1\eta_1} e^{iX_2\eta_2} \,\hat G_\text{F}(Y,\eta_1,\eta_2) \,, \\
\psi_{\BtoBDis} &\equiv   \int \frac{\ud\eta_1\ud\eta_2}{(-\eta_1)^{1+\alpha_1+\dev}(-\eta_2)^{1+\alpha_2+\dev}}\,e^{iX_1\eta_1} e^{iX_2\eta_2}\, \hat G_\text{D}(Y,\eta_1,\eta_2)\,.
\ea
Here, $Y$ denotes the exchange energy and $X_i = \sum_a k_a$ is the total energy flowing into vertex $i$.  
To avoid clutter, we have dropped the overall normalizations. 
As above, we define the independent basis elements 
\begin{align}\label{eq:pm psi 2site}
\psi^{(\pmr\pmb)}_{\BtoB} &=  \int \ud\eta_1\ud\eta_2 \,\frac{e^{iX_1\eta_1}e^{iX_2\eta_2} }{(-\eta_1)^{1+\alpha_1+\dev}(-\eta_2)^{1+\alpha_2+\dev}}  \Big[ \bar h^{\pmr}_Y (\eta_1)  h^{\pmb}_Y (\eta_2)\, \theta_{12} +  h^{\pmr}_Y (\eta_1)  \bar h^{\pmb}_Y (\eta_2)\, \theta_{21} \Big]\, , \\
\psi^{(\pmr\pmb)}_{\BtoBDis}&=    \int\frac{\ud\eta_1}{(-\eta_1)^{1+\alpha_1+\dev}}\,e^{iX_1\eta_1}h_Y^{\pmr}(\eta_1) \int \frac{\ud\eta_2}{(-\eta_2)^{1+\alpha_2+\dev}}\,e^{iX_2\eta_2}h_Y^{\pmb}(\eta_2) \,,
\end{align}
where $\theta_{ij} \equiv \theta(\eta_i-\eta_j)$ is the Heaviside theta function.
As its name suggests, the disconnected piece factorizes into two separate integrals, 
\be
\psi^{(\pmr\pmb)}_{\BtoBDis}= \psi^{\pmr}_c(\alpha_1) \psi^{\pmb}_c(\alpha_2)\,,
\ee
where each factor corresponds to the contact diagram~\eqref{eq:contact3pts}. Both the connected and the disconnected pieces satisfy differential equations.
\begin{itemize}
\item The differential equation for the disconnected basis elements is
\ba\label{eq: exchange disc dif}
\ud \psi^{(\pmr\pmb)}_{\BtoBDis} &= \Big[\alpha_1 \ell^{\pmr}_1 + \alpha_2 \ell^{\pmb}_2\Big]\, \psi^{(\pmr\pmb)}_{\BtoBDis} \\
&\quad+ \dev (\ell^{\pmr}_1-\ell_0) \, \psi^{(\mpr\pmb)}_{\BtoBDis}  + \dev (\ell^{\pmb}_2-\ell_0) \, \psi^{(\pmr\mpb)}_{\BtoBDis}  \,,\\
\ea
where ${\rm d} \equiv \sum_i {\rm d} X_i \partial_{X_i} + {\rm d} Y \partial_Y$ and we defined the dlog-forms of the letters $\ell_i^{\pm}\equiv \dl (X_i\pm Y)$ and $ \ell_0\equiv \dl Y$.
This follows from the 
fact that the disconnected piece is a product of two contact diagrams and each of them satisfies~\reef{eq: contact final}.

\item
The differential equation for the connected basis elements is
\ba\label{eq: exchange time dif}
\ud \psi^{(\pmr\pmb)}_{\BtoB} &= \Big[\alpha_1 \ell^{\pmr}_1 + \alpha_2 \ell^{\pmb}_2\Big]\, \psi^{(\pmr\pmb)}_{\BtoB} \\
&\quad+ \dev (\ell^{\pmr}_1-\ell_0)\,  \psi^{(\mpr\pmb)}_{\BtoB} + \dev (\ell^{\pmb}_2-\ell_0) \, \psi^{(\pmr\mpb)}_{\BtoB} - \frac{\pmr 1 \mpb 1}{2} (\ell_1^{\pmr}-\ell_2^{\pmb})\, J_{\Col}^{(c)}\,.
\ea
Notice that this involves a new source function 
\ba\label{eq: col int}
 J_{\Col}^{(c)} \equiv  \int \frac{\ud\eta}{(-\eta)^{1+\alpha_1+\alpha_2}} \,e^{i(X_1+X_2)\eta} \,.
\ea
This term arises from collapsing the bulk-to-bulk propagator, which produces a contact diagram with conformally coupled external lines. 
The differential of the new source function is
\ba
\label{equ:Collapse}
\ud J_{\Col}^{(c)}&=(\alpha_1+\alpha_2)L\, J_{\Col}^{(c)}\,,\\
\ea
where $L\equiv \dl (X_1+X_2)$. Hence, the system closes with the addition of this function.
\end{itemize} 

\vskip 4pt
It is again useful to group the system of first-order differential equations into matrix form:
\be 
\label{eq:1excmatrixform}
\ud \begin{bmatrix}
 \psi^{(\pr\pb)}_{\BtoB} \\[4pt] 
   \psi^{(\mr\mb)}_{\BtoB} \\[4pt]
  \psi^{(\mr\pb)}_{\BtoB} \\[4pt]  
  \psi^{(\pr\mb)}_{\BtoB} \\[4pt]  
 J_{\Col}^{(c)}
\end{bmatrix} = A \cdot  \begin{bmatrix}
 \psi^{(\pr\pb)}_{\BtoB} \\[4pt] 
   \psi^{(\mr\mb)}_{\BtoB} \\[4pt]
  \psi^{(\mr\pb)}_{\BtoB} \\[4pt]  
  \psi^{(\pr\mb)}_{\BtoB} \\[4pt]  
 J_{\Col}^{(c)}
\end{bmatrix} \,,
\ee
where the connection matrix is 
\beq\label{eq: A of 2site}
\begin{aligned}
A = &  \begin{bmatrix}
\alpha_1\hs \ell_1^+  + \alpha_2\hs \ell_2^+   &\quad 0 & \quad 0 & \quad 0 & \quad 0  \\
0 & \quad \alpha_1 \hs \ell_1^- + \alpha_2 \hs \ell_2^- & \quad 0 & \quad 0 &\quad 0  \\
0 & \quad 0 & \quad \alpha_1\hs \ell_1^-  + \alpha_2\hs \ell_2^+   & \quad 0    &\quad \ell_1^- - \ell_2^+ \\
0 & \quad 0  & \quad 0&\quad \alpha_1\hs \ell_1^+  + \alpha_2\hs \ell_2^-   &\quad   \ell_2^--\ell_1^+ \\
0 &\quad 0 & \quad 0 & \quad 0 & (\alpha_1+\alpha_2) L
\end{bmatrix} \\[4pt]
&+ \dev \begin{bmatrix}
0 &\quad 0&\quad \ell_1^+-\ell_0 & \quad \ell_2^+-\ell_0  & \quad 0  \\
0 &\quad 0&\quad \ell_2^- -\ell_0 & \quad \ell_1^- -\ell_0  & \quad 0  \\
\ell_1^- - \ell_0 & \quad \ell_2^+ -\ell_0 & \quad 0  &\quad 0 &\quad 0 \\
\ell_2^- - \ell_0 & \quad \ell_1^+ -\ell_0 & \quad 0  &\quad 0 &\quad 0 \\
0 &\quad 0 & \quad 0 & \quad 0 & \quad0
\end{bmatrix} .
\end{aligned}
\eeq
We see that the contact solution $J_{\Col}^{(c)}$ appears as a source in the equations for $\psi^{(\mr\pb)}_{\BtoB}$ and $\psi^{(\pr\mb)}_{\BtoB}$, but not for $\psi^{(\pr\pb)}_{\BtoB}$ and $\psi^{(\mr\mb)}_{\BtoB}$. 
When $\xi = 0$, the matrix is upper triangular, but 
$\dev \ne 0$ induces  additional mixings between the basis functions. 
We can reassemble the time-ordered part of the wavefunction as
\ba
 \psi_{\BtoB} &=  \psi^{(\pr\pb)}_{\BtoB} +  \psi^{(\mr\mb)}_{\BtoB} +  \psi^{(\mr\pb)}_{\BtoB} +  \psi^{(\pr\mb)}_{\BtoB} \,. 
\ea
Note that the collapsed term is only a source term in differential equations and does not show up explicitly in $\psi_{\BtoB}$. 

\vskip 4pt
\begin{eBox3}
{\bf Derivation}:\ To derive equations~\eqref{eq: exchange time dif} and~\eqref{equ:Collapse}, we proceed in two steps, following the same logic as in Section~\ref{ssec:contact}.
\begin{itemize}
\item \textit{Shifted vertex factor}: We first take the derivatives with respect to $X_1$ and $X_2$.
This shifts the powers of $\eta_1$ and $\eta_2$, respectively, and we get
\begin{align}\label{eq:X1}
\partial_{X_1} \psi^{(\pmr\pmb)}_{\BtoB} (\alpha_1,\alpha_2) &= -i \psi^{(\pmr\pmb)}_{\BtoB}(\alpha_1-1,\alpha_2)\,, \\
\partial_{X_2} \psi^{(\pmr\pmb)}_{\BtoB} (\alpha_1,\alpha_2) &= -i \psi^{(\pmr\pmb)}_{\BtoB}(\alpha_1,\alpha_2-1)\,.
\label{eq:X2}
\end{align}
 The derivative with respect to the exchange energy acts on the functions $h_Y^\pm(\eta)$. Using~\reef{eq: h dif k}, we get
\ba\label{eq: Y dev exchange}
\partial_Y \psi^{(\pmr\pmb)}_{\BtoB}(\alpha_1,\alpha_2)=& \;\mpr\; i  \psi^{(\pmr\pmb)}_{\BtoB} (\alpha_1-1,\alpha_2) \;\mpb\; i  \psi^{(\pmr\pmb)}_{\BtoB} (\alpha_1,\alpha_2-1) \\
&- \frac{\xi}{Y} \left( \psi^{(\mpr\pmb)}_{\BtoB} (\alpha_1,\alpha_2)+ \psi^{(\pmr\mpb)}_{\BtoB} (\alpha_1,\alpha_2)\right) .
\ea
By substituting~\reef{eq:X1} and \reef{eq:X2}, one can write this as
\ba\label{eq: Y der exchange}
\partial_Y \psi^{(\pmr\pmb)}_{\BtoB} =&\; \pmr\; \partial_{X_1} \psi^{(\pmr\pmb)}_{\BtoB} \; \pmb\;  \partial_{X_2}  \psi^{(\pmr\pmb)}_{\BtoB} - \frac{\xi}{Y} \left( \psi^{(\mpr\pmb)}_{\BtoB} + \psi^{(\pmr\mpb)}_{\BtoB}\right) ,
\ea
where we have suppressed the dependence on $\alpha_i$ because all the functions have the same indices $(\alpha_1,\alpha_2)$.

\item \textit{Integration-by-parts}: To obtain a relation between $\psi^{(\pmr\pmb)}_{\BtoB} (\alpha_1-1,\alpha_2)$ and $\psi^{(\pmr\pmb)}_{\BtoB} (\alpha_1,\alpha_2)$, we consider the total derivative
\be
\begin{aligned}
0= \int &\frac{\ud\eta_2}{(-\eta_2)^{1+\alpha_2+\dev}} \,  e^{iX_2\eta_2}\, \partial_{\eta_1} \bigg[ \frac{\ud\eta_1}{(-\eta_1)^{\alpha_1+\dev}}
e^{iX_1\eta_1} \\
& \times \Big(\bar h^{\pmr}_Y (\eta_1)  h^{\pmb}_Y (\eta_2)\,\theta_{12} +  h^{\pmr}_Y (\eta_1)  \bar h^{\pmb}_Y (\eta_2)\, \theta_{21}  \Big) \bigg]\,,
\end{aligned}
\ee
which leads to 
\ba\label{eq: two-site manipulations}
0=\;& (\alpha_1+\dev)\psi^{(\pmr\pmb)}_{\BtoB} (\alpha_1,\alpha_2) \\
&+i X_1 \psi^{(\pmr\pmb)}_{\BtoB} (\alpha_1-1,\alpha_2)\\
&\;\pmr\;i Y \psi^{(\pmr\pmb)}_{\BtoB} (\alpha_1-1,\alpha_2)- \dev \Big[ \psi^{(\pmr\pmb)}_{\BtoB} (\alpha_1,\alpha_2) - \psi^{(\mpr\pmb)}_{\BtoB} (\alpha_1,\alpha_2)  \Big]\\
&- \half (\pmr 1 \, \mpb 1) \hs J_{\Col}^{(c)}\,,
\ea
where we have defined
\ba
\label{equ:source}
 J_{\Col}^{(c)}&= - \int \frac{\ud\eta}{(-\eta)^{1+\alpha_1+\alpha_2+2\dev}} e^{i(X_1+X_2)\eta}  \Big[\bar h^{+}_Y (\eta)  h^{-}_Y(\eta) -  h^{+}_Y (\eta)  \bar h^{-}_Y (\eta) \Big] \\
 &= \int \frac{\ud\eta}{(-\eta)^{1+\alpha_1+\alpha_2 }} e^{i(X_1+X_2)\eta} \,.
\ea
The new function $J_{\Col}^{(c)}$ arises from the derivative of the Heaviside function, $\partial_{\eta_i} \theta_{ij}=-\partial_{\eta_j} \theta_{ij}= \delta({\eta_i-\eta_j})$, which effectively collapses the internal propagator. In the second line of \eqref{equ:source}, we used the relation
\be\label{eq: wronskian}
h^+_{Y}(\eta) \bar h^-_{Y}(\eta)-\bar h^+_{Y}(\eta)  h^-_{Y}(\eta)=  (-\eta)^{2\dev}\,,
\ee
which follows from the  Wronskian of the Hankel functions. 

\vskip 4pt
Combining \eqref{eq:X1} and \eqref{eq: two-site manipulations}, we get
\be\label{eq:X1 der exchange}
\partial_{X_1} \psi^{(\pmr\pmb)}_{\BtoB} =\frac{1}{X_1\,\pmr\, Y}\left(\alpha_1\, \psi^{(\pmr\pmb)}_{\BtoB}  + \dev \, \psi^{(\mpr\pmb)}_{\BtoB} - \half (\pmr 1\, \mpb 1)\hs J_{\Col}^{(c)} \right) ,
\ee
where we again left the $\alpha_i$ dependence implicit. Similarly, the $X_2$-derivative is
\be\label{eq:X2 der exchange}
\partial_{X_2} \psi^{(\pmr\pmb)}_{\BtoB} =\frac{1}{X_2\,\pmb\, Y}\left(\alpha_2\,\psi^{(\pmr\pmb)}_{\BtoB}  + \dev \, \psi^{(\pmr\mpb)}_{\BtoB} - \half (\mpr 1\, \pmb 1)\hs J_{\Col}^{(c)} \right) .
\ee
The $Y$-derivatives can then be found using~\reef{eq: Y der exchange}.
Combining these partial derivatives, we obtain equation~\eqref{eq: exchange time dif} in terms of the total kinematic differential.

\vskip 4pt
Finally, the derivatives of the collapsed source term can be derived directly from its definition~\eqref{equ:source}:
\ba
\partial_{X_1} J_{\Col}^{(c)}=\partial_{X_2} J_{\Col}^{(c)}&= \frac{\alpha_1+\alpha_2}{X_1+X_2} J_{\Col}^{(c)}\,,\\
\partial_Y J_{\Col}^{(c)}&= 0\,.
\ea
Combining these partial derivatives, we obtain equation~\eqref{equ:Collapse} in terms of the total differential.
The explicit solution for this source term is
\be
 J_{\Col}^{(c)}= c_{\Col} \hs (X_1+X_2)^{\alpha_1+\alpha_2}\,, \quad {\rm with}\quad c_{\Col}\equiv \hs i^{\alpha_1+\alpha_2}\hs \Gamma(-\alpha_1-\alpha_2) \,,
 \label{equ:contact-source}
\ee
where the overall normalization is found by direct integration of~\eqref{equ:source}. 
\end{itemize}
\end{eBox3}

\vskip 10pt
In~\cite{Arkani-Hamed:2015bza,Arkani-Hamed:2018kmz}, a {\it second-order} differential equation was derived for the massive exchange graph:
\be
\Big[\left(X_1^2-Y^2\right)\partial_{X_1}^2 +2 X_1(1-\alpha_1)\partial_{X_1}- (\alpha_1+\dev)(1-\alpha_1+\dev) \Big] \psi_{e}^{(2)} = d(Y)   (X_1+X_2)^{\alpha_1+\alpha_2-1}\,,
\label{equ:Ex2}
\ee
where $d(Y)$ is independent of $X_1,X_2$. This equation can be viewed as a consequence of the wave equation obeyed by the bulk Green's functions.
 Let us confirm that this equation can be recovered from our system of first-order equations.

\vskip 4pt
Consider the following combination of basis functions
\be
\psi^{\pmr}_{\BtoB} \equiv  \psi^{(\pmr\pb)}_{\BtoB}  +  \psi^{(\pmr\mb)}_{\BtoB}\,.
\ee
Using~\reef{eq:X1 der exchange}, it is easy to verify that these functions, along with the collapsed term $J_{\Col}^{(c)}$, satisfy a closed system of first-order differential equations involving just three functions:
\be
\partial_{X_1} \psi^{\pmr}_{\BtoB} =\frac{1}{X_1\,\pmr\, Y}\Big[\alpha_1\,\psi^{\pmr}_{\BtoB}  + \dev \, \psi^{\mpr}_{\BtoB} \,\pmr\, J_{\Col}^{(c)} \Big]\,.
\label{equ:X1psi}
\ee
It is furthermore convenient to define the connected wavefunction  $\psi_{\BtoB} =  \psi^{\pr}_{\BtoB}+ \psi^{\mr}_{\BtoB}$, as well as the auxiliary function $\tilde \psi_{\BtoB} \equiv  \psi^{\pr}_{\BtoB}- \psi^{\mr}_{\BtoB}$. Equation \eqref{equ:X1psi} then implies
\ba
\label{eq:2ndorderfirstordersys}
\partial_{X_1}\psi_{\BtoB} &= \frac{1}{X_1^2-Y^2}\left[ (\alpha_1+\dev) X_1 \psi_{\BtoB}- (\alpha_1-\dev)Y \tilde \psi_{\BtoB} -2 Y J_{\Col}^{(c)} \right] ,\\
\partial_{X_1}\tilde \psi_{\BtoB} &= \frac{1}{X_1^2-Y^2}\left[ (\alpha_1-\dev)X_1 \tilde \psi_{\BtoB} -(\alpha_1-\dev) Y \psi_{\BtoB}+2 X_1 J_{\Col}^{(c)} \right] .
\ea
By taking another $X_1$-derivative, we can eliminate the function $\tilde \psi_{\BtoB}$ and its derivative.
This implies the following equation for the time-ordered contribution of the wavefunction:
\be
\Big[\left(X_1^2-Y^2\right)\partial_{X_1}^2 +2 X_1(1-\alpha_1)\partial_{X_1}- (\alpha_1+\dev)(1-\alpha_1+\dev) \Big] \psi_{\BtoB} = -2 Y \partial_{X_1}  J_{\Col}^{(c)}\,.
\ee
By substituting the source term $J_{\Col}^{(c)}$ given by   \eqref{equ:contact-source}, we obtain equation \eqref{equ:Ex2} with $d(Y) \equiv -2Y (\alpha_1+\alpha_2) \hs c_{\Col}$. The disconnected contribution  $\psi_{\BtoBDis}$ is annihilated by the same differential operator and so satisfies the same equation without the inhomogeneous source term on the right-hand side. Putting these pieces together, we find that the full wavefunction satisfies~\eqref{equ:Ex2}, as expected.

\subsection{Double Exchange}
\label{sec:Double Exchange}

The single-exchange example that we just discussed already displays essentially all of the complexities of the general case. Nevertheless, in order to see the general pattern, it is useful to consider an additional example.
Next, we therefore present results for a three-site graph involving two exchanges of massive fields and an arbitrary number of conformally coupled external legs: 
\begin{align}
\psi_e^{(3)} &\ \equiv\ \raisebox{-26pt}{
\begin{tikzpicture}[line width=1. pt, scale=2]
\draw[line width=2.pt,color=darkgray] (-0.75,0) -- (0.75,0);
\draw[line width=1.pt,color=lightgray] (0,0) -- (0,0.6);
\draw[line width=1.pt,color=lightgray] (-0.75,0) -- (-0.95,0.6);
\draw[line width=1.pt,color=lightgray] (-0.75,0) -- (-0.55,0.6);
\draw[line width=1.pt,color=lightgray] (0.75,0) -- (0.95,0.6);
\draw[line width=1.pt,color=lightgray] (0.75,0) -- (0.55,0.6);
\draw[lightgray, line width=2.pt] (-1.2,0.6) -- (1.2,0.6);
\draw[fill=purple3, purple3] (0,0) circle (.03cm);
\draw[fill=Red, Red] (-0.75,0) circle (.03cm);
\draw[fill=Blue, Blue] (0.75,0) circle (.03cm);
\draw  (-0.75,0.0) node[below]  {\small $X_{1}$};  
\draw  (0,0.0) node[below]  {\small $X_{2}$};  
\draw  (0.75,0.0) node[below]  {\small $X_{3}$};  
\draw  (-0.375,0.0) node[above]  {\small $Y$};  
\draw  (0.375,0.0) node[above]  {\small $Y'$};  
\end{tikzpicture} } \,.
\label{equ:psi3e}
\end{align}
The total energy flowing from vertex $i$ to the boundary is denoted by $X_i$ and the energies exchanged between the vertices are labeled $Y$ and $Y'$. 
Since the propagators do not necessarily correspond to fields with the same mass, we use $\dev$ and $\dev'$ for the weight parameters of each propagator. We also define $\dev_1 \equiv \dev$, $\dev_2\equiv \dev+\dev'$ and $\dev_3 \equiv \dev'$, which appear as powers in the vertex factors.

\vskip 4pt
For simplicity,
we consider only the fully time-ordered contribution. The disconnected contributions can be related to the single-exchange and contact examples we have already considered.
Using the Feynman rules presented in Section~\ref{sec:background}, the connected wavefunction coefficient \eqref{equ:psi3e} is given by the following integral
\beq
\psi_e^{(3)} =  \int \left[ \prod_{i=1}^3 \frac{\ud\eta_i }{(-\eta_i)^{1+\alpha_i+\dev_i}}\,e^{iX_i\eta_i} \right] \,\hat G_\text{F}(Y,\eta_1,\eta_2) \hs \hat G_\text{F}(Y',\eta_2,\eta_3)\,.
\label{equ:psi3e2}
\eeq
We separate this into $16$ basis functions by substituting the mode functions $g_Y(\eta) = h^{+}_Y(\eta) + h^{-}_Y(\eta)$ and $\bar g_Y(\eta) = \bar h^{+}_Y(\eta) + \bar h^{-}_Y(\eta)$.
Each bulk-to-bulk propagator then separates into four parts:
\beq
\hat G_\text{F}^{(\pmr \pmp)}(Y,\eta_1,\eta_2)  \equiv \bar h^{\pmr}_Y (\eta_1)  h^{\pmp}_Y (\eta_2)\, \theta_{12} +  h^{\pmr}_Y (\eta_1)  \bar h^{\pmp}_Y (\eta_2)\, \theta_{21} \,,
\eeq
so that~\eqref{equ:psi3e2} can be expressed as the sum of the following functions
\beq
\psi^{(\pmr\pmpl)(\pmpr\pmb)} =  \int \left[ \prod_{i=1}^3 \frac{\ud\eta_i }{(-\eta_i)^{1+\alpha_i+\dev_i}}\,e^{iX_i\eta_i} \right] \,\hat G_\text{F}^{(\pmr\pmpl)}(Y,\eta_1,\eta_2) \hs \hat G_\text{F}^{(\pmpr\pmb)}(Y',\eta_2,\eta_3)\,.
\eeq
Applying the same procedure as in the previous examples, one can straightforwardly derive the differential of these basis functions:
\ba
{\rm d} \psi^{(\pmr\pmpl)(\pmpr\pmb)} &=\Big[\alpha_1 \ell^{\pmr}_{1}   + \alpha_2\ell^{\pmpl\pmpr}_{2}  + \alpha_3\ell^{\pmb}_{3}  \Big]\, \psi^{(\pmr\pmpl)(\pmpr\pmb)} \\[2pt]
&\quad + \dev \hs(\ell^{\pmr}_{1} -L_{12}) \hs \psi^{(\mpr\pmpl)(\pmpr\pmb)}+ \dev\hs(\ell^{\pmpl\pmpr}_{2}-L_{12})  \psi^{(\pmr\mppl)(\pmpr\pmb)} \\[3pt]
&\quad+ \dev'(\ell^{\pmb}_{3} -L_{23}) \hs \psi^{(\pmr\pmpl)(\pmpr\mpb)} + \dev' (\ell^{\pmpl\pmpr}_{2}-L_{23}) \hs \psi^{(\pmr\pmpl)(\mppr\pmb)}\\[2pt]
&\quad - \frac{(\textcolor{red3} {\pm \,1} \,\textcolor{purple3}{\mp_1 \, 1} )}{2} (\ell^{\pmr}_{1} - \ell^{\pmpl\pmpr}_{2}) \hs J^{(c)(\pmpr\pmb)}  -  \frac{(\textcolor{blue3} {\pm \,1} \,\textcolor{purple3}{\mp_2 \, 1} )}{2}(\ell^{\pmb}_{3} - \ell^{\pmpl\pmpr}_{2}) \hs J^{(\pmr\pmpl)(c)}\,,
\label{equ:3site}
\ea
where we have defined the following shorthand for the dlog-forms that appear:
\ba\label{eq: 3-site der}
\ell_1^{\pmr} &=\dl (X_1\,\pmr\,Y)\,, &  \qquad L_{12}&= \dl  Y\,,\\
\ell_3^{\pmb} &=\dl (X_3\,\pmb\,Y')\,, &  \qquad L_{23}&= \dl  Y'\,,\\
\ell_2^{\pmpl\pmpr} &=\dl (X_2\,\pmpl\,Y\,\pmpr\,Y')\,.
\ea
Despite the apparent complexity of~\eqref{equ:3site}, there is a simple pattern in these equations.
The source terms in the first line are simply the original functions, while the functions appearing in the second and third lines (proportional to $\dev$ and $\dev'$) can be obtained from all possible sign flips of a single entry in the superscript of $\psi^{(\pmr\pmpl)(\pmpr\pmb)}$.  The letters multiplying these source functions are correlated with these sign flips. Finally, in the last line, new functions $J^{(c)(\pmpr\pmb)}$ and  $J^{(\pmr\pmpl)(c)}$ appear, which arise from collapsing either the left or right propagator.  

\vskip 4pt
The functions with a single collapsed propagator are closely related to the functions appearing in the two-site wavefunction discussed in Section~\ref{sec:singleexc}. For example, we can write $J^{(\pmr\pmpl)(c)}$ as
\ba
J^{(\pmr\pmpl)(c)} 
 &= \raisebox{-26pt}{
\begin{tikzpicture}[line width=1. pt, scale=2]
\draw[line width=2.pt,darkgray] (0,0) -- (1,0);
\draw[line width=1.pt,lightgray] (0,0) -- (-0.25,0.55);
\draw[line width=1.pt,lightgray] (0,0) -- (0.25,0.55);
\draw[line width=1.pt,lightgray] (1,0) -- (0.75,0.55);
\draw[line width=1.pt,lightgray] (1,0) -- (1,0.55);
\draw[line width=1.pt,lightgray] (1,0) -- (1.25,0.55);
\draw[lightgray, line width=2.pt] (-0.5,0.55) -- (1.5,0.55);
\draw[fill=Red,Red] (0,0) circle (.03cm);
\draw[fill=purple3,purple3] (1,0) circle (.03cm);
\node[scale=1] at (0,-.2) {$X_1$};
\node[scale=1] at (1,-.2) {$X_2+X_3$};
\node[scale=1,above] at (0.5,0) {$Y$};
\end{tikzpicture}
}
=  \psi^{(\pmr\pmpl)}(X_1,X_2+X_3)\,. 
\ea
The differential of the collapsed source functions then follows directly from equation~\eqref{eq: exchange time dif} for the two-site chain. For instance,  the differential of $J^{(\pmr\pmpl)(c)} $ is
\beq
\begin{aligned}
{\rm d}  J^{(\pmr\pmpl)(c)} &=  \left[\alpha_1 \ell^{\pmr}_1 + (\alpha_2+\alpha_3) \ell^{\pmpl}_{23} \right] J^{(\pmr\pmpl)(c)} \\[4pt]
&\quad+ \dev(\ell^{\pmr}_1- L_{12})\,  \psi^{(\mpr\pmpl)(c)}+ \dev \left(\ell^{\pmpl}_{23} -  L_{12} \right)  J^{(\pmr\mppl)(c)}  \\
&\quad  -  \half(\pmr 1 \mppl 1)\left(\ell_1^{\pmr}-\ell^{\pmpl}_{23} \right) J^{(c)(c)}\,,
\end{aligned}
\label{equ:dpsi2c}
\eeq
where a new letter $ \ell^{\pmpl}_{23} \equiv \dl (X_2+X_3 \,\pmpl\, Y)$ has appeared because the line connecting vertices $2$ and $3$ was collapsed to define this source function.

\vskip 4pt
In the last line of~\eqref{equ:dpsi2c}, a new source function appears $J^{(c)(c)}$, which arises from collapsing two propagators:
\be
J^{(c)(c)}= \raisebox{-30pt}{
\begin{tikzpicture}[line width=1. pt, scale=2]
\draw[lightgray, line width=1.pt] (0,0) -- (0.0,0.55);
\draw[lightgray, line width=1.pt] (0,0) -- (0.2,0.55);
\draw[lightgray, line width=1.pt] (0,0) -- (-0.2,0.55);
\draw[lightgray, line width=1.pt] (0,0) -- (0.4,0.55);
\draw[lightgray, line width=1.pt] (0,0) -- (-0.4,0.55);
\draw[lightgray, line width=2.pt] (-0.55,0.55) -- (0.55,0.55);
\draw[fill=black] (0,0) circle (.03cm);
\node[scale=1] at (0,-.2) {$X_1+X_2+X_3$};
\end{tikzpicture}} \ .
\ee
The differential of this function is
\be
{\rm d} J^{(c)(c)}  = (\alpha_1+\alpha_2+\alpha_3)L\, J^{(c)(c)} \,,
\ee
where $L \equiv \dl  (X_1+X_2+X_3)$.

\vskip4pt
The differential equations that we found are more structured than one would have anticipated. Indeed, we saw that the equations obtained by integration-by-parts can be understood in terms of sign flips and collapsing propagators of the underlying Feynman graph. This provides a simple ``binary" interpretation of the differential equations for these functions, which is simply related to the pattern of sign flips. This pattern can be straightforwardly extended to arbitrary graphs.
In order to do so, it is convenient to encode the differential equations pictorially, which makes 
the combinatorial features of these equations more manifest.

\newpage
\section{Circling Back to the Boundary}
\label{sec:Flow}

Looking at the differential equations derived in Section~\ref{sec:DE} suggests a simple organizing structure. 
In~\cite{Arkani-Hamed:2023kig, Baumann:2025qjx}, a similar pattern was discovered for the differential equations of conformally coupled fields in  power-law cosmologies. 
This became especially vivid when both the basis functions and the letters of the differential equations were represented combinatorially in terms of graph tubings. These tubings ``evolve" as one takes derivatives, subject to simple rules. This recasting of the evolution of the wavefunction as dynamics through the combinatorics of kinematic variables was called  ``kinematic flow"~\cite{Arkani-Hamed:2023kig,Arkani-Hamed:2023bsv}.

\vskip4pt
In this section, we will show that a similar kinematic flow description also exists for fields with generic masses in de Sitter space. In fact, there is a sense in which the construction is even more natural for the massive case. As we saw in Section~\ref{sec:DE}, the differential equations are essentially determined by patterns of sign flips, and so it is not very surprising that this procedure has a simple combinatorial description.
Like in~\cite{Baumann:2025qjx}, the differential equations separate into independent sectors graded by the number of disconnected propagators.  For simplicity, we will therefore restrict our discussion to fully connected graphs
(with only Feynman propagators). The equations for the contributions with disconnected propagators can be obtained from the equations for the fully time-ordered subgraphs.

\subsection{Marked Graphs and Tubings}
\label{ssec:tubings}

We start by introducing the concept of ``truncated graphs." This means removing all conformally coupled (gray) external lines and only keeping track of the energies associated to these lines. For example, we would truncate the following graphs:
\beq
\raisebox{-25pt}{
\begin{tikzpicture}[line width=1. pt, scale=2]
\draw[lightgray, line width=1.pt] (0,0) -- (0.0,0.55);
\draw[lightgray, line width=1.pt] (0,0) -- (0.25,0.55);
\draw[darkgray, line width=2.pt] (0,0) -- (-0.25,0.55);
\draw[lightgray, line width=2.pt] (-0.5,0.55) -- (2.6,0.55);
\node[scale=1] at (-0.25,0.68) {\small $k_1$};
\node[scale=1] at (0,0.68) {\small $k_2$};
\node[scale=1] at (0.25,0.68) {\small $k_3$};
\draw[fill=black] (0,0) circle (.03cm);
\begin{scope}[xshift=1.1cm]
\draw[line width=2.pt,darkgray] (0,0) -- (1,0);
\draw[line width=1.pt,lightgray] (0,0) -- (-0.25,0.55);
\draw[line width=1.pt,lightgray] (0,0) -- (0.25,0.55);
\draw[line width=1.pt,lightgray] (1,0) -- (0.75,0.55);
\draw[line width=1.pt,lightgray] (1,0) -- (1.25,0.55);
\node[scale=1] at (-0.25,0.68) {\small $k_1$};
\node[scale=1] at (0.25,0.68) {\small $k_2$};
\node[scale=1] at (0.75,0.68) {\small $k_3$};
\node[scale=1] at (1.25,0.68) {\small $k_4$};
\draw[fill=Red,Red] (0,0) circle (.03cm);
\draw[fill=Blue,Blue] (1,0) circle (.03cm);
\node[scale=1] at (0,-.15) {\phantom{$X_1$}};
\end{scope}
\end{tikzpicture}
} 
\qquad \sim \qquad
\raisebox{-25pt}{
\begin{tikzpicture}[line width=1. pt, scale=2]
\draw[darkgray, line width=2.pt] (0,0) -- (0,0.55);
\draw[lightgray, line width=2.pt] (-0.25,0.55) -- (0.25,0.55);
\draw[fill=black] (0,0) circle (.03cm);
\node[scale=1] at (0,-.15) {$X$};
\node[scale=1] at (0,0.68) {\small $k$};
\begin{scope}[xshift=1.cm]
\draw[line width=2.pt,darkgray] (0,0.25) -- (1,0.25);
\draw[fill=Red,Red] (0,0.25) circle (.03cm);
\draw[fill=Blue,Blue] (1,0.25) circle (.03cm);
\node[Red,scale=1] at (0,0.1) {$X_1$};
\node[Blue,scale=1] at (1,0.1) {$X_2$};
\node[scale=1] at (0.5,0.13) {$Y$};
\end{scope}
\end{tikzpicture}
} \ ,
\eeq
where 
\beq
\begin{aligned}
k&=k_1, &\qquad Y&\equiv |{\bf k}_1+{\bf k}_2|\,,\\
X&=k_2+k_3\,, & X_1&\equiv k_1+k_2\,,\quad X_2&\equiv k_3+k_4\,.
\end{aligned}
\eeq
This is natural  from the bulk perspective because the mode functions of conformally coupled scalars are plane waves, so the energies flowing in/out of vertices just add together. 
From these truncated graphs, we then create ``marked graphs" by adding crosses on all the remaining lines (internal and external). The letters (singularities) of the differential equations, as well as the basis functions, can then be represented graphically as ``tubings" of these marked graphs.

\subsubsection*{Letters}
The letters of the differential equations are in one-to-one correspondence with ``tubes"---connected proper subgraphs, represented by 
encircling the vertices and crosses---on the marked graph. 
To each tube we assign a combination of kinematic variables given by the sum of vertex energies enclosed by the tube and the energies of the lines piercing the tube. 
For tubes that intersect a line and enclose the corresponding cross, one flips the sign of the energy associated to that line.
As an illustration, the differential equations for the two-site chain (with a massive propagator) exhibit the following singularities:\footnote{In~\cite{Arkani-Hamed:2023kig, Baumann:2025qjx}, each tube had to include at least one vertex creating 5 of the 6 letters in (\ref{equ:Sing1}).  In the case of a massive propagator, the differential equation has the additional letter$\Lm$ corresponding to a singularity at vanishing internal energy $Y$. For conformally coupled fields, the singularities associated with the internal energies can be removed by a rescaling of the wavefunction, so this letter was absent in~\cite{Arkani-Hamed:2023kig, Baumann:2025qjx}. }
\beq
\begin{aligned}
	&\Lap \ \equiv\ \ud\log(X_1+Y)\,,  &&\Lam\ \equiv\ \ud\log(X_1-Y)\,,\\
	&\Lbp \ \equiv\ \ud\log(X_2+Y)\,, &&\Lbm\ \equiv\ \ud\log(X_2-Y)\,,\\
	&\Lab \ \equiv\ \ud\log(X_1+X_2)\,, &&\Lm\ \equiv\ \ud\log Y\,.
\end{aligned}
\label{equ:Sing1}
\eeq
In a small abuse of terminology and notation, we are relating tubes of the marked graph directly to the dlog-forms of letters that will appear in the differential equations (as opposed to the letters themselves).

\vskip 4pt
A novelty of the massive case is that external lines can also be marked. For example, a single massive external leg would lead to the following singularities:  
\beq
\LConp
  \equiv\ \ud\log(X+k)\,,
\quad
\LConm
 \equiv\ \ud\log(X-k)\,, 
 \quad
\LConc
 \equiv\ \ud\log k\,.
 \label{equ:Sing2}
\eeq
The singularities for more complicated graphs are the obvious generalizations of the singularities (tubes) shown in (\ref{equ:Sing1}) and (\ref{equ:Sing2}).

\subsubsection*{Functions}
The functions in the differential system are also in one-to-one correspondence with graphical objects. Each function can be represented by a ``complete tubing" of the marked graph, where each vertex is included in exactly one tube.
The crosses on massive propagators can be in multiple overlapping tubes. It is best to illustrate this with some examples.

\begin{itemize}
\item {\it Single exchange:}
The two-site chain (with a massive internal propagator) has the following five tubings and  associated functions: 
\begin{align}
\begin{split}
\psi^{(++)}:\ \ \raisebox{4pt}{\psipp}
 \hspace{1.1cm}   & 
\psi^{(-+)}:\ \ \raisebox{4pt}{\psimp}
 \hspace{1.1cm}   
J^{(c)}:\ \ \raisebox{4pt}{\psic} \\
\psi^{(--)}:\ \ \raisebox{4pt}{\psimm} 
 \hspace{1.1cm}   &\psi^{(+-)}:\ \raisebox{4pt}{\psipm} 
\end{split}
\label{equ:internal}
\end{align}
This is a graphical encoding of all possible assignments of a pair of $+$ or $-$ signs to the internal line, matching the possible propagator splits discussed in Section~\ref{sec:masterI}.  We
assign a minus sign when a tube encircles the corresponding cross and a plus sign when it does not.
Completely enclosing an internal line like in $J^{(c)}$ corresponds to collapsing the propagator.  

\item {\it Contact diagram:}
As a simple example with a massive external line, we consider a
contact diagram with a single massive leg. It has the following graph tubings and associated basis functions: 
\beq
\psi^{+}_c: \ 
\FConp
 \hspace{1.5cm} 
 \psi^{-}_c: \
\FConm
\label{equ:external}
\eeq
We considered the equations satisfied by these functions in Section~\ref{ssec:contact}.
\end{itemize}
Of course, we can also have diagrams with a combination of massive internal and external legs.
The graph tubings in that case are obtained by a straightforward combination of the tubings shown in~(\ref{equ:internal}) and~(\ref{equ:external}).

\subsubsection*{Conformal limit}

In~\cite{Baumann:2025qjx}, similar graph tubings were introduced to represent the basis functions for theories with conformally coupled fields in a power-law cosmology. In the case of massive propagators, we have additional basis functions and their graphical representations have a slightly different meaning.  To prevent confusion, we briefly highlight the difference in notation and discuss the conformal limit of the basis functions introduced in this work.

\begin{itemize}
\item {\it Single exchange:} We first look at the two-site graph with a single massive exchange.
In that case, the basis functions were represented graphically in~\eqref{equ:internal} and algebraically in 
\eqref{eq:pm psi 2site}:
\beq
\label{eq:conformalexcdiag}
\psi^{(\pmr\pmb)} =  \int \ud\eta_1\ud\eta_2 \,\frac{e^{iX_1\eta_1}e^{iX_2\eta_2} }{(-\eta_1)^{1+\alpha_1+\dev}(-\eta_2)^{1+\alpha_2+\dev}}  \Big[ \bar h^{\pmr}_Y (\eta_1)  h^{\pmb}_Y (\eta_2)\, \theta_{12} +  h^{\pmr}_Y (\eta_1)  \bar h^{\pmb}_Y (\eta_2)\, \theta_{21} \Big] \,.
\eeq
We recall that in the conformal limit  $h_k^- = \bar h_k^+ = 0$, so  that $\psi^{(++)} =\psi^{(--)} \to 0$.
This means that the new connected basis functions associated to the tubings \raisebox{3pt}{\scalebox{0.7}{$\psipp$}} and \raisebox{3pt}{\scalebox{0.7}{$\psimm$}} are absent for conformally coupled fields, as required.\footnote{Note that, in~\cite{Baumann:2025qjx}, the tubing  \raisebox{3pt}{\scalebox{0.7}{$\psipp$}} was used to represent the {\it disconnected} part of the wavefunction. In this paper, the same tubing has a different meaning.}

\vskip 4pt
Next, we consider the functions with mixed signs, $\psi^{(-+)}$ and $\psi^{(+-)}$, corresponding to the tubings \raisebox{3pt}{\scalebox{0.7}{$\psimp$}} and \raisebox{3pt}{\scalebox{0.7}{$\psipm$}}, respectively.
 It is interesting to see that, for generic masses, these functions contain
a mixture of the two possible time orderings. In the conformal limit, however, half of each function vanishes and we get 
\beq
\begin{aligned}
\psi^{(+-)} &\longrightarrow \, \psi_{\includegraphics{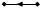}}\,,\\
\psi^{(-+)} &\longrightarrow\, \psi_{\includegraphics{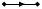}}\,,
\end{aligned}
\eeq
so that the basis functions reduce to their (anti)time-ordered  contributions.

\vskip 4pt
We see that two phenomena happen in the conformal limit. First, the size of the basis decreases because some functions identically vanish. Second, the remaining functions have a definite time ordering, which importantly is not the case away from the conformally coupled limit. These two features persist in more general examples.

\item 
{\it Contact diagram:}  The basis functions for a contact diagram with a single massive leg were represented graphically in \eqref{equ:external} and algebraically in \eqref{eq:contact3pts}: 
\beq
\psi^{\pm}_c = \int \frac{\ud\eta}{ (-\eta)^{1+\alpha+\dev}}\, e^{iX\eta} \,  h^{\pm}_k (\eta)\,.
\eeq
It is trivial to see that in the conformal limit, where $h^-_k \to 0$ and $h^+_k \to e^{ik\eta}$, we have $\psi^-_c \to 0$ and only $\psi_c^+$ survives. This is consistent with the fact that, in the conformal case, we have a unique basis function with energy $X+k$. 
\end{itemize}

\subsection{Massive Kinematic Flow}
\label{sec:kflowalgorithm}

All of the ingredients of the problem now have a graphical representation, so it is useful to write 
the differential equations derived in Section~\ref{sec:DE} in graphical form.
For example, the differential of the function $\psi^{(-+)} \equiv \psi_{\includegraphics[scale=0.6]{Figures/Tubings/Functions/psi-.pdf}}$ in~\eqref{equ:internal} is
\beq
\begin{aligned}
{\rm d}\psi_{\includegraphics[scale=0.6]{Figures/Tubings/Functions/psi-.pdf}} &\, = \, \Big(
\alpha_1   \, \includegraphics[scale=0.9,valign=c]{Figures/Tubings/two/Lam.pdf}   \, +\, \alpha_2     \includegraphics[scale=0.9,valign=c]{Figures/Tubings/two/Lbp.pdf}  \Big) \,  \psi_{\includegraphics[scale=0.6]{Figures/Tubings/Functions/psi-.pdf}} 
  \, +\, 
  \Big(
 \includegraphics[scale=0.9,valign=c]{Figures/Tubings/two/Lam.pdf}   \, -\,     \includegraphics[scale=0.9,valign=c]{Figures/Tubings/two/Lbp.pdf}  \Big)  \, J_{\includegraphics[scale=0.6]{Figures/Tubings/Functions/psic.pdf}}  \\
 &\qquad + \dev\, \Big[ \Big(\includegraphics[scale=0.9,valign=c]{Figures/Tubings/two/Lam.pdf}  -  \Lm\Big) \psi_{\includegraphics[scale=0.6]{Figures/Tubings/Functions/psipp.pdf}} 
 + \Big(\includegraphics[scale=0.9,valign=c]{Figures/Tubings/two/Lbp.pdf}  -  \Lm\Big) \psi_{\includegraphics[scale=0.6]{Figures/Tubings/Functions/psimm.pdf}} 
  \Big]\,.
\end{aligned}
\label{equ:psi-}
\eeq
The first line on the right-hand side coincides with the conformally coupled case~\cite{Baumann:2025qjx}, while the second line contains the additional contributions from the physics of the massive particle.
An essentially similar equation exists for the differential of $\psi_{\includegraphics[scale=0.6]{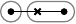}}$, where all the graphs are reflected across the central cross (corresponding to interchanging the kinematic variables $X_1\leftrightarrow X_2$).

\vskip 4pt
The three functions that appear as sources in (\ref{equ:psi-}) have the following differentials:
\begin{align}
{\rm d}\psi_{\includegraphics[scale=0.6]{Figures/Tubings/Functions/psipp.pdf}} &\, = \, \Big(
\alpha_1   \, \includegraphics[scale=0.9,valign=c]{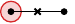}   \, +\, \alpha_2     \includegraphics[scale=0.9,valign=c]{Figures/Tubings/two/Lbp.pdf}  \Big) \,  \psi_{\includegraphics[scale=0.6]{Figures/Tubings/Functions/psipp.pdf}}   \\
 &\qquad + \dev\, \Big[ \Big(\includegraphics[scale=0.9,valign=c]{Figures/Tubings/two/Lap.pdf}  -  \Lm\Big) \psi_{\includegraphics[scale=0.6]{Figures/Tubings/Functions/psi-.pdf}} 
 + \Big(\includegraphics[scale=0.9,valign=c]{Figures/Tubings/two/Lbp.pdf}  -  \Lm\Big) \psi_{\includegraphics[scale=0.6]{Figures/Tubings/Functions/psi+.pdf}} 
  \Big]\,, \nonumber \\[10pt]
  {\rm d}\psi_{\includegraphics[scale=0.6]{Figures/Tubings/Functions/psimm.pdf}} &\, = \, \Big(
\alpha_1   \, \includegraphics[scale=0.9,valign=c]{Figures/Tubings/two/Lam.pdf}   \, +\, \alpha_2     \includegraphics[scale=0.9,valign=c]{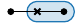}  \Big) \,  \psi_{\includegraphics[scale=0.6]{Figures/Tubings/Functions/psimm.pdf}}   \\
 &\qquad + \dev\, \Big[ \Big(\includegraphics[scale=0.9,valign=c]{Figures/Tubings/two/Lam.pdf}  -  \Lm\Big) \psi_{\includegraphics[scale=0.6]{Figures/Tubings/Functions/psi+.pdf}} 
 + \Big(\includegraphics[scale=0.9,valign=c]{Figures/Tubings/two/Lbm.pdf}  -  \Lm\Big) \psi_{\includegraphics[scale=0.6]{Figures/Tubings/Functions/psi-.pdf}} 
  \Big]\,, \nonumber \\[10pt]
{\rm d} 
J_{\includegraphics[scale=0.6]{Figures/Tubings/Functions/psic.pdf}}  &\, = \, (\alpha_{1}+\alpha_2)\, 
 \includegraphics[scale=0.9,valign=c]{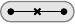}\ J_{\includegraphics[scale=0.6]{Figures/Tubings/Functions/psic.pdf}} \,.
\end{align}
Again, the contributions in the first line of each expression are the same as in the conformally coupled case~\cite{Baumann:2025qjx}, while the second line contains new terms coming from the massive propagator.
This pictorial representation efficiently encodes the pattern of signs and letters appearing in~\eqref{eq: exchange time dif}.

\subsubsection*{Flow rules}

In Section~\ref{sec:DE}, we saw that the system of differential equations satisfied by the functions $\psi^{(\pm\cdots \pm)}$ are simply related to the signs of the functions and the pattern of collapsing internal lines. These equations can be derived by systematically applying integration-by-parts identities and shift relations. We have also seen that the results can be given a simple graphical representation in terms of tubings of the corresponding graphs. We now explain that the equations can also be derived from a set of simple rules that govern the growth of graph tubings through the differential system.
  
\vskip 4pt
The starting point is the graph tubing associated to a specific ``parent function" of interest. As an illustration, we will take the function $\psi_{\includegraphics[scale=0.6]{Figures/Tubings/Functions/psi-.pdf}}$, whose differential appears in~\eqref{equ:psi-}.  This differential follows from three simple steps. 
\begin{itemize}
\item[1.] {\bf Activation}: Each tube of the graph tubing gets ``activated" and becomes a letter in the differential equation.  
The letter is multiplied by the function corresponding to the original graph times the sum of the parameters $\alpha_i$ associated to its enclosed vertices. This generates the first term on the right-hand side of 
\eqref{equ:psi-}: 
\beq
{\rm d}\psi_{\includegraphics[scale=0.6]{Figures/Tubings/Functions/psi-.pdf}} \, 
\supset\, \Big(
\alpha_1   \, \includegraphics[scale=0.9,valign=c]{Figures/Tubings/two/Lam.pdf}   \, +\, \alpha_2     \includegraphics[scale=0.9,valign=c]{Figures/Tubings/two/Lbp.pdf}  \Big) \,  \psi_{\includegraphics[scale=0.6]{Figures/Tubings/Functions/psi-.pdf}} \ .
 \label{equ:example3}
 \eeq

\item[2.] {\bf Merger}: If the graph tubing has tubes that are adjacent to each other, these tubes can ``merge" to form a larger tube. 
The function associated to this new tubing will appear as a source function in the differential equation. This source function multiplies the {\it difference} of the two letters corresponding to the two tubes involved in the merger. The letter containing the cross on the merged edge appears with a plus sign. This explains the second term on the right-hand side of 
\eqref{equ:psi-}:
\beq
{\rm d}\psi_{\includegraphics[scale=0.6]{Figures/Tubings/Functions/psi-.pdf}} \, \supset \, \Big(
\alpha_1   \, \includegraphics[scale=0.9,valign=c]{Figures/Tubings/two/Lam.pdf}   \, +\, \alpha_2     \includegraphics[scale=0.9,valign=c]{Figures/Tubings/two/Lbp.pdf}  \Big) \,  \psi_{\includegraphics[scale=0.6]{Figures/Tubings/Functions/psi-.pdf}} 
  \, +\, 
  \Big(
 \includegraphics[scale=0.9,valign=c]{Figures/Tubings/two/Lam.pdf}   \, -\,     \includegraphics[scale=0.9,valign=c]{Figures/Tubings/two/Lbp.pdf}  \Big)  \, J_{\includegraphics[scale=0.6]{Figures/Tubings/Functions/psic.pdf}} \ .
 \eeq
 If the theory contains only conformally coupled fields, this generates the full differential~\cite{Baumann:2025qjx}. 
 To capture massive propagators, however, we need one additional rule.

 \item[3.] {\bf Mixing}:  Tubes that are pierced by massive propagators (lines with crosses on them) can {\it grow} and {\it shrink} to produce new source functions.  The new source functions multiply the {\it difference} of the letter corresponding to the
 changed tube and the letter of the relevant internal line, times the parameter $\xi$ describing the massive field.
 This completes the derivation of the differential in~\eqref{equ:psi-}:
 \beq
\begin{aligned}
{\rm d}\psi_{\includegraphics[scale=0.6]{Figures/Tubings/Functions/psi-.pdf}} &\, = \, \Big(
\alpha_1   \, \includegraphics[scale=0.9,valign=c]{Figures/Tubings/two/Lam.pdf}   \, +\, \alpha_2     \includegraphics[scale=0.9,valign=c]{Figures/Tubings/two/Lbp.pdf}  \Big) \,  \psi_{\includegraphics[scale=0.6]{Figures/Tubings/Functions/psi-.pdf}} 
  \, +\, 
  \Big(
 \includegraphics[scale=0.9,valign=c]{Figures/Tubings/two/Lam.pdf}   \, -\,     \includegraphics[scale=0.9,valign=c]{Figures/Tubings/two/Lbp.pdf}  \Big)  \, J_{\includegraphics[scale=0.6]{Figures/Tubings/Functions/psic.pdf}}  \\
 &\qquad + \dev\, \Big[ \Big(\includegraphics[scale=0.9,valign=c]{Figures/Tubings/two/Lam.pdf}  -  \Lm\Big) \psi_{\includegraphics[scale=0.6]{Figures/Tubings/Functions/psipp.pdf}} 
 + \Big(\includegraphics[scale=0.9,valign=c]{Figures/Tubings/two/Lbp.pdf}  -  \Lm\Big) \psi_{\includegraphics[scale=0.6]{Figures/Tubings/Functions/psimm.pdf}} 
  \Big]\,.
\end{aligned}
\eeq
The first term in the second line arises from the shrinking of the left tube, while the second term comes from the growth of the right tube. 
\end{itemize}
By systematically applying these three rules, one can derive the differential equations for arbitrary Feynman graphs with massive propagators.

\subsection{Selected Examples}

To make the application of these rules more concrete, we now illustrate them in a few  examples.
The examples in this section involve specific tree diagrams, but the algorithm works for arbitrary graphs (including diagrams with loops), without modification. We provide an explicit application to the simplest one-loop case in Appendix~\ref{app:loop}.

\subsubsection*{Double exchange}

As a slightly more complex example than the single exchange considered in Section~\ref{sec:kflowalgorithm}, let us describe the
 double exchange of massive fields, with conformally coupled external lines, as studied in Section~\ref{sec:Double Exchange}.
First, we note that the list of letters is identical to the conformally coupled case~\cite{Baumann:2025qjx}
\beq
\begin{aligned}
	&\includegraphics[scale=0.9,valign=c]{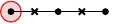} \ \equiv\ \ud\log(X_1+Y ) \,, &\quad& \includegraphics[scale=0.9,valign=c]{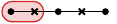}\ \equiv\ \ud\log(X_1-Y )\,,\\
	&\includegraphics[scale=0.9,valign=c]{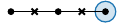} \ \equiv\ \ud\log(X_3+Y' )\,, &&\includegraphics[scale=0.9,valign=c]{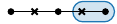}\ \equiv\ \ud\log(X_3-Y' )\,,\\
	&\includegraphics[scale=0.9,valign=c]{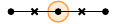} \ \equiv\ \ud\log(X_2+Y +Y' ) \,,&& \includegraphics[scale=0.9,valign=c]{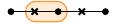}\ \equiv\ \ud\log(X_2-Y +Y' )\,,\\
	& && \includegraphics[scale=0.9,valign=c]{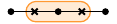}\ \equiv\ \ud\log(X_2-Y -Y' )\,,\\
	& \includegraphics[scale=0.9,valign=c]{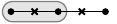}\ \equiv\ \ud\log(X_1+X_2+Y' )\,,&& \includegraphics[scale=0.9,valign=c]{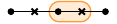}\ \equiv\ \ud\log(X_2+Y -Y' )\,,\\
	&\includegraphics[scale=0.9,valign=c]{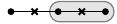}\ \equiv\ \ud\log(X_2+X_3+Y )\,, && \includegraphics[scale=0.9,valign=c]{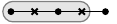}\ \equiv\ \ud\log(X_1+X_2-Y' )\,,\\
	& \includegraphics[scale=0.9,valign=c]{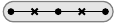}\ \equiv\ \ud\log(X_1+X_2+X_3)\,, \quad && \includegraphics[scale=0.9,valign=c]{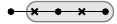}\ \equiv\ \ud\log(X_2+X_3-Y )\,,
\end{aligned}\label{equ:3Site-Letters}
\eeq
 with two additional letters that encircle a single cross:
\beq
\begin{aligned}
\dlogYL	
	 \ &\equiv\ \ud\log Y\,,  
&
\hspace{.7cm}
\dlogYR	
	 \ &\equiv\ \ud\log Y' \,.
\end{aligned}
\eeq
In  the conformally coupled case, the latter two letters were removed by rescaling the wavefunction.

\vskip 4pt
To demonstrate how the kinematic flow rules generate the differential equations, we consider the basis element $\psi^{(\pr\mmpl)(\pppr\mb)}$ which has the graphical representation $\psi_{\TSiteF{1}{-1}{1}{-1}{0.6}} $. Following the rules stated above, one finds the differential 
\beq
\begin{aligned}
&{\rm d}\psi_{\TSiteF{1}{-1}{1}{-1}{0.6}} = \Big(\alpha_1 \,\TSiteL{1}{0}{0}{0}{0.85}+  \alpha_2 \TSiteL{0}{-1}{1}{0}{0.85}  + \alpha_3 \TSiteL{0}{0}{0}{-1}{0.85}\Big)\psi_{\TSiteF{1}{-1}{1}{-1}{0.6}}\\
&\quad+ \Big(\TSiteL{0}{-1}{1}{0}{0.85}- \TSiteL{1}{0}{0}{0}{0.85} \Big) J_{\TSiteFCL{1}{-1}{0.6}}+ \Big( \TSiteL{0}{0}{0}{-1}{0.85}-\TSiteL{0}{-1}{1}{0}{0.85} \Big) J_{\TSiteFCR{1}{-1}{0.6}}\\
&\quad+ \dev\, \Big[\Big( \TSiteL{1}{0}{0}{0}{0.85} - \dlogYL \Big)\psi_{\TSiteF{-1}{-1}{1}{-1}{0.6}}+\Big( \TSiteL{0}{-1}{1}{0}{0.85} - \dlogYL\Big)\psi_{\TSiteF{1}{1}{1}{-1}{0.6}} \Big]\\
&\quad+ \dev^\prime \Big[\Big( \TSiteL{0}{0}{0}{-1}{0.85} - \dlogYR \Big)\psi_{\TSiteF{1}{-1}{1}{1}{0.6}}+\Big( \TSiteL{0}{-1}{1}{0}{0.85} - \dlogYR\Big)\psi_{\TSiteF{1}{-1}{-1}{-1}{0.6}} \Big]\,,
\end{aligned}
\eeq
where the first line corresponds to activation, the second and third lines correspond to merging of tubes on the left and right propagators, respectively, and finally the fourth and fifth lines correspond to shrinking/growth of tubes on the left and right propagators, respectively. As expected, this precisely matches with the differential found in~\reef{equ:3site} for the basis function  $\psi^{(\pr\mmpl)(\pppr\mb)}$.  
This introduces a number of new functions as sources, including ones corresponding to collapsing propagators. We can obtain their differentials via the same procedure.

\subsubsection*{Contact diagram}
As an example with massive external lines, we consider a 
three-point contact diagram with massive legs, corresponding to the Feynman diagram:
\be
\psi_c \ = \  \raisebox{-16pt}{ \begin{tikzpicture}[line width=1. pt, scale=2]
\node[scale=1] at  (0.0,0.68)  {\small $k_2$};
\node[scale=1] at  (-0.25,0.68)  {\small $k_1$};
\node[scale=1] at  (0.25,0.68)  {\small $k_3$};
\draw[darkgray, line width=2.pt] (0,0) -- (0.0,0.55);
\draw[darkgray, line width=2.pt] (0,0) -- (0.25,0.55);
\draw[darkgray, line width=2.pt] (0,0) -- (-0.25,0.55);
\draw[lightgray, line width=2.pt] (-0.5,0.55) -- (0.5,0.55);
\draw[fill=black] (0,0) circle (.03cm);
\end{tikzpicture}}\ .
\ee
For each leg, with energy $k_i$, we allow for an arbitrary mass parametrized by $\dev_i$.
First, note that by expressing the mode functions in terms of the two auxiliary functions $h^\pm_k (\eta)$, the three-point wavefunction can be written as a sum over $2^3=8$ basis elements, each of which takes the form:
\be
\psi^{(\pm_1\pm_2\pm_3)}= \int \frac{\ud\eta}{ (-\eta)^{1+\alpha+\dev}} \,  \prod_{i=1}^3h^{\pm_i}_{k_i} (\eta)\,,
\ee
where we defined $\dev \equiv \dev_1+\dev_2+\dev_3$. As an illustration, we derive the differential equation for one of the basis functions 
\be
\psi^{(-++)}\ = \
\raisebox{-16pt}{
\begin{tikzpicture}[line width=1. pt, scale=2]
\oneMinusmpp
\oneBody
\end{tikzpicture}} \ .
\ee
Following the flow algorithm, the differential of this function is 
\ba
{\rm d}\psi_{\scalebox{0.5}{\raisebox{-32pt}{\begin{tikzpicture}[line width=1. pt, scale=2]
\oneMinusmpp
\oneBody
\end{tikzpicture}}}}
 \, = \,\alpha \, \scalebox{0.7}{\raisebox{-16pt}{\begin{tikzpicture}[line width=1. pt, scale=2]
\oneMinusLetter{1}
\oneBody
\end{tikzpicture}}} \, \psi_{\scalebox{0.5}{\raisebox{-32pt}{\begin{tikzpicture}[line width=1. pt, scale=2]
\oneMinusmpp
\oneBody
\end{tikzpicture}}}} 
 & +\dev_1\left[\scalebox{0.7}{\raisebox{-16pt}{\begin{tikzpicture}[line width=1. pt, scale=2]
\oneMinusLetter{1}
\oneBody
\end{tikzpicture}}}
\ -\
\scalebox{0.7}{\raisebox{-14pt}{\begin{tikzpicture}[line width=1. pt, scale=2]
\singleCross{1}
\oneBody
\end{tikzpicture}}}\right]\psi_{\scalebox{0.5}{\raisebox{-32pt}{\begin{tikzpicture}[line width=1. pt, scale=2]
\oneppp
\oneBody
\end{tikzpicture}}}}
\\
 & +\dev_2\left[\scalebox{0.7}{\raisebox{-16pt}{\begin{tikzpicture}[line width=1. pt, scale=2]
\oneMinusLetter{1}
\oneBody
\end{tikzpicture}}}
\ -\
\scalebox{0.7}{\raisebox{-14pt}{\begin{tikzpicture}[line width=1. pt, scale=2]
\singleCross{0}
\oneBody
\end{tikzpicture}}}
\right]
\psi_{\scalebox{0.5}{\raisebox{-32pt}{\begin{tikzpicture}[line width=1. pt, scale=2]
\twoMinus{0}{1}
\oneBody
\end{tikzpicture}}}}\\
 & +\dev_3\left[\scalebox{0.7}{\raisebox{-16pt}{\begin{tikzpicture}[line width=1. pt, scale=2]
\oneMinusLetter{1}
\oneBody
\end{tikzpicture}}}
\ -\
\scalebox{0.7}{\raisebox{-14pt}{\begin{tikzpicture}[line width=1. pt, scale=2]
\singleCross{-1}
\oneBody
\end{tikzpicture}}}
\right]
\psi_{\scalebox{0.5}{\raisebox{-32pt}{\begin{tikzpicture}[line width=1. pt, scale=2]
\twoMinus{-1}{1}
\oneBody
\end{tikzpicture}}}}\ . 
\ea
The first term (proportional to $\alpha$) arises from the standard activation of the tube to produce the corresponding letter. The second term (proportional to $\dev_1$) has a new source function created by the shrinking tube, while the  third and fourth terms (proportional to $\dev_2$ and $\dev_3$)  come from growing the tube onto the second and third legs, respectively.

\vskip4pt
One can follow the same procedure for the other basis elements. The resulting system of first-order equations can be shown to imply the conformal Ward identities satisfied by the three-point function of massive fields~\cite{Coriano:2013jba,Bzowski:2013sza}.

\section{Massive Solutions to Massive Problems}
\label{sec:Solutions}

In the previous two sections, we derived differential equations for wavefunction coefficients with massive propagators and identified an interesting combinatorial structure in these equations.
In this section, we consider the systematics of obtaining solutions to these equations.

\vskip4pt
For concreteness, we will focus on the simplest case of a single massive exchange between conformally coupled fields. The relevant differential equations for the connected component are
\ba
\label{eq:sec5diffeq}
\ud \psi^{(\pmr\pmb)}_{\BtoB} &= \Big[\alpha_1\, \dl (X_1\,\pmr\, Y) + \alpha_2\, \dl (X_2\,\pmb\, Y)\Big]\, \psi^{(\pmr\pmb)}_{\BtoB} - \frac{\pmr 1 \mpb 1}{2} \hs \dl \left(\frac{X_1\,\pmr\, Y}{X_2\,\pmb\, Y}\right)
J_{\Col}^{(c)}\\[2pt]
&\quad + \dev \,\bigg[ \dl \left(\frac{X_1\,\pmr\, Y}{Y}\right)\,  \psi^{(\mpr\pmb)}_{\BtoB} +   \dl \left(\frac{X_2\,\pmb\, Y}{Y}\right) \psi^{(\pmr\mpb)}_{\BtoB}\bigg]  \,,
\ea
where $J_{\Col}^{(c)} = c_{\Col} (X_1 + X_2)^{\alpha_1 + \alpha_2}$,
with a constant coefficient $c_{\Col}$.
Different values of $\alpha_i$ correspond to different $n$-point functions, where the number of fields involved in each vertex determines $\alpha_i$ via~\eqref{eq:paramsaxi}. We are interested in the four-point function in $d=3$, so that $\alpha_i\to 0$.
While the correlator for this process can be written as a generalized hypergeometric series for any  mass parameter~\cite{Arkani-Hamed:2018kmz},  the first-order equations~\eqref{eq:sec5diffeq} are especially well suited to an analysis in the parametric regimes where the mass of the exchanged particle is taken either to infinity or to the conformal value. 
We will therefore derive solutions to these equations in the limits of light ($\dev \to 0$) and heavy ($\dev \to \infty$) particle exchange.

\subsection{Light Particle Exchange}
We first consider the case of light particle exchange, where the mass of the exchanged particle is close to the conformally coupled value $\nu = \tfrac{1}{2}$. 
The solution can then be organized as a perturbative expansion in the small parameter $\dev \to 0$, so that each function  in~\eqref{eq:sec5diffeq} is 
\beq
\psi^{(\pm\pm)}_{\BtoB} = \sum_{n=0}^\infty \dev^{n}  \psi^{(\pm\pm)}_{n}\,.
\eeq
We then solve 
the equations order by order in $\xi$. At each order in the expansion, the functions satisfy differential equations of the form
\be
\Big[\ud-\alpha_1\, \dl (X_1\,\pm\, Y) - \alpha_2\, \dl (X_2\,\pm\, Y)\Big]\, \psi_n^{(\pm\pm)} = S^{\pm\pm}(\psi_{n-1})\,,
\ee
where the sources on the right-hand side depend on the functions at one lower order. The structure of these sources is such that the $++$ and $--$ functions vanish at leading order, while the $+-$ and $-+$ functions vanish at the first subleading order. This pattern persists at higher order, so that each order in $n$ depends only on half of the $\pm\pm$ functions. In order for the $\alpha_i\to 0$ limit to be regular, we have to combine these functions with the disconnected wavefunction $\psi^{(\raisebox{1.25pt}{\resizebox{0.025\textwidth}{!}{\BtoBDis}})}$, which can also be expanded in $\xi$. All told, the wavefunction takes the form
\beq
\psi_n = \left\{ 
\begin{array}{ll}
 \displaystyle \psi_n^{(+-)}+\psi_n^{(-+)}-\frac{1}{\alpha}\,\psi^{(\raisebox{1.25pt}{\resizebox{0.025\textwidth}{!}{\BtoBDis}})}_n & \quad n = {\rm even}\,,  \\[12pt]
\displaystyle \psi_n^{(++)}+\psi_n^{(--)} -\frac{1}{\alpha}\,\psi^{(\raisebox{1.25pt}{\resizebox{0.025\textwidth}{!}{\BtoBDis}})}_n & \quad n = {\rm odd}\,.
\end{array}
\right.
\eeq
In the following, we will describe in detail how to find these solutions at leading and first subleading order.

\subsubsection*{Leading order}

At leading order, equation~\eqref{eq:sec5diffeq} simplifies to 
\be
\ud \psi^{(\pmr\pmb)}_0 -\Big[\alpha_1\, \dl (X_1\,\pmr\, Y) + \alpha_2\, \dl (X_2\,\pmb\, Y)\Big]\, \psi^{(\pmr\pmb)}_0 + \frac{\pmr 1 \mpb 1}{2} \dl \left(\frac{X_1\,\pmr\, Y}{X_2\,\pmb\, Y}\right)
J^{(c)} = 0\,.
\ee
These equations separate into two qualitatively different cases.
\begin{itemize}
\item The functions $\psi_0^{(++)}$ and $\psi_0^{(--)}$ satisfy homogeneous equations
\begin{align}
\ud \psi_0^{(++)} - \Big[\alpha_1\, \dl (X_1+ Y)  + \alpha_2\, \dl (X_2+ Y)\Big]\psi_0^{(++)} &= 0\,,\\[1pt]
\ud \psi_0^{(--)} - \Big[ \alpha_1 \, \dl (X_1- Y) + \alpha_2 \, \dl (X_2- Y)\Big]\psi_0^{(--)} &= 0\,,
\end{align}
which have the following power-law solutions 
\begin{align}
\label{eq:psi0pp}
\psi_0^{(++)} = \,c_{++}(X_1+Y)^{\alpha_1}(X_2+Y)^{\alpha_2}\,,\\
\psi_0^{(--)} = \,c_{--}(X_1-Y)^{\alpha_1}(X_2-Y)^{\alpha_2}\,,
\end{align}
where the normalizations are arbitrary. In fact, since $\psi_0^{(--)}$ has unphysical singularities at $X_{i}-Y=0$, we will set its overall coefficient to zero. The solution $\psi_0^{(++)}$ is degenerate with the disconnected solution, which we discuss below. It is also straightforward to check from~\eqref{eq:conformalexcdiag} that the integrals that compute these contributions vanish in the $\xi\to 0$ limit.

\item The functions $\psi^{(+-)}_0$ and $\psi^{(-+)}_0$ satisfy inhomogeneous equations
\begin{align}
\ud \psi_0^{(+-)} - \Big[\alpha_1\,\dl (X_1+ Y)  + \alpha_2\, \dl (X_2- Y)\Big]\psi_0^{(+-)} &= \dl \left(\frac{X_2- Y}{X_1+Y}\right)\hs J^{(c)}\,,\\
\ud \psi_0^{(-+)} - \Big[  \alpha_1\, \dl (X_1- Y)  + \alpha_2\, \dl (X_2+ Y) \Big]\psi_0^{(-+)} &= \dl \left(\frac{X_1- Y}{X_2+ Y}\right)\hs J^{(c)}\,,
\end{align}
where $J^{(c)}$ is given by~\eqref{equ:contact-source}.
Imposing regularity as $X_1\to Y$ and $X_2\to Y$, we find
\be
\label{eq:connectedsmallmass}
\psi_0^{(+-)}(X_1,X_2,Y) = - c_{\Col}\, \frac{1}{\alpha_2} \frac{(X_1+X_2)^{1+\alpha_1+\alpha_2}}{X_1+Y}\,{}_2F_1\bigg[\begin{array}{c}
1\ ,\  1+\alpha_1 \\
1 - \alpha_2
\end{array} \bigg\rvert  -\frac{X_2-Y}{X_1+Y}\,\bigg] \,.
\ee
The function $\psi_0^{(-+)}$ takes the same form, with $\alpha_1\leftrightarrow \alpha_2$ and $X_1\leftrightarrow X_2$. Using hypergeometric identities, it can be verified that the sum $\psi_0^{(+-)}+\psi_0^{(-+)}$ reproduces the solution obtained in~\cite{Baumann:2025qjx}. 
\end{itemize}
\vskip4pt
In addition to these connected solutions, we require the disconnected contribution, which is given by a product of the contact solutions~\eqref{eq:contact3pt} for each of the two vertices. In the limit $\xi\to 0$, the leading disconnected solution is
\be
\label{eq:disconnectedsmallmass}
\psi^{(\raisebox{1.25pt}{\resizebox{0.025\textwidth}{!}{\BtoBDis}})}_0 = - 2c_{\Col}(X_1+Y)^{\alpha_1}(X_2+Y)^{\alpha_2}\,,
\ee
which happens to take the same functional form as~\eqref{eq:psi0pp}. 
It is convenient to normalize the disconnected contribution by $c_{\Col}$. This choice is useful for two reasons: first, $c_{\Col}$ depends on $\alpha_1$ and $\alpha_2$, which simplifies taking the limit in which these parameters are scaled to zero; second, it allows us to keep track of how the solution depends on the normalization of $J^{(c)}$.

\vskip4pt
By combining~\eqref{eq:connectedsmallmass} and~\eqref{eq:disconnectedsmallmass}, we obtain the wavefunction for a single massive exchange with generic parameters $\alpha_1$ and $\alpha_2$. 
Since we are interested in the four-point function, we want to take the limit  $\alpha \equiv \alpha_1 = \alpha_2 \to 0$. However, this limit has to be taken with care, since the individual components of the wavefunction are actually singular in this limit. In fact, requiring this limit to be regular fixes the relative normalization between the connected and disconnected components. Putting these pieces together, we obtain
\be
\label{eq:linearcombination}
\boxed{\psi_0 = \psi_0^{(+-)}+\psi_0^{(-+)}-\left(\frac{1}{\alpha}+\frac{\pi^2}{6}\alpha\right)\psi^{(\raisebox{1.25pt}{\resizebox{0.025\textwidth}{!}{\BtoBDis}})}_0}\ ,
\ee
which is finite in the limit $\alpha \to 0$. Substituting the limiting forms of~\eqref{eq:connectedsmallmass} and~\eqref{eq:disconnectedsmallmass}, we obtain the four-point function coming from the exchange of a conformally coupled field\footnote{Note that $c_{\Col} \to-1/ (2\alpha)$ in this limit, so that $\psi_0$ is finite.} 
\be
\label{eq:dilog}
\psi_0  = -2c_{\Col}\alpha\left[ 
{\rm Li}_2\left(-\frac{X_1-Y}{X_2+Y}\right)
+{\rm Li}_2\left(-\frac{X_2-Y}{X_1+Y}\right)
+\frac{1}{2}\log^2\left(\frac{X_1+Y}{X_2+Y}\right)
-\, \frac{\pi^2}{6} \right].
\ee
The linear combination~\eqref{eq:linearcombination} involves $1/\alpha^2$ and $1/\alpha$ divergences, but these cancel, so that $\psi_0$ is finite in the limit $\alpha \to 0$, and the final result has transcendentality two. The $\pi^2/6$ term ensures that $\psi_0$ vanishes in the soft limit $Y\to 0$, as required by the original time integral.

\subsubsection*{Polylogs and symbols}

We have seen that, at leading order, the differential equations can be solved for arbitrary values of~$\alpha$. At subleading order, however,  this is generally not possible. For simplicity, we therefore expand around $\alpha = 0$.  The various contributions to the wavefunction can then be expressed in terms of (generalized) {\it polylogarithms}. 

\vskip 4pt
Functions of the type that will appear can be written as iterated integrals
\begin{equation}
\label{eq:iterated}
  F^{(n)} = \int \ud\log R_1 \circ \dots \circ \ud\log R_n =\int_a^b \left(\int _a^t\ud\log R_1 \circ \dots \circ \ud\log R_{n-1}\right)\ud\log R_{n}(t)\, ,
\end{equation}
where $n$ is the transcendentality, $R_i$ are rational functions, and $a$ and $b$ are rational numbers. The second equality implicitly defines the composition $\circ$ of the iterated integrals as in~\cite{Goncharov:2010jf}. Much of the important information about the function $F^{(n)} $, including the locations of its branch points and its differential, can be reconstructed from the rational factors $R_i$. 
A convenient way to capture this information is through the so-called {\it symbol}~\cite{goncharov2001multiple,goncharov2009simple,Goncharov:2010jf}.  (See~\cite{Duhr:2011zq,Duhr:2014woa} for a review, and~\cite{Arkani-Hamed:2017fdk,Hillman:2019wgh,Arkani-Hamed:2023kig} for related applications in the cosmological context.)

\vskip4pt
The symbol is essentially a  simple way to keep track of the factors in the iterated integral~\eqref{eq:iterated} and their ordering. Concretely, the symbol of~\eqref{eq:iterated} is
\be
\mathcal{S}(F^{(n)}) = R_1\otimes \cdots\otimes R_n\,,
\ee
where $\otimes$ denotes the tensor product taken in the multiplicative group of rational functions.
Terms separated by
$\otimes$ should implicitly be thought of as being arguments of a logarithm, so that $\otimes$ has the same multilinearity properties as products of logarithms.
The leftmost entries of the symbol encode the branch points of the corresponding function. The discontinuities across these branch cuts are obtained by deleting the entry corresponding to the branch point. The resulting symbol is the symbol of the discontinuity. Similarly, the symbol also has information about the differential of the corresponding function. In this case, one reads from right to left. Concretely, we produce the differential of a symbol by replacing the last entries by $\ud\log$ of their arguments.

\vskip4pt
In the end, the symbol encodes most of the information about polylogarithmic functions, but with a much more economical presentation. 
As an example, the symbol of~\eqref{eq:dilog} is
\be
{\cal S}(\psi_0) = \frac{X_1+Y}{X_1+X_2}\otimes\frac{X_2-Y}{X_2+Y}+\frac{X_2+Y}{X_1+X_2}\otimes\frac{X_1-Y}{X_1+Y}\,.
\ee
The symbol does project out some of the information present in the original function; in particular, terms of lower transcendentality times zeta values are lost (for example, the constant $\pi^2/6$ in~\eqref{eq:dilog} is absent in the symbol).
In most cases of interest, it is possible to integrate the symbol to recover the full function by imposing appropriate boundary conditions. In the case of~\eqref{eq:dilog}, this boundary condition is that the function vanishes as $Y\to 0$~\cite{Hillman:2019wgh}.

\subsubsection*{Subleading order}

After this digression on polylogs and symbols, it is now straightforward to continue the analysis at subleading order. 
We need to take into account that the lower-order solutions $\psi_0^{(+-)}$ and  $\psi_0^{(-+)}$ are multiplied by $\dev$ in the second line of \eqref{eq:sec5diffeq} and can therefore act as sources for $\psi_1^{(\pm\pm)}$. There are again two qualitatively different cases.
\begin{itemize}
\item The functions $\psi_1^{(+-)}$ and $\psi_1^{(-+)}$ satisfy homogeneous equations
\begin{align}
\ud\psi_1^{(+-)} - \Big[\alpha_1 \,\dl(X_1+Y)+\alpha_2\, \dl(X_2-Y)\Big]\psi_1^{(+-)}  &= 0\,,\\
\ud\psi_1^{(-+)} - \Big[\alpha_1\, \dl(X_1-Y)+\alpha_2\, \dl(X_2+Y)\Big]\psi_1^{(-+)}  &= 0\,,
\end{align}
which lead to 
\begin{align}
\label{eq:psi1ppm}
\psi_1^{(+-)} = c_{+-}\,(X_1+Y)^{\alpha_1}(X_2-Y)^{\alpha_2}\,,\\
\psi_1^{(-+)} = c_{-+}\,(X_1-Y)^{\alpha_1}(X_2+Y)^{\alpha_2}\,,
\end{align}
where $c_{+-}$ and $c_{-+}$ are integration constants (that eventually will be set to zero).

\item
The functions $\psi_1^{(++)}$ and $\psi_1^{(--)}$ satisfy inhomogeneous equations
\begin{align}
\label{eq:orderxipp}
\ud\psi_1^{(++)} - \Big[\alpha_1\, \dl(X_1+Y)+\alpha_2 \, \dl(X_2+Y)\Big]\psi_1^{(++)} &= S^{++}\,, \\
\ud\psi_1^{(--)} - \Big[\alpha_1\, \dl(X_1-Y)+\alpha_2\, \dl(X_2-Y)\Big]\psi_1^{(--)} &= S^{--} \,,
\label{eq:orderximm}
\end{align}
where the sources are 
\be
S^{\pm \pm} \equiv
 \dl \left(\frac{X_1 \pm Y}{Y}\right)\,  \psi^{(-+)}_0+   \dl \left(\frac{X_2 \pm  Y}{Y}\right) \psi^{(+-)}_0\,.
\ee
For general $\alpha_i$, these source functions are Gauss hypergeometric functions, so one expects the solutions $\psi_1^{(\pm\pm)}$ to be expressible in terms of generalized hypergeometric functions. However, we have not been able to obtain a closed-form expression in this case. Fortunately, no such formula is required in the limit $\alpha_i \to 0$.

We therefore solve the equations~\eqref{eq:orderxipp} and~\eqref{eq:orderximm} perturbatively as $\alpha \equiv \alpha_i\to 0$.  We first consider $\psi_1^{(++)}$. To order $\alpha^0$, its solution is 
\be
\psi_1^{(++)} = \frac{c_{\Col}}{\alpha}\, f^{(-2)} +c_{\Col}\,f^{(-1)} +c_{\Col}\alpha\,f^{(0)}-c_{+}\frac{c_{\Col}}{\alpha^2}\,,
\label{equ:524}
\ee
where the first two coefficient functions are
\be
\label{eq:falphaexan}
\begin{aligned}
f^{(-2)} &= \log\left(\frac{4Y^2}{X_1^+ X_2^+}\right) - c_+ \log(X_1^+ X_2^+)\,, \\[2pt]
f^{(-1)}  &=  
\frac{1}{2}\log^2\left(\frac{X_1^+}{X_2^+}\right)-\log\Big(X_1^+ X_2^+\Big)\log\left(\frac{X^+_1 X_2^+}{4Y^2}\right) -\frac{1}{2}c_+ \log^2(X_1^+X_2^+)\,,
\end{aligned}
\ee
with $X_{12}\equiv X_1+X_2$ and $X_i^\pm \equiv X_i \pm Y$. The terms proportional to $c_+$ in (\ref{equ:524}) and (\ref{eq:falphaexan}) reflect the freedom to add the 
homogeneous solution $-c_+ \hs c_{\Col}(X_1^+X_2^+)^\alpha/\alpha^2$, and these terms arise just from expanding this solution in $\alpha$.
Finally, the function $f^{(0)}$ is a transcendentality-three combination of polylogs. Its symbol is
\begin{align}
\nonumber
{\cal S}(f^{(0)}) &=X_{12}\otimes X_1^-\otimes \frac{(X_1^+)^2}{4 Y^2}+X_{12}\otimes X_1^+\otimes \frac{4 Y^2}{(X_2^+)^2}+X_1^+\otimes X_2^-\otimes \frac{4 Y^2}{(X_2^+)^2}\\\nonumber
   &+X_1^+\otimes X_1^+\otimes \frac{4 Y^2}{X_1^+
   (X_2^+)^3}+2 Y\otimes X_1^+ X_2^+\otimes X_1^+ X_2^++X_1^+\otimes X_2^+\otimes
   \frac{1}{(X_1^+)^3 (X_2^+)^3}\\[4pt]
   &+X_1^+ X_2^+\otimes 2
   Y\otimes  X_1^+ X_2^+
  -\frac{c_+}{2} \,X_1^+X_2^+\otimes X_1^+X_2^+\otimes X_1^+X_2^+ +(X_1\leftrightarrow X_2)
   \,,
   \label{eq:symbolpp}
\end{align}
where $(X_1\leftrightarrow X_2)$ indicates we should add the same expression with $X_1$ and $X_2$ interchanged everywhere.
 The integration constants in the particular solution
in~\eqref{eq:falphaexan} and~\eqref{eq:symbolpp} are fixed by two considerations: 1) we fix constants so that the only symbol entries are $X_{12}, X_1^\pm, X_2^\pm$ and $2Y$ and 2) we impose that the solution at each order in $\alpha$ has uniform transcendentality.  The solution $f^{(0)}$ can be shifted by an arbitrary transcendental constant that does not affect its symbol.\footnote{We expect that this constant can be determined by requiring the final solution to vanish as $Y\to 0$.}

\vskip 4pt
Similarly,  $\psi_1^{(--)}$ can be written as
\be
\psi_1^{(--)} =\frac{c_{\Col}}{\alpha}\,g^{(-2)} +c_{\Col} \,g^{(-1)} +c_{\Col}\alpha\, g^{(0)}+c_- \frac{c_{\Col}}{\alpha^2}\,,
\ee
where the first two coefficient functions are
\be
\begin{aligned}
g^{(-2)} &= \log\left(\frac{4Y^2}{X_1^- X_2^-}\right)+ c_- \log(X_1^- X_2^-)\\[2pt]
g^{(-1)}  &=  2\hs {\rm Li}_2\left(-\frac{X_1^-}{2Y}\right)+2\hs{\rm Li}_2\left(-\frac{X_2^-}{2Y}\right)+2\log^2(2Y)-\frac{1}{2}\left(1-c_- \right)\log^2(X_1^-X_2^-)\,,
\end{aligned}
\ee
and the coefficient $c_-$ again parameterizes the freedom to add the homogeneous solution $c_- c_{\Col}(X_1^-X_2^-)^\alpha/\alpha^2$.
As before, the function 
$g^{(0)}$ has transcendentality three and is rather  complex, so we just display its symbol
\begin{align}
\nonumber
{\cal S}(g^{(0)}) &=X_{12}\otimes X_1^-\otimes \frac{(X_2^-)^2}{4 Y^2}
+X_{12}X_1^+\otimes X_1^+\otimes \frac{4Y^2}{(X_1^-)^2}
   +X_{1}^+\otimes X_1^-\otimes \frac{1}{(X_1^-)^2(X_2^-)^2}\\[3pt]\nonumber
   &+X_1^+\otimes X_2^-\otimes \frac{4 Y^2}{(X_1^-)^2}+\frac{1}{2}X_1^-X_2^-\otimes X_1^-X_2^-\otimes \frac{1}{X_1^-X_2^-}+X_1^+X_2^+\otimes 2Y\otimes X_1^-X_2^-
   \\[2pt]
   &+2Y\otimes X_1^-X_2^-\otimes X_1^-X_2^-+\frac{1}{2}c_- \,X_1^-X_2^-\otimes X_1^-X_2^-\otimes X_1^-X_2^-+(X_1\leftrightarrow X_2)\,.
\label{eq:g0exp}
\end{align}
which can be integrated back into the full function by imposing appropriate boundary conditions.

\end{itemize}
To construct the full solution at $O(\dev)$, we also require the disconnected wavefunction at this order. Fortunately, we actually know the disconnected solution to all orders in $\dev$. It is just given by a product of the three-point contact solution~\eqref{eq:contact3pt} for the left and right vertices
\be
\begin{aligned}
\psi^{(\raisebox{1.25pt}{\resizebox{0.025\textwidth}{!}{\BtoBDis}})} =
-2c_{\Col}(2Y)^{\alpha_1+\alpha_2}\,\times\, & {}_2F_1\bigg[\begin{array}{c}
-\alpha_1-\dev, 1-\alpha_1+\dev\\[-1pt]
1-\alpha_1
\end{array}\bigg\rvert - \frac{X_1^-}{2Y}\bigg] \\
\times\, & {}_2F_1\bigg[\begin{array}{c}
-\alpha_2-\dev, 1-\alpha_2+\dev\\[-1pt]
1-\alpha_2
\end{array}\bigg\rvert - \frac{X_2^-}{2Y}\bigg] \,.
\end{aligned}
\label{equ:hyper}
\ee
At leading order, this indeed reproduces~\eqref{eq:disconnectedsmallmass}. Expanding~\eqref{equ:hyper} to $O(\dev)$, and taking the limit $\alpha\to 0$, we get
\be
\psi^{(\raisebox{1.25pt}{\resizebox{0.025\textwidth}{!}{\BtoBDis}})}_1 =  2c_{\Col} \left( f^{(-2)}+\alpha\,h^{(-1)} + \alpha^2\,h^{(0)}\right) ,
\ee
where the leading coefficient $f^{(-2)}$ is the same as in~\eqref{eq:falphaexan}, $h^{(-1)}$ is given by
\be
\begin{aligned}
h^{(-1)} =  &\ {\rm Li}_2\left(-\frac{X_1^-}{2Y}\right)+{\rm Li}_2\left(-\frac{X_2^-}{2Y}\right)+\log(2Y)\log(2YX_1^+X_2^+)\\
&\ -\frac{1}{2}\log^2(X_1^+X_2^+)-\log(X_1^+)\log(X_2^+)\,,
\end{aligned}
\ee
and $h^{(0)}$ is a fairly complicated transcendentality-three function, with symbol 
\begin{align}
\nonumber
{\cal S}(h^{(0)}) &= 
X_1^+\otimes X_1^+\otimes \frac{4 Y^2}{X_1^-X_1^+ X_2^+{}^2}+X_1^+\otimes X_1^-\otimes \frac{1}{X_1^-
   X_2^+}+X_1^+\otimes X_2^+\otimes \frac{4 Y^2}{X_1^- X_1^+{}^2 X_2^- X_2^+{}^2}\\[3pt]
   &\hspace{.325cm}+2 Y\otimes X_1^+X_2^-\otimes X_1^+
   X_2^-+\frac{1}{2}X_1^+X_2^+\otimes 2 Y\otimes X_1^-
   X_1^+X_2^-X_2^+ + (X_1\leftrightarrow X_2)
   \,.
\label{eq:h0exp}
\end{align}

We now have all the ingredients that are needed to construct the wavefunction at $O(\dev)$. Adding together the connected and disconnected solutions, we get
\be
\boxed{\psi_1 = \psi_1^{(++)}+\psi_1^{(--)} -\frac{1}{\alpha}\,\psi^{(\raisebox{1.25pt}{\resizebox{0.025\textwidth}{!}{\BtoBDis}})}_1 
}\ .
\ee
We find that, upon setting $c_+ = c_- = 1$, this is the unique linear combination (up to an additive constant) that remains regular in the $\alpha \to 0$ limit, with no folded singularities. We can write the resulting finite wavefunction as 
\be 
\psi_1 = c_{\Col}\alpha \left(f^{(0)}+g^{(0)}-2h^{(0)}\right)
\ ,
\ee
where the functions $f^{(0)}$, $g^{(0)}$ and $h^{(0)}$ are given by~\eqref{eq:symbolpp},~\eqref{eq:g0exp} and~\eqref{eq:h0exp}, respectively, with $c_+ = c_- = 1$. The factor $c_{\Col}\alpha$ is $-1/2$ as $\alpha\to 0$, and we can write the symbol of the resulting first-order wavefunction as 
\be
\begin{aligned}
{\cal S}(\psi_1)&=X_{12}\otimes X_1^-\otimes \frac{4 Y^2}{X_1^+X_2^-}+X_{12}\otimes X_1^+\otimes \frac{X_1^-X_2^+}{4 Y^2}+X_1^+\otimes X_2^-\otimes \frac{X_1^-X_2^+}{4 Y^2}\\
&\hspace{.35cm}+X_1^+\otimes X_2^+\otimes
   \frac{4 Y^2}{X_1^-X_2^-}+2 Y\otimes X_1^-\otimes \frac{X_2^+}{X_2^-}+2 Y\otimes X_1^+\otimes \frac{X_2^-}{X_2^+}\\
   &\hspace{.45cm}+X_1^+\otimes X_1^-\otimes \frac{X_2^-}{X_2^+}+(X_1\leftrightarrow X_2)\,,
\end{aligned}
\ee
which is manifestly free of folded singularities (i.e.~no $X_1^-$ or $X_2^-$ in the leftmost slot). We have checked that this matches the symbol of the solution obtained in~\cite{Gasparotto:2024bku}. 

\vskip4pt
It is possible to continue this procedure to higher orders in $\dev$. At $O(\dev^n)$, we would find a solution of transcendentality $n+2$, fixed by the same boundary conditions of regularity as $\alpha\to 0$ and absence of folded singularities. It would be interesting to characterize the polylogarithmic functions that appear to all orders.

\subsection{Heavy Particle Exchange}

It is also interesting to consider the limit where the mass of the exchanged particle goes to infinity, so that $\dev\to\infty$.
In this limit, the structure of the differential equations simplifies substantially and encodes the emergence of the effective field theory (EFT) expansion.

\vskip4pt
To begin, we write the set of equations in~\eqref{eq:sec5diffeq} in the following matrix form 
\beq
\label{eq:largemassequation}
\ud   I -(A^{(0)} + \nu \, A^{(1)})\cdot  I=  J\,,
\eeq
where we have introduced
\be
\label{eq:IandJdef}
I \equiv 
\left[
\begin{array}{c}
\psi^{(++)}\\
\psi^{(--)}\\
\psi^{(+-)}\\
\psi^{(-+)}\\
\end{array}
\right]\,,
\hspace{1cm}
J \equiv 
\left[
\begin{array}{c}
0\\
0\\
(\ell_2^+-\ell_1^-)\hs J^{(c)}\\
(\ell_1^+-\ell_2^-)\hs J^{(c)}\\
\end{array}
\right]\,,
\ee
so that the contact solution $J^{(c)}$, defined by~\eqref{equ:contact-source}, appears as a source in the equations for $\psi^{(\pm\pm)}$. It is convenient to extract the components proportional to $\nu$, so that
the relevant connection matrices in (\ref{eq:largemassequation}) are
\beq
\begin{aligned}
A^{(0)} &\equiv   \begin{bmatrix}
\alpha_1\hs \ell_1^+  + \alpha_2\hs \ell_2^+   &  0 & \frac{1}{2}(\ell_0-\ell_1^+) &   \frac{1}{2}(\ell_0-\ell_2^+)      \\[2pt]
0 &   \alpha_1 \hs \ell_1^- + \alpha_2 \hs \ell_2^- &  \frac{1}{2}(\ell_0-\ell_2^-) &   \frac{1}{2}(\ell_0-\ell_1^-)   \\[2pt]
\frac{1}{2}(\ell_0-\ell_1^-)&   \frac{1}{2}(\ell_0-\ell_2^+)  &   \alpha_1\hs \ell_1^-  + \alpha_2\hs \ell_2^+   &   0     \\[2pt]
\frac{1}{2}(\ell_0-\ell_2^-) & \frac{1}{2}(\ell_0-\ell_1^+ )  &   0&  \alpha_1\hs \ell_1^+  + \alpha_2\hs \ell_2^-   
\end{bmatrix} , \\[6pt]
A^{(1)} &\equiv  \begin{bmatrix}
0 &\quad 0&\quad \ell_1^+-\ell_0 & \quad \ell_2^+-\ell_0   \\
0 &\quad 0&\quad \ell_2^- -\ell_0 & \quad \ell_1^- -\ell_0    \\
\ell_1^- - \ell_0 & \quad \ell_2^+ -\ell_0 & \quad 0  &\quad 0  \\
\ell_2^- - \ell_0 & \quad \ell_1^+ -\ell_0 & \quad 0  &\quad 0 
\end{bmatrix} ,
\end{aligned}
\eeq
where as before $\ell_i^{\pm}\equiv \dl (X_i\pm Y)$ and $ \ell_0\equiv \dl Y$.
The choice to write the equation directly in terms of $\nu$ organizes the large-mass limit in a more convenient way than $\dev$, because this is the parameter that we are scaling to be large in this limit.

\vskip4pt
Equation~\eqref{eq:largemassequation} is form-valued, and so actually implies three separate equations. To make this explicit, it is useful to momentarily restore indices to write 
\beq
\label{eq:components}
\ud Z^i \Big( \partial_{Z^i}I^a -(A_i^{(0)}{}^{\hs ab} + \nu \, A_i^{(1)}{}^{\hs ab})\,  I^b \Big)=  \ud Z^i\,J_i^a\,.
\eeq
This is separately true for each kinematic differential $\ud Z_i = \{\ud X_{1,2},  \ud Y\}$, implying three equations. They are not independent, and solving any one of them provides a vector of functions that satisfies them all. For concreteness, we will explicitly solve the $X_1$-equation, and then comment on the relation to the other equations.

\vskip4pt
We want to solve~\eqref{eq:components} perturbatively as $\nu\to\infty$.\footnote{For unitary dS representations, $\nu$ goes to infinity along the imaginary axis. Since we are expanding in $\nu^{-1}$, we are capturing the part of the answer that is complex analytic in this parameter and so this subtlety does not matter. However, it would be very interesting to understand how to capture the nonperturbative contributions in~$\nu$ from this perspective, which carry the signatures of particle production.}
To facilitate this, we expand
\be
I = \sum_{n=0}^\infty \nu^{-n}\,I_n\,,
\ee
and solve for the component functions of the vector $I_n$ at each order in $\nu$.

\subsubsection*{Leading order}  
At leading order, equation~\eqref{eq:components} becomes trivial
\be
 - \nu \, A^{(1)}\cdot  I_0= 0\,,
\ee
and just sets $I_0 = 0$. At order $\nu$, the equation is again algebraic and reads
\be
- \, A^{(1)}_{X_1}\cdot  I_1=  J_{X_1}\,,
\ee
where $A^{(1)}_{X_1}$ is the $X_1$-component (the coefficient of $\ud X_1$) of $A^{(1)}$. 
By inverting the matrix $A^{(1)}_{X_1}$, we obtain the first-order solution
\be
\label{eq:order1largemeq}
I_1 = - \big(A_{X_1}^{(1)}\big)^{-1}\cdot J_{X_1} = 
\left[
\begin{array}{c}
J^{(c)}\\
- J^{(c)}\\
0\\
0\\
\end{array}
\right] .
\ee
We see that, at this order, the individual solutions for the basis elements are nonzero, but when we sum them to obtain the actual wavefunction, we find that $\psi_1 = 0$. One can readily check that~\eqref{eq:order1largemeq} also solves the $X_2$ and $Y$-equations. 

\vskip4pt
At next order, we have to solve the equation
\be
\label{eq:order1eqlargexi}
\frac{1}{\nu} \bigg((\partial_{X_1} -A^{(0)}_{X_1}) \cdot I_1 - A^{(1)}_{X_1}\cdot  I_2 \bigg) = 0\,.
\ee
By inverting $A^{(1)}$, this admits an algebraic solution
\be
I_2 = \big(A^{(1)}_{X_1}\big)^{-1}\cdot\big(\partial_{X_1} -A^{(0)}_{X_1}\big) \cdot I_1\,.
\ee
Upon substituting in the explicit expressions, we find
\be
I_2 =
\left[
\begin{array}{r}
\frac{1}{2} J^{(c)}\\
-\frac{1}{2} J^{(c)}\\
D_{X_1}^+ J^{(c)}\\
-D_{X_1}^- J^{(c)}\\
\end{array}
\right] ,
\ee
where,  for future convenience, we have defined the following two differential operators 
\be
D_{X_1}^\pm  = (X_1\pm Y)\partial_{X_1}-\alpha_1\,.
\ee
When we sum the entries of $I_2$, we get 
\be
\boxed{
\psi_2 = 2Y\partial_{X_1} J^{(c)}
}\ ,
\label{equ:psi2}
\ee
which is nonzero (unlike $\psi_1$).

\vskip 4pt
Our case of interest is the four-point function of conformal scalars  arising from exchanging a massive particle. We should therefore evaluate~\eqref{equ:contact-source} and \eqref{equ:psi2} as $\alpha_1,\alpha_2\to 0$. Taking this limit, we find
\be
\psi_2 = -\frac{2Y}{X_1+X_2}\,.
\label{eq:conformal4ptfunction}
\ee
Notice that this wavefunction only has the total energy singularity of a contact interaction, and indeed is precisely the four-point function of conformally coupled scalars with a $\phi^4$ interaction. Physically, this is a satisfying result: in the infinite-mass limit, the exchange of a massive particle reduces to an effective point-like contact interaction.

\subsubsection*{Subleading order}
It is interesting to continue to solve~\eqref{eq:largemassequation} iteratively at subleading order. As we will see, this reproduces the EFT expansion from integrating out the massive particle. First, we note that the recursive structure of~\eqref{eq:order1eqlargexi} persists at each order 
\be
\frac{1}{\nu^n} \bigg((\partial_{X_1} -A^{(0)}_{X_1}) \cdot I_n - A^{(1)}_{X_1}\cdot  I_{n+1} \bigg) = 0\,,
\ee
so that a formal solution is
\be
I_{n+1} = \big(A^{(1)}_{X_1}\big)^{-1}\cdot\big(\partial_{X_1} -A^{(0)}_{X_1}\big) \cdot I_n\,.
\ee
This equation can be written in a suggestive way by introducing the matrix
\be
M_{X_1} \equiv \big(A^{(1)}_{X_1}\big)^{-1}\cdot\big(\partial_{X_1} -A^{(0)}_{X_1}\big) 
=
\left[
\begin{array}{cccc}
\tfrac{1}{2} & 0& D_{X_1}^- & 0   \\
0 & \tfrac{1}{2}&0 & D_{X_1}^+   \\
D_{X_1}^+ & 0 & \tfrac{1}{2} & 0 \\
0 & D_{X_1}^- & 0 & \tfrac{1}{2}
\end{array}
\right] ,
\ee
so that $I_n = M_{X_1}^{n-1}\cdot I_1$. 
It is convenient at this point to consider  separately the cases where $n$ is even or odd. To do this, we notice that the action of $M^2_{X_1}$ on a general vector is relatively simple:
\be
M_{X_1}^{2}. \left[
\begin{array}{c}
F_1\\
F_2\\
F_3\\
F_4
\end{array}
\right] 
=
\left[
\begin{array}{c}
({\cal C}_{X_1}-D_{X_1}^+)F_1+D_{X_1}^-F_3\\
({\cal C}_{X_1}-D_{X_1}^-)F_2+D_{X_1}^+F_4\\
({\cal C}_{X_1}-D_{X_1}^-)F_3+D_{X_1}^+F_1\\
({\cal C}_{X_1}-D_{X_1}^+)F_4+D_{X_1}^-F_2\\
\end{array}
\right] 
,
\ee
where we have defined the second-order differential operator
\be
\begin{aligned}
{\cal C}_{X_1} &=D_{X_1}^\pm D_{X_1}^\mp+D_{X_1}^\mp+\frac{1}{4}\\
&= (X_1^2-Y^2)\partial_{X_1}^2+2(1-\alpha_1)X_1\partial_{X_1}+\left(\alpha_1-\frac{1}{2}\right)^2\,.
\end{aligned}
\ee
Adding up the various components, we find the extremely simple recursion relation between wavefunctions at different orders in $\nu$:
\be
\psi_{n+2} = {\cal C}_{X_1}\psi_{n}\,.
\ee
Since $\psi_1=0$, we find that the odd contributions vanish, $\psi_{2n+1} = 0$, while the even contributions are given by $\psi_{2n} = {\cal C}_{X_1}^{n-1}\psi_{2}$. Putting this all together, we find that the wavefunction in the large mass limit is
\be
\label{eq:EFTexpansion}
\boxed{\psi_{\rm EFT} = \sum_{n=1}^\infty \nu^{-2n}\,{\cal C}_{X_1}^{n-1}\left(2Y\partial_{X_1} J^{(c)}\right) }\ .
\ee
Notice that we have written $\psi_{\rm EFT}$ in terms of the differential operator ${\cal C}_{X_1}$. One can easily check that the four-point function~\eqref{eq:conformal4ptfunction} satisfies $({\cal C}_{X_1}-{\cal C}_{X_2})\psi_2 = 0$, so we could have just as well written the answer in terms of $X_2$ (and indeed this is the expression we would have obtained by solving the $X_2$-equations). Alternatively, we can straightforwardly check that these expressions solve the $X_2$ and $Y$-equations as well. It is worth noting that the nontrivial compatibility between these equations relies on the fact that $\psi$ satisfies $({\cal C}_{X_1}-{\cal C}_{X_2})\psi = 0$, which is equivalent to the conformal Ward identities arising from de Sitter symmetry~\cite{Arkani-Hamed:2018kmz}. It is rather interesting that these Ward identities are an emergent consequence of the first-order system~\eqref{eq:largemassequation}.

\vskip4pt
The expression~\eqref{eq:EFTexpansion} is a remarkably compact formula for the all-orders EFT expansion in the large $\nu$ limit. We see that, in this limit, the differential equations rather nicely organize themselves into algebraic equations that can be solved iteratively. It is further interesting to note that~\eqref{eq:EFTexpansion} can be formally summed to obtain
\be
\psi_{\rm EFT} = -\frac{1}{{\cal C}_{X_1}-\nu^2}\, \psi_2\,,
\label{eq:formalresum}
\ee
where 
$\psi_2$ is the contact solution~\eqref{eq:conformal4ptfunction}. By acting with the differential operator ${\cal C}_{X_1}-\nu^2$ on both sides, we obtain a second-order equation satisfied by the {\it full} wavefunction. (To compare this with~\cite{Arkani-Hamed:2018kmz}, note that ${\cal C}_{X_1} \equiv \Delta_u + \frac{1}{4}$, where $\Delta_u$ is the differential operator introduced in~\cite{Arkani-Hamed:2018kmz}.)
We can think of the EFT solution as formally inverting this differential operator, which removes the zero mode that contains the nonperturbative in $\nu^{-1}$ pieces due to cosmological particle production in de Sitter. 
This is reflected in the fact that the series~\eqref{eq:EFTexpansion} is only asymptotic in $\nu^{-1}$ (see also~\cite{DuasoPueyo:2025lmq}). It would be very interesting to understand whether these nonperturbative corrections can be deduced from the series~\eqref{eq:EFTexpansion} using techniques from resurgence.

\subsection{Collapsed Limit}
It is instructive to analyze the first-order system in a special kinematic regime. In this subsection, we study the collapsed limit 
$Y \to 0$ and show that the system can be solved exactly for arbitrary mass. This makes the emergence of particle production particularly transparent.

\vskip4pt
It is most convenient to start from the equations~\eqref{eq:2ndorderfirstordersys}:
\ba
\partial_{X_1}\psi_{\BtoB} &= \frac{1}{X_1^2-Y^2}\left[ (\alpha_1+\dev) X_1 \psi_{\BtoB}- (\alpha_1-\dev)Y \tilde \psi_{\BtoB} -2 Y \hs g^2\hs c_{\Col} (X_1+X_2)^{\alpha_1+\alpha_2} \right] ,\\
\partial_{X_1}\tilde \psi_{\BtoB} &= \frac{1}{X_1^2-Y^2}\left[ (\alpha_1-\dev)X_1 \tilde \psi_{\BtoB} -(\alpha_1-\dev) Y \psi_{\BtoB}+2 X_1\hs g^2\hs c_{\Col} (X_1+X_2)^{\alpha_1+\alpha_2} \right] .
\label{equ:EQUATIONS}
\ea
Here, $\psi_{\BtoB}$ denotes the connected part of the wavefunction and 
$\tilde \psi_{\BtoB}$  is the corresponding auxiliary combination introduced earlier.
 To simplify notation, we will drop the subscripts, i.e.~$\psi \equiv \psi_{\BtoB}$ and $\tilde \psi \equiv \tilde \psi_{\BtoB}$.  The normalization $c_{\Col} $ is defined in~\eqref{equ:contact-source}, and we have restored the coupling $g^2$ for clarity.  
In order to simplify the system, it is helpful to define
\beq
\begin{aligned}
\hat \psi &\equiv \frac{1}{2Y}\hs \psi\,,\\[4pt]
\hat G &\equiv \frac{1}{2Y}  \left[(\alpha_1-\dev) \hs Y\hs \tilde \psi +2 Y \hs g^2 \hs c_{\Col} \hs (X_1+X_2)^{\alpha_1+\alpha_2}\right] ,
\end{aligned}
\eeq
so that (\ref{equ:EQUATIONS}) becomes
\beq
\begin{aligned}
\partial_{X_1} \hat \psi &= \frac{1}{X_1^2-Y^2}\Big[ (\alpha_1+\dev) X_1 \hat \psi - \hat G \Big] ,\\
\partial_{X_1} \hat G &= \frac{(\alpha_1-\dev)}{X_1^2-Y^2}\Big[ X_1 \hat G -(\alpha_1-\dev) Y^2 \hat \psi \Big] + g^2\hs c_{\Col} (\alpha_1+\alpha_2)(X_1+X_2)^{\alpha_1+\alpha_2-1}.
\end{aligned}
\eeq
Taking the limits $\alpha_1,\alpha_2 \to 0$ and $Y\to 0$, 
we then obtain the reduced system
\beq
\begin{aligned}
\partial_{X_1}\hat\psi &= \frac{\dev}{X_1}\hat \psi - \frac{1}{X_1^2}\hat G \, ,\\
\partial_{X_1}\hat G &= -\frac{\dev}{X_1} \hat G  -\frac{g^2}{X_1+X_2} \,,
\end{aligned}
\label{eq:firstorderequationY0}
\eeq
where we used $c_{\Col} \to - (\alpha_1 +\alpha_2)^{-1}$.
We see that the $Y\to 0$ limit greatly reduces the complexity of the system. In particular, $\hat \psi$ has dropped out of the equation for $\hat G$, so the system is now in upper-triangular form: $\hat G$ obeys a sourced homogeneous equation and is determined independently, and then $\hat \psi$ follows by a single integration. 

\vskip 4pt
The only remaining dimensionless kinematic variable in the limit $Y=0$ is the ratio $r \equiv X_2/X_1$, and so it is convenient to express $\hat \psi$ and $\hat G$ as 
\beq
\begin{aligned}
\hat \psi(X_1,X_2)  &= (X_1X_2)^{-1/2} \, \uppsi(r)\, , \\
\hat G(X_1,X_2)  &= \left(\frac{X_2}{X_1}\right)^{-1/2} \,{\cal G}(r)\, .
\end{aligned}
\eeq	
Defining $r \equiv e^\rho$, the equations~\eqref{eq:firstorderequationY0} then take the form
\beq
\begin{aligned}
\big(\partial_\rho+\nu\big)\uppsi(\rho) &= {\cal G}(\rho)\,,\\
\big(\partial_\rho-\nu\big) {\cal G}(\rho) &= \frac{g^2}{2}{\rm sech}\Big(\frac{\rho}{2}\Big)\, ,
\end{aligned}
\label{eq:y02}
\eeq
where recall $\nu = \dev + \frac{1}{2}$. 
For generic $\nu$, the solutions of the first-order system (\ref{eq:y02}) can be written in terms of exponential and Gauss hypergeometric functions
\beq
\begin{aligned}
{\cal G}(\rho) &= 2\nu\, A\,e^{\nu\rho} + g^2\, F_-(\rho)\,,\\[2pt]
\hspace{-2cm}\uppsi(\rho)&= Ae^{\nu\rho}+ Be^{-\nu\rho} + \frac{g^2}{2 \nu}\Big[ F_-(\rho) - F_+(\rho) \Big]\, , 
\end{aligned}
\eeq
where $A$ and $B$ are integration constants and
\beq
F_\pm(\rho) \equiv \frac{e^{\rho/2}}{\frac{1}{2}\pm \nu} \, {}_2F_1\bigg[\begin{array}{c}
1 \,,\,  \tfrac{1}{2} \pm \nu\\[-1pt]
\tfrac{3}{2} \pm \nu
\end{array}\bigg\rvert  -e^{\rho}\,\bigg] \,.
\eeq
In order to interpret this solution, it is useful to notice that the equations~\eqref{eq:y02} can be combined into
\be
\boxed{\Big(\partial_\rho^2 -\nu^2\Big) \uppsi=  \frac{g^2}{2}{\rm sech}\left(\frac{\rho}{2}\right)}\ .
\ee
For heavy fields, where $\nu = i\mu$ (with $\mu \in\mathbb{R}$), this is the equation of a {\it forced harmonic oscillator} in the kinematic variable $\rho =\log(X_2/X_1)$~\cite{Arkani-Hamed:2018kmz}.
This makes manifest that the solution will display the characteristic oscillations of a massive field.
Here, we have seen this oscillator solution emerge from the first-order system.

\section{Conclusions}
\label{sec:Conclusions}

Adding mass to scalar particles in flat space is a relatively mild modification to the structure of the theory, where the mode functions and (tree-level) analytic structure of observables do not change.
In contrast, massive particles in de Sitter space are substantially different from massless  ones.
Perhaps the most relevant feature is that the mode functions become Hankel functions, which leads to a significantly  more intricate analytic structure of the associated correlation functions.

\vskip4pt
In this paper, we have shown that despite this apparent complexity, the differential equations satisfied by massive correlators are governed by remarkably simple combinatorial structures that are very similar to their conformally coupled counterparts. 
As in the conformal case, the differential equations can be viewed as a flow through the space of kinematic variables, visualized in terms of graphical tubings and governed by simple rules. There are also interesting new features in the massive case:
1) Massive propagators require additional basis functions, corresponding to new graph tubings, and 2) Tubes pierced by massive propagators can shrink to produce new source functions (while in the conformally coupled case they only grow).  
Accounting for these effects, the kinematic flow rules generate the differential equations for arbitrary Feynman graphs. 

\vskip4pt
There are many phenomenologically important questions in which integrals of this type arise. For example, characterizing the signatures of massive particles during inflation requires understanding processes involving massive exchanges~\cite{Chen:2009zp,Baumann:2011nk,Noumi:2012vr,Arkani-Hamed:2015bza,Lee:2016vti,Arkani-Hamed:2018kmz,Baumann:2019oyu,Sleight:2019hfp,Wang:2019gbi,Bodas:2020yho,Pimentel:2022fsc,Jazayeri:2022kjy}. To demonstrate the utility of these techniques, we solved the simplest such case (involving a single massive exchange) in the parametric limits of large and small masses, which are particularly easy to take at the level of the differential equations. The large-mass limit is closely connected to the effective field theory expansion in de Sitter space, and it would be interesting to understand the non-perturbative corrections that arise from cosmological particle production. Furthermore, in Appendix~\ref{app:massless}, we describe a dictionary between the equations for massive fields in de Sitter space and massless fields in a power-law cosmology. The small-mass limit is mapped to the de Sitter limit, so a systematic study of this limit may shed some light on the infrared structure of massless fields in de Sitter space, where the deviation from de Sitter serves as a physical regulator. 

\vskip4pt
Finally, it is conceptually interesting that massive correlators share so many similarities with their simpler conformally coupled cousins. Aside from the existence of a kinematic flow, the massive wavefunction can be written as a twisted integral of a universal rational integrand (see Appendix~\ref{app:combin}). This is therefore an ideal representation to study the properties of generic Feynman graphs (including features like  the Passarino--Veltman reduction and the evaluation of loop momentum integrals). Studying the properties of the {\it integrand} will also be interesting.  
In the conformal case, the integrand is closely related to ``cosmological polytopes"~\cite{Arkani-Hamed:2017fdk}. In the massive setting, this suggests the existence of corresponding ``massive cosmological polytopes", which represent new geometric structures worthy of further exploration. 
Relatedly, we saw in several places the emergence of second-order conformal Ward identities  from the first-order system of differential equations. From the bulk perspective, these Ward identities follow from the symmetries of de Sitter space. Clarifying how this spacetime picture emerges from the combinatorial structure of the first-order system may therefore illuminate the broader question of how spacetime itself emerges from more fundamental principles.

 \paragraph{Acknowledgments} Thanks to Santiago Ag\"u\'i Salcedo, Nima Arkani-Hamed, Shounak De, Saba Etezad-Razavi, Yuhan Fu, Daniel Glazer, Thomas Grimm, Callum Jones, Guilherme Pimentel, Andrzej Pokraka, Giulio Salvatori, Simon Telen, and Chen Yang for helpful discussions. We would additionally like to thank Harry Goodhew for discussions and collaboration on related topics. We are especially grateful to Paolo Benincasa for inspiring and extremely helpful discussions about the content of this paper, and for collaboration on related topics.
 
 \vskip 4pt
 \noindent
The research of DB is funded by the European Union (ERC,  \raisebox{-2pt}{\includegraphics[height=0.9\baselineskip]{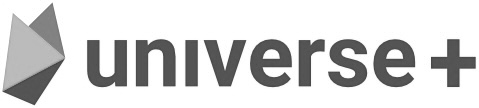}}, 101118787). DB is further supported by a Yushan Professorship at National Taiwan University (NTU) funded by the Ministry of Education (MOE) NTU-112V2004-1. He also holds the Chee-Chun Leung Chair of Cosmology at NTU.  AJ is supported by DOE award DE-SC0025323 and by the Kavli Institute for Cosmological Physics at the University of Chicago. HL is supported in part by DOE award DE-SC0013528.
KSV is supported by the European Research Council under the European Union's Seventh Framework Programme (FP7/2007-2013), ERC Grant agreement ADG 834878. KSV is also supported  by the Dutch Research Council (NWO  \raisebox{-3pt}{\includegraphics[height=0.9\baselineskip]{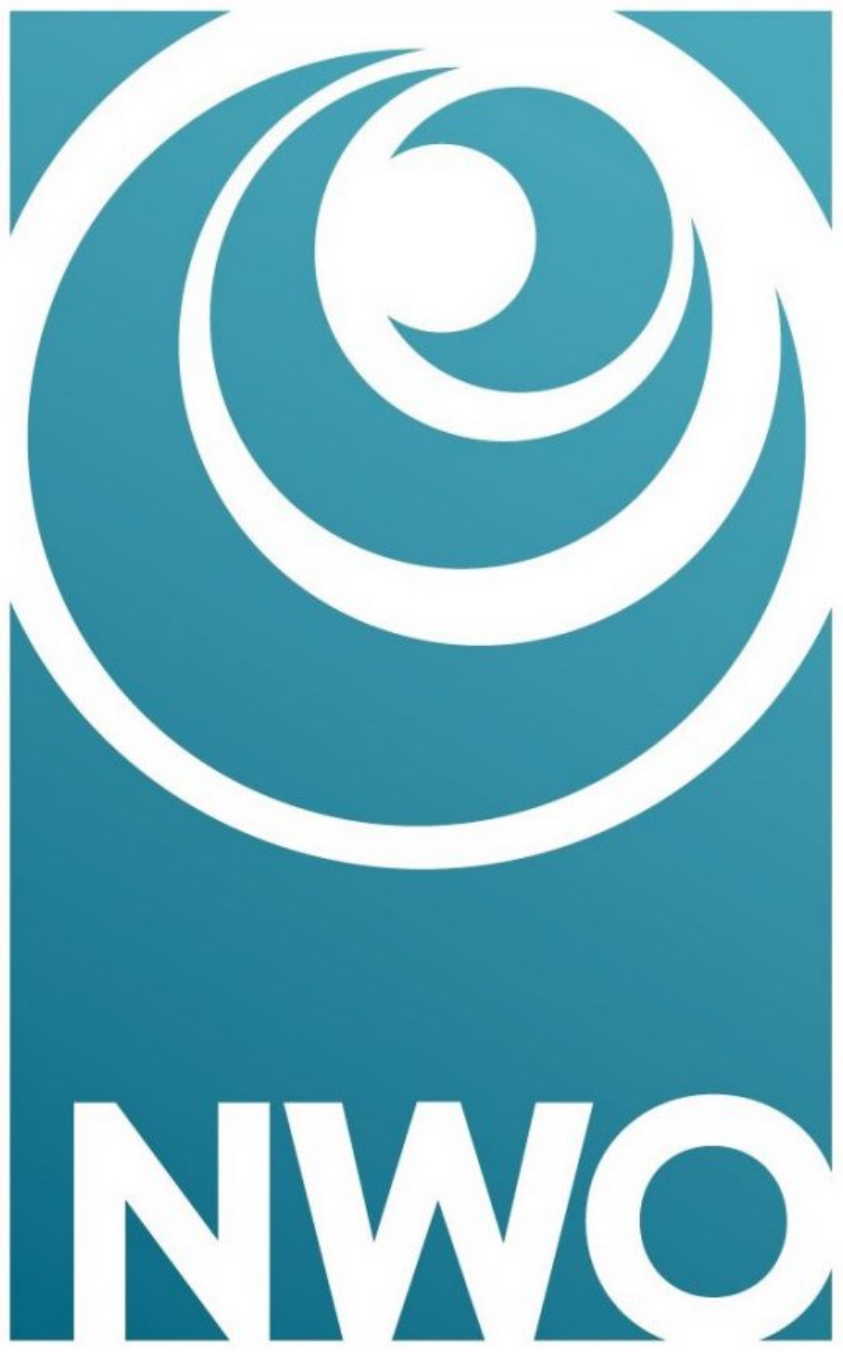}}) under {\hypersetup{urlcolor=black}\href {https://doi.org/10.61686/PRTOQ68396}{the grant}} for the project \textit{Constraining the Space of Cosmological Theories} with file number VI.Veni.242.438 of the research programme NWO Talent Programme, Veni.

\newpage
\appendix

\section{Massless Integrals}
\label{app:massless}

In the main text, we focused on massive fields in exact de Sitter spacetime. 
However, all of our formulas also apply to correlation functions of massless scalar fields in a power-law cosmology, after a simple relabeling of parameters. In this appendix, we provide a dictionary between these two situations. (See also~\cite{Benincasa:2022gtd}.)

\subsection{Power-Law Cosmology}

We begin by considering the following action of a massive scalar field in a $(d+1)$-dimensional curved spacetime 
\be
S = \int\ud^{d+1}x\sqrt{-g}\Bigg(\frac{1}{2}\phi\left[\square -\frac{d-1}{8d}R \right]\phi-\frac{m^2}{2}\phi^2+\frac{\kappa}{2}R\phi^2- \sum_{p=3}^{\infty}\frac{\lambda_p}{p!} \phi^p\Bigg)\,,
\label{eq:massivescalarFLRW}
\ee
where we allowed for a coupling to the Ricci scalar $R$, with
\beq
\kappa =\begin{cases} \ \ \, 0 & \text{conformal coupling}\,, \\[6pt]  {\displaystyle \frac{d-1}{4d}} & \text{minimal coupling}\,. \end{cases} 
\eeq
We assume that the cosmological background is described by a flat Robertson--Walker metric, with scale factor $a(\eta)$ and conformal Hubble parameter ${\cal H}\equiv a'/a$. By a Weyl rescaling, $g_{\mu\nu} = a^2\eta_{\mu\nu}$ and
$\phi = a^{-(d-1)/2}\varphi$, the theory can then be cast
as a time-dependent theory in flat Minkowski space:
\be
S = \int\ud^{d+1}x \,\Bigg(-\frac{1}{2} \eta^{\mu \nu}\partial_\mu \varphi \partial_\nu \varphi -\frac{m^2(\eta)}{2}\varphi^2-\sum_{p=3}^{\infty}\frac{\lambda_p(\eta)}{p!} \varphi^p
\Bigg)\, ,
\label{eq:flatspacefieldtheory}
\ee
where the time-dependent mass and couplings are
\beq 
\begin{aligned}
m^2(\eta) &= m^2 a^2-   2d\hs  \kappa \bigg({\cal H}'+\frac{d-1}{2}{\cal H}^2
\bigg)\,,\\
\lambda_p(\eta)&= \lambda_p\hs a^{2+\frac{1}{2} (d-1)(2-p)}\,.
\end{aligned}
\eeq
Note that for $d=3$, the coupling becomes $\lambda_p(\eta) = \lambda_p\hs a^{4-p}$, which is time-independent for the conformal value $p=4$.

\vskip 4pt
We let the cosmological background be a power law in conformal time:
\be\label{eq: FRW scale factor}
a(\eta)= \left( \frac{\eta}{\eta_0}\right)^{-(1+\varepsilon)}\,.
\ee
For $\varepsilon>-1$, the evolution corresponds to 
 accelerated expansion, while, for $\varepsilon<-1$,  we have decelerated collapse. Several cosmologies can be described by this power-law ansatz: $\varepsilon=0$ (de Sitter), $\varepsilon=-1$ (Minkowski), $\varepsilon=-2$ (radiation-dominated) and  $\varepsilon=-3$ (matter-dominated).
 
 \vskip 4pt
The linear equation of motion for the field (in Fourier space) then is
\be
\varphi_k''+\Big(k^2+m^2(\eta)\Big)\varphi_k = 0\,,
\label{eq:powerlawfrwscalareom}
\ee
where the effective mass is
\be
\label{eq:effectivemass}
m^2(\eta) = m^2 \left(\frac{\eta}{\eta_0}\right)^{-2(1+\varepsilon)}- \frac{\kappa\,(1+\varepsilon)}{\eta^2}\Bigg(\frac{4d+2d(d-1)(1+\varepsilon)}{2}\Bigg)\,.
\ee
Equations of the form~\eqref{eq:powerlawfrwscalareom} can be solved in terms of Bessel functions if the mass scales like $\eta^{-2}$. In these cases, we can define $m^2(\eta) \equiv \mu^2(\varepsilon,m)\hs \eta^{-2}$, so that the Bunch--Davies solution  is
\be
\hat f_k(\eta) = \frac{\sqrt \pi}{2}e^{-i\frac{\pi\nu }{2}}(-\eta)^{1/2}H_\nu^{(2)}(-k\eta)\,,\qquad \nu \equiv \sqrt{\frac{1}{4}- \mu^2(\varepsilon,m)}\,,
\label{eq:hankelsoln}
\ee
where $H_\nu^{(2)}$ is a Hankel function of the second kind.
Looking at~\eqref{eq:effectivemass}, we see that there are two scenarios with this time dependence for the mass~\cite{Benincasa:2022gtd}.

\begin{itemize}
\item {\boldsymbol{$\varepsilon=0$}:} This is the case of a scalar field of generic mass in an exact de Sitter spacetime, which we considered in the main text. It is convenient to define $\eta_0 \equiv -H^{-1}$. We then have
\be
\nu =\begin{cases}  {\displaystyle \sqrt{\frac{1}{4}-\frac{m^2}{H^2}}}  &  \text{conformal coupling}\,, \\[12pt]  {\displaystyle  \sqrt{\frac{d^2}{4}-\frac{m^2}{H^2}}}  & \text{minimal coupling}\,. \end{cases} 
\ee
Note that, for the conformal coupling, the mass is shifted, so that the formula for $\nu$ is independent of the spacetime dimension.

\item  {\boldsymbol{$m=0$}:} This is the case of a massless field in an arbitrary power-law cosmology ($\varepsilon \ne 0$).
Note that the scaling $m^2(\eta) \propto \eta^{-2}$ follows for any value of the non-minimal coupling parameter $\kappa$. For the conformal and minimal couplings, the parameter $\nu$ simplifies to
\be
\nu = \begin{cases}  {\displaystyle \frac{1}{2}}  &  \text{conformal coupling}\,, \\[12pt]  {\displaystyle \frac{1}{2}+\frac{d-1}{2}(1+\varepsilon)}  & \text{minimal coupling}\,. \end{cases} 
\label{equ:nu}
\ee
The conformally coupled case was considered in detail in~\cite{Arkani-Hamed:2017fdk,Arkani-Hamed:2023kig,Arkani-Hamed:2023bsv,Baumann:2025qjx,De:2023xue,Fan:2024iek,He:2024olr,De:2024zic,Glew:2025ypb,Capuano:2025ehm,McLeod:2026jpz,Fu:2026dqb}. 
Here, we focus on the minimally coupled case.

\end{itemize}
\noindent
Given that the mode functions in each of these cases can be written in terms of Hankel functions, we can now relate them to the constructions in the main text.

\subsection{Massless Fields}

We now describe how the differential equations derived in the main text can be applied to the case of massless fields in a general power-law cosmology. It is useful to briefly summarize the massive construction first.

\vskip 4pt
Recall that in Section~\ref{sec:background}, we worked in terms of the original field $\phi$ and not the Weyl-transformed field $\varphi =  a^{(d-1)/2}\phi$. The mode function for the field $\phi$ is then~\eqref{eq: modefunction}, with vertex factors $V_i = i\lambda_{p_i} (-H\eta)^{-(d+1)}$, for each vertex $i$.
In order to simplify the structure of the differential equations, we then defined $g_k$ in~\eqref{equ:g} by extracting an overall factor as
\be
f_k(\eta) = (-\eta)^{d/2-\nu}  \times i H^{\frac{1}{2}(d-1)} \hs e^{-i\frac{\pi\nu}{2}}  \frac{g_k(\eta)}{\sqrt{2k}}\,.
\label{eq:appmodefmassive}
\ee
Since this overall factor involves powers of $\eta$, we absorbed these factors into an effective vertex factor
\be
V_{i,\rm eff} \equiv  i \lambda_{p_i} H^{-(d+1)}(-\eta)^{-(1+\alpha_i + \dev_i)}\,,
\label{equ:Veff}
\ee
where the parameters $\alpha_i$ and $\xi_i$ were defined in~\eqref{eq:paramsaxi}. In Section~\ref{sec:DE}, we then derived differential equations satisfied by the wavefunction in terms of these parameters.

\vskip 4pt
We now consider the analogous construction for a massless field. In that case, the mode function of $\phi$ is 
\begin{align}
f_k(\eta) 
&= i \eta_0^{-\frac{1}{2}(d-1)(1+\varepsilon)}e^{-i\frac{\pi\nu}{2}} \frac{g_k(\eta)}{\sqrt{2k}}\,,
\end{align}
which is of the form~\eqref{eq:appmodefmassive} if we write $\eta_0 = H^{-1}$ and account for additional factors of $\eta_0$ required by dimensional analysis. The substantive difference is that we do not need to extract any factors of $\eta$, so the effective vertex factor in this case is the original vertex factor
\be
V_i = i\lambda_{p_i} \left( \frac{\eta}{\eta_0}\right)^{-(d+1)(1+\varepsilon)}\,.
\label{equ:Vp}
\ee
Comparing this to (\ref{equ:Veff}), we see that the massless field in a power-law cosmology  maps to the case of a massive field in de Sitter if we
set  $1+\alpha_i +\xi_i = (d+1)(1+\varepsilon)$.

\vskip 4pt
In the massive case, we had two parameters $\alpha_i$ (which depends on $d$ and $p_i$, the number of lines connected to the vertex $i$) and $\dev_i$ (which depends on the masses of the fields entering the vertex). In the massless case, we instead have the parameters $\varepsilon$, $d$, and $p_i$.
It is easy to define a dictionary between those sets of parameters. We first use  (\ref{equ:nu}) to express $\dev \equiv \nu - \frac{1}{2}$ in terms $\varepsilon$. We then substitute this into $\alpha_i = (d+1)(1+\varepsilon) - 1- \xi_i$ and use the fact that $\dev$ is the same for all fields to write $\dev_i = p_i\dev$. This leads to the following map:
\begin{align}
\dev_i &= \frac{d-1}{2}(1+\varepsilon) \, p_i\,,\\
\alpha_i &= (1+\varepsilon) \left(d+1-\frac{d-1}{2}\,p_i \right) -1\,.
\end{align}
With these replacements, the differential equations and the kinematic flow derived in the main text can be applied directly to  massless fields in a power-law cosmology.

\newpage
\section{Combinatorics and Geometry}
\label{app:combin}

In Section~\ref{sec:DE}, we derived first-order differential equations for 
the wavefunction coefficients by viewing them as time integrals that sum over the spacetime evolution of the theory.
An interesting complementary approach, summarized in Section~\ref{sec:twistedint}, is to 
cast the integrals of interest as twisted integrals of rational functions in an auxiliary space of energy variables.  
 In this appendix, we elaborate further on this alternative perspective, which makes contact with geometry and combinatorics.

\subsection{Integrals and Combinatorics}
\label{sec:intcomb}

The key idea is to use the following integral representation of the Hankel function (see also~\cite{Gasparotto:2024bku})
\be
H_\nu^{(2)}(x) = \frac{i}{\sqrt{\pi}}\frac{2^{1+\nu}x^{-\nu}}{\Gamma[\tfrac{1}{2}-\nu]}\int_1^{\infty}\ud t \,e^{-ixt}(t^2-1)^{-\nu-\frac{1}{2}}\, ,
\label{eq:besselID}
\ee
which expresses the Hankel function as an integral over a rational function times an exponential.\footnote{Strictly speaking, this representation is only valid for ${\rm Re}\,\nu <1/2$. However, since our ultimate goal is to derive differential equations for $\psi$, we will adopt the philosophy that the parameter $\nu$ can be analytically continued directly at the level of the differential equations. Alternatively, we can use the fact that $H_\nu^{(2)}(x) = e^{i\pi\nu}H_{-\nu}^{(2)}(x)$ to find the equivalent expression
\be
H_\nu^{(2)}(x) = \frac{i}{\sqrt{\pi}}\frac{2^{1-\nu}e^{i\nu\pi}x^\nu}{\Gamma[\nu+\tfrac{1}{2}]}\int_1^{\infty}\ud t \,e^{-ixt}(t^2-1)^{\nu-\frac{1}{2}}\, ,
\label{eq:besselID2}
\ee
which is valid for ${\rm Re}\,\nu > -1/2$.
}
Using this representation (and its complex conjugate for $H^{(1)}_\nu(x)$), we can write the propagators in~\eqref{equ:scale-prop}  as 
\beq
\begin{aligned}
\hat K(k,\eta) &={\cal N}_k\, \int_1^{\infty}\ud s\, (s^2-1)^{-\nu-\frac{1}{2}}\,{\cal K}(k,\eta) \,,\\
\hat G_{\rm F}(k,\eta,\eta') & = 2k \hs \lvert {\cal N}_k\rvert^2\,\int_1^{\infty}\ud t_1\ud t_2\, (t_1^2-1)^{-\nu-\frac{1}{2}}(t_2^2-1)^{-\nu-\frac{1}{2}}\,{\cal G}_{\rm F}(k,\eta,\eta')\,,\\
\hat G_{\rm D}(k,\eta,\eta') & = 2k\hs \lvert {\cal N}_k\rvert^2\,\int_1^{\infty}\ud t_1\ud t_2\, (t_1^2-1)^{-\nu-\frac{1}{2}}(t_2^2-1)^{-\nu-\frac{1}{2}}\,{\cal G}_{\rm D}(k,\eta,\eta')\,,
\end{aligned}
\eeq
where we have defined the normalization constant 
\be
{\cal N}_k \equiv \frac{2^{\nu+\frac{1}{2}}}{\Gamma[\tfrac{1}{2}-\nu]}k^{-\xi}\,,
\ee
along with the reduced propagators
\beq
\label{eq:reduced}
\begin{aligned}
{\cal K}(k,\eta) &\equiv e^{i (ks) \eta}\,,\\
{\cal G}_{\rm F}(k,\eta,\eta') &\equiv \frac{1}{2k}\left(e^{-ik( t_1\eta-t_2\eta')}\theta(\eta-\eta')+e^{-ik( t_2\eta'-t_1\eta)}\theta(\eta'-\eta)\right) , \\
{\cal G}_{\rm D}(k,\eta,\eta')  &\equiv \frac{1}{2k}e^{ik( t_1\eta+t_2\eta')}\,.
\end{aligned}
\eeq
The benefit of these various redefinitions is now apparent: the reduced propagators take exactly the same form as those of a scalar field in flat space. The novel feature is that each internal line has two energy variables $kt_1$ and $kt_2$, which are integrated over.

\vskip 4pt
The reason for introducing the integral representation of the Hankel function is that the integrals over time simplify. 
Recall that each vertex has a factor
\beq
V_i =  i \lambda_i (-\eta_i)^{-(1+\gamma_i)} = -\lambda_i\,\frac{i^{\gamma_i}}{\Gamma[1+\gamma_i]}\int_0^{\infty}\ud x\, x^{\gamma_i} e^{ix \eta_i}\,,
\label{eq:energyintegral}
\eeq
where $\gamma_i  \equiv \alpha_i+\xi_i$, with $\alpha_i$ and $\xi_i$ defined in~\eqref{eq:paramsaxi}. In the second equality of~\eqref{eq:energyintegral}, we have expressed the polynomial factor in (conformal) time as an energy integral~\cite{Arkani-Hamed:2017fdk}.\footnote{Note that in some cases 
$\alpha_i$ will be positive. In these cases, the integer part of this combination should be written as derivatives with respect to vertex energies, with the remainder parametrized as an energy integral.}
This will allow us to perform all the time integrals and write the wavefunction as an integral of a rational integrand.
In practice, we will be interested in the limit $\eta_* \to 0$, where the constants appearing in~\eqref{eq: F C N} simplify for real values of $\nu$. In particular, we then have $C_k^{({\rm D})} = C_k^{({\rm F})}$. This simplification makes it easy to recombine these pieces at the level of the integrand. In cases involving principal series fields, the relative factor between the Feynman and disconnected propagators is more complicated, and we will denote it abstractly by $A$, keeping in mind that $A=1$ for real $\nu$.

\subsubsection*{Universal integrand}

The wavefunction coefficient  $\psi$ 
 for arbitrary internal and external states becomes a (twisted) integral over the parameters $s, t, x$ of a {\it universal integrand}, $\uppsi$. This universal integrand can be obtained via the usual Feynman rules with the building blocks~\eqref{eq:reduced} and \eqref{eq:energyintegral}. The physical states and interactions of interest then dictate how we twist this integrand and integrate it.

\vskip4pt
Let us illustrate this with the example of a two-site massive exchange: 
\be
\begin{aligned}
\uppsi^{(2)} &=
\raisebox{-12pt}{
\begin{tikzpicture}[line width=1. pt, scale=2]
\draw[fill=black] (0,0) -- (1,0);
\draw[fill=black] (0,0) circle (.03cm);
\draw[fill=black] (1,0) circle (.03cm);
\draw[fill=white] (.5,0) circle (.0275cm);
\node[scale=.85] at (0,-.15) {$X_1$};
\node[scale=.85] at (1,-.15) {$X_2$};
\node[scale=.85] at (.5,-.15) {$Y$};
\node[scale=.85] at (.25,.12) {$t_{L}$};
\node[scale=.85] at (.75,.12) {$t_{R}$};
\end{tikzpicture}
} \\[2pt]
 &=-\int\ud\eta_1\ud\eta_2\, e^{iX_1\eta_1}e^{iX_2\eta_2}\Big[{\cal G}_{\rm F}(\eta_1,\eta_2)- A\,{\cal G}_{\rm D}(\eta_1,\eta_2)\Big]\,,
\label{eq:integrandexc}
\end{aligned}
\ee
where we have labeled the graph by the energies $X_a$ ($a=1,2$) of each external vertex, its internal vertex energy $Y$, and the variables $t_{L,R}$ associated to the internal line. Since the integrand does not depend on the number of external lines emanating from a vertex, it makes sense to truncate the external lines as we have done.
We have also marked the internal line with a white circle to note the fact that the internal energy can change in the exchange (which will eventually be integrated over).
Performing the time integrals, we obtain
\be
\uppsi^{(2)} = \uppsi^{(2)}_{\rm F} - A \,\uppsi^{(2)}_{\rm D} \,,
\label{eq:singleexchangemassive}
\ee
where the time-ordered and disconnected pieces are  
\beq
\begin{aligned}
\uppsi^{(2)}_{\rm F} &\equiv  \frac{1}{2Y} \bigg[\frac{1}{X_1+X_2+(t_L-t_R)Y}\frac{1}{X_1+t_LY} +\frac{1}{X_1+X_2+(t_R-t_L)Y}\frac{1}{X_2+t_RY}\bigg]\,,\\
\uppsi^{(2)}_{\rm D} &\equiv   \frac{1}{2Y}\, \frac{1}{X_1+t_LY}\frac{1}{X_2+t_RY}\,.
\end{aligned}
\eeq
How we integrate this universal integrand depends on the situation we are interested in. 
In all cases of interest, we integrate over the internal  variables $t_b$ ($b=L,R$), weighted by the twist factors $(t_b^2-1)^{-\nu-\frac{1}{2}}$, and shift the external variables $X_a \mapsto X_a+x_a$ and then integrate over $x_a$ with the appropriate twist.
If in addition the external lines are themselves massive, we write the shift of the external energies as $x_a +s_1+\cdots +s_n$, and then also integrate over the $s$ variables (again with the appropriate twist). As an example, the four-point function of conformally coupled scalar fields exchanging a field of general mass in four-dimensional de Sitter space is given by the following integral (up to an overall $Y$-dependent normalization factor)
\beq
\psi^{(2)} \sim \int \bigg(\prod_{a=1,2} \ud x_a \,x_a^\xi \prod_{b=L,R} \ud t_b \,(t_b^2-1)^{-\nu-\frac{1}{2}}  \bigg)\,\uppsi(X_a+x_a,t_b Y)\,,
\label{eq:psi4 int rep}
\eeq
where the integrand is a rational function with linear factors. The integrand therefore naturally has an associated hyperplane arrangement and the machinery of twisted cohomology can be applied. We will explore some aspects of this in Section~\ref{sec:polytopes}.

\vskip4pt
It is straightforward to cast the wavefunction involving arbitrary states as a similar twisted integral by using the propagators in~\eqref{eq:reduced}. However, performing the time integrals to obtain a rational integrand quickly becomes fairly tedious, so it is useful to organize the calculation into an (equivalent) combinatorial mnemonic.

\subsubsection*{Combinatorics}

The time integrals that compute the universal massive integrand have a nice combinatorial structure that we can exploit to 
give a direct algorithm for its construction.

\vskip4pt
\noindent
{\it Graphs and maximal tubings:} The integrand is a rational function, and so it can naturally be related to tubings of a (labeled) graph. 
For the massive integrands, the natural graph is the underlying Feynman diagram that we use to compute $\uppsi$. It is convenient to mark the internal lines of the graph with a white circle as in~\eqref{eq:integrandexc} to keep track of the fact that the internal energy flowing through the graph is different on the two sides of the marking.
In addition, it will be convenient to distinguish between time-ordered internal lines (which we denote by solid lines) and disconnected internal lines (which we denote by dotted lines).

\vskip4pt
With this understanding, the terms in the time-integral representation of $\uppsi$ are in one-to-one correspondence with the maximal tubings of these graphs.\footnote{A {\it tube}~\cite{carr2005coxetercomplexesgraphassociahedra,devadoss2009realization} is a proper non-empty subset of a graph, so that the nodes contained in the tube yield a connected subgraph. Tubes are {\it compatible} in the sense of~\cite{carr2005coxetercomplexesgraphassociahedra,devadoss2009realization} if they do not intersect and are not adjacent. A {\it tubing} is a collection of pairwise compatible tubes. For clarity, we will call such a tubing a {\it math tubing} (since it relies on the notion of compatibility common in the mathematics literature) in order to distinguish it from a slightly different notion of tubing which will appear in Section~\ref{sec:Flow}.
Since a tubing is a proper subset, it can contain at most $n-1$ tubes for a $n$-vertex graph. Such a tubing with $n-1$ tubes is a {\it maximal tubing}. It is important to specify that the white dots on internal lines are {\it not} vertices, and so by convention we always leave them outside tubes when the tube does not fully enclose an internal line.}
This correspondence between tubings and the time-integral representation of the wavefunction is a consequence of the fact that a maximal tubing induces an orientation on the graph, obtained by drawing arrows that point away from each tube, starting from the smallest tube and working outwards.
For example, one maximal tubing of a three-site chain is
\be
 \begin{tikzpicture}[baseline=(current  bounding  box.center)]
\draw[fill=black,decoration={markings,
    mark=at position 0.3 with {\arrow[Blue,scale=2]{stealth}},mark=at position 0.825 with {\arrow[Blue,scale=2]{stealth}}},
    postaction=decorate] (-1,0) -- (1,0);
\draw[fill=black] (0,0) circle (.5mm);
\draw[fill=black] (-1,0) circle (.5mm);
\draw[fill=black] (1,0) circle (.5mm);
\draw[color=black, line width=0.8pt] (-1,0) ellipse (.185cm and .185cm);
\node [
        draw, color=black, line width=0.8pt,
        rounded rectangle,
        minimum height = 1.4em,
        minimum width = 4.75em,
        rounded rectangle arc length = 180,
    ] at (-.5,0)
    {};
\end{tikzpicture}\ ,
\ee
where we have also drawn the induced orientation of the underlying graph. All we then need to reproduce the time-integral representation of $\uppsi$ is a rule to associate a rational function to each maximal tubing, which can be done by slightly adapting the construction of~\cite{Arkani-Hamed:2017fdk}. One nice feature of this construction is that there are often multiple maximal tubings that induce the same orientation on the graph. The simplest example arises for the three-site chain:
\be
\raisebox{3pt}{
 \begin{tikzpicture}[baseline=(current  bounding  box.center)]
\draw[fill=black,decoration={markings,
    mark=at position 0.3 with {\arrow[Blue,scale=2]{stealth[reversed]}},mark=at position 0.825 with {\arrow[Blue,scale=2]{stealth}}},
    postaction=decorate] (-1,0) -- (1,0);
\draw[fill=black] (0,0) circle (.5mm);
\draw[fill=black] (-1,0) circle (.5mm);
\draw[fill=black] (1,0) circle (.5mm);
\draw[color=black, line width=0.8pt] (0,0) ellipse (.185cm and .185cm);
\node [
        draw, color=black, line width=0.8pt,
        rounded rectangle,
        minimum height = 1.4em,
        minimum width = 4.75em,
        rounded rectangle arc length = 180,
    ] at (-.5,0)
    {};
\end{tikzpicture}
}
~~~~~{\rm and}~~~~~
\raisebox{3pt}{
 \begin{tikzpicture}[baseline=(current  bounding  box.center)]
\draw[fill=black,decoration={markings,
    mark=at position 0.3 with {\arrow[Blue,scale=2]{stealth[reversed]}},mark=at position 0.825 with {\arrow[Blue,scale=2]{stealth}}},
    postaction=decorate] (-1,0) -- (1,0);
\draw[fill=black] (0,0) circle (.5mm);
\draw[fill=black] (-1,0) circle (.5mm);
\draw[fill=black] (1,0) circle (.5mm);
\draw[color=black, line width=0.8pt] (0,0) ellipse (.185cm and .185cm);
\node [
        draw, color=black, line width=0.8pt,
        rounded rectangle,
        minimum height = 1.4em,
        minimum width = 4.75em,
        rounded rectangle arc length = 180,
    ] at (.5,0)
    {};
\end{tikzpicture}
}
\label{equ:TUBING}
\ee
We see that both tubings lead to the same pattern of arrows. This is expected from the time integral. For this time ordering, there are two integration regions, depending on whether the leftmost or rightmost vertex is the latest, and so it is satisfying that the combinatorial construction naturally splits up these two contributions.

\vskip4pt
\noindent
{\it Rational function:} We now describe the algorithm to produce $\uppsi$, before illustrating it with some simple examples. 
Starting from the Feynman graph (with truncated external lines), we consider all maximal tubings. We then associate a linear factor to each tube in a given maximal tubing. 
Specifically, to each tube we assign the sum of energies of all (black) vertices enclosed along with the sum of energies carried by lines piercing the tube. For tubes that enclose an internal white dot, we also add $(t_L-t_R) Y$ or $(t_R-t_L) Y$, depending on the direction that the arrow is pointing (the direction the arrow points away from gets the $+$ sign). We divide by each of these linear factors. For example, the tubing in (\ref{equ:TUBING}) leads to
\be
\raisebox{3pt}{
 \begin{tikzpicture}[baseline=(current  bounding  box.center)]
\draw[fill=black,decoration={markings,
    mark=at position 0.31 with {\arrow[Blue,scale=2]{stealth[reversed]}},mark=at position 0.8 with {\arrow[Blue,scale=2]{stealth}}},
    postaction=decorate] (-1,0) -- (1,0);
\draw[fill=black] (0,0) circle (.5mm);
\draw[fill=black] (-1,0) circle (.5mm);
\draw[fill=black] (1,0) circle (.5mm);
\draw[color=Red, line width=0.8pt] (0,0) ellipse (.185cm and .185cm);
\node [
        draw, color=purple3, line width=0.8pt,
        rounded rectangle,
        minimum height = 1.4em,
        minimum width = 4.75em,
        rounded rectangle arc length = 180,
    ] at (.5,0)
    {};
        \draw[color=Blue,fill=white,line width=.1mm] (-.5,0) circle (.4mm);
    \draw[color=Blue,fill=white,line width=.1mm] (.5,0) circle (.4mm);
\end{tikzpicture}
}
 =  
 \frac{1}{{\color{purple3} X_2+X_3+t^R_{12}Y_{12}+(t_{23}^L-t_{23}^R)Y_{23}}}\frac{1}{{\color{Red} X_2+t_{12}^RY_{12}+t_{23}^LY_{23}}}\,.
\ee
To obtain the wavefunction, we also associate a “universal tube” to each
 maximal tubing  that encloses the entire graph, and assign it a linear factor following the same rules.
 Each maximal tubing gives a contribution to $\uppsi$ that is $1$ divided by the product of these linear factors, along with a factor of $1/(2Y_i)$ for each internal line. We sum up all the contributions for the fully connected graph, and then proceed to dash each of the internal lines in all possible ways, and repeat the procedure with each of these disconnected graphs. The wavefunction integrand $\uppsi$ is the sum of all of these pieces with weight $(-A)$ for each disconnected internal line.

\vskip4pt
\noindent
{\it An example:}
Let us illustrate the procedure  for the case of the single massive exchange.
 The underlying Feynman graph is the same as in~\eqref{eq:integrandexc}. We first consider the connected graph, which has two maximal tubings:
\begin{align}
\raisebox{2pt}{
 \begin{tikzpicture}[baseline=(current  bounding  box.center)]
\draw[fill=black,decoration={markings,
    mark=at position 0.59 with {\arrow[Blue,scale=2]{stealth}}},
    postaction=decorate] (-0.6,0) -- (0.6,0);
\draw[fill=black] (-0.6,0) circle (.5mm);
\draw[fill=black] (0.6,0) circle (.5mm);
\draw[color=Red, line width=0.6pt] (-0.6,0) ellipse (.185cm and .185cm);
\draw[color=Blue,fill=white,line width=.1mm] (0,0) circle (.4mm);
    \node [
        draw, color=Blue, line width=0.6pt,
        rounded rectangle,
        minimum height = 1.4em,
        minimum width = 5.25em,
        rounded rectangle arc length = 180,
    ] at (0,0)
    {};
\end{tikzpicture}
}
&= \frac{1}{2Y}\,\frac{1}{{\color{Blue} X_1+X_2+(t_L-t_R)Y}}\frac{1}{{\color{Red} X_1+t_L Y}}\,, \\[4pt]
\raisebox{2pt}{
\begin{tikzpicture}[baseline=(current  bounding  box.center)]
\draw[fill=black,decoration={markings,
    mark=at position 0.6 with {\arrow[Blue,scale=2]{stealth[reversed]}}},
    postaction=decorate] (-0.6,0) -- (0.6,0);
\draw[fill=black] (-0.6,0) circle (.5mm);
\draw[fill=black] (0.6,0) circle (.5mm);
\draw[color=Red, line width=0.6pt] (0.6,0) ellipse (.185cm and .185cm);
\draw[color=Blue,fill=white,line width=.1mm] (0,0) circle (.4mm);
    \node [
        draw, color=Blue, line width=0.6pt,
        rounded rectangle,
        minimum height = 1.4em,
        minimum width = 5.25em,
        rounded rectangle arc length = 180,
    ] at (0,0)
    {};
\end{tikzpicture}
}
&= \frac{1}{2Y}\,\frac{1}{{\color{Blue} X_1+X_2+(t_R-t_L)Y}}\frac{1}{{\color{Red} X_2+t_RY}}\,,
\end{align}
where we have written the corresponding rational functions. In addition, there is a single disconnected graph with a dashed internal line:
\be
\raisebox{2pt}{
 \begin{tikzpicture}[baseline=(current  bounding  box.center)]
\draw[fill=black,dotted,thick] (-0.6,0) -- (0.6,0);
\draw[fill=black] (-0.6,0) circle (.5mm);
\draw[fill=black] (0.6,0) circle (.5mm);
\draw[color=Red, line width=0.6pt] (-0.6,0) ellipse (.185cm and .185cm);
\draw[color=Red, line width=0.6pt] (0.6,0) ellipse (.185cm and .185cm);
\draw[fill=white,line width=.2mm] (0,0) circle (.4mm);
\end{tikzpicture}
}
= \frac{1}{2Y}\, \frac{1}{{\color{Red} X_1+t_LY}}\frac{1}{{\color{Red} X_2+t_RY}}\,,
\ee
where we have associated a universal tube to each of the subgraphs generated by disconnecting the internal line. Summing up these three contributions with a relative weight of $(-A)$ for the disconnected piece reproduces~\eqref{eq:singleexchangemassive}.

\vskip4pt
The same procedure can be applied to arbitrary graphs. For a three-site chain, there are five maximal tubings. A representative example is
\be
\begin{aligned}
\raisebox{3pt}{
 \begin{tikzpicture}[baseline=(current  bounding  box.center)]
\draw[fill=black,decoration={markings,
    mark=at position 0.31 with {\arrow[Blue,scale=2]{stealth[reversed]}},mark=at position 0.8 with {\arrow[Blue,scale=2]{stealth}}},
    postaction=decorate] (-1,0) -- (1,0);
\draw[fill=black] (0,0) circle (.5mm);
\draw[fill=black] (-1,0) circle (.5mm);
\draw[fill=black] (1,0) circle (.5mm);
\draw[color=Red, line width=0.8pt] (0,0) ellipse (.185cm and .185cm);
\node [
        draw, color=purple3, line width=0.8pt,
        rounded rectangle,
        minimum height = 1.4em,
        minimum width = 4.75em,
        rounded rectangle arc length = 180,
    ] at (.5,0)
    {};
    \node [
        draw, color=Blue, line width=0.8pt,
        rounded rectangle,
        minimum height = 1.9em,
        minimum width = 8em,
        rounded rectangle arc length = 180,
    ] at (0,0)
    {};
        \draw[color=Blue,fill=white,line width=.1mm] (-.5,0) circle (.4mm);
    \draw[color=Blue,fill=white,line width=.1mm] (.5,0) circle (.4mm);
\end{tikzpicture}
}
 &=  
\,\frac{1}{4Y_{12}Y_{23}} \frac{1}{{\color{Blue} X_1+X_2+X_3+(t_{12}^R-t_{12}^L)Y_{12}+(t_{23}^L-t_{23}^R)Y_{23}}}
\\
&\hspace{.65cm}
\times
\frac{1}{{\color{purple3} X_2+X_3+t_{12}^R Y_{12}+(t_{23}^L-t_{23}^R)Y_{23}}}\frac{1}{{\color{Red} X_2+t_{12}^RY_{12}+t_{23}^LY_{23}}}\,.
\end{aligned}
\ee
In addition, there are four singly disconnected contributions, such as
\be
\begin{aligned}
\raisebox{3pt}{
 \begin{tikzpicture}[baseline=(current  bounding  box.center)]
 \draw[fill=black,dotted,thick] (-1,0) -- (0,0);
\draw[fill=black,decoration={markings,
    mark=at position 0.6 with {\arrow[Blue,scale=2]{stealth}}},
    postaction=decorate] (0,0) -- (1,0);
\draw[fill=black] (0,0) circle (.5mm);
\draw[fill=black] (-1,0) circle (.5mm);
\draw[fill=black] (1,0) circle (.5mm);
\draw[color=Red, line width=0.8pt] (0,0) ellipse (.185cm and .185cm);
\draw[color=Blue, line width=0.8pt] (-1,0) ellipse (.185cm and .185cm);
\node [
        draw, color=Blue, line width=0.8pt,
        rounded rectangle,
        minimum height = 1.4em,
        minimum width = 4.75em,
        rounded rectangle arc length = 180,
    ] at (.5,0)
    {};
        \draw[color=black,fill=white,line width=.2mm] (-.5,0) circle (.4mm);
    \draw[color=Blue,fill=white,line width=.1mm] (.5,0) circle (.4mm);
\end{tikzpicture}
}
 &\ =  \
\,\frac{1}{4Y_{12}Y_{23}} \frac{1}{{\color{Blue} X_1+t_{12}^LY_{12}}}\frac{1}{{\color{Red} X_2+t_{12}^RY_{12}+t_{23}^LY_{23}}} \hspace{2.8cm}\phantom{x}\\
&\hspace{.85cm} \times \frac{1}{{\color{Blue} X_2+X_3+t_{12}^RY_{12}+(t_{23}^L-t_{23}^R)Y_{23}}}\,.
\end{aligned}
\ee
Finally, there is one fully disconnected contribution, which has a single maximal tubing (with each vertex circled). As before, adding up all these contributions (weighted by $(-A)$ for each disconnected line) reproduces the universal integrand obtained by integrating the Feynman rules.

\vskip4pt
This procedure can be given a formal proof along the lines of~\cite{Arkani-Hamed:2017fdk}, which computes bulk time integrals directly by successively collapsing internal lines in a way compatible with the graph orientation and combining the energies. The end result matches the output of this combinatorial exercise.
An interesting feature of this massive case is that the time-integral representation does not produce any spurious singularities (in contrast to the conformally coupled case).

\subsection{Massive Geometries}
\label{sec:polytopes}
One of the nice features of the above representation is that it casts the integrals of interest in a (twisted) Euler form. Since the factors appearing in the integrand are linear, the integrals are naturally associated to hyperplane arrangements~\cite{zaslavsky1975facing,stanley2004introduction,orlik1992arrangements}. As such, these integrals can be given a geometric interpretation, and the tools of positive geometry and twisted cohomology can be systematically employed. Most basically, this makes it clear that the massive wavefunction is part of a finite-dimensional vector space of integrals, and therefore is part of a first-order system of differential equations. 

\begin{figure}[t!]
	\centering
\includegraphics[scale=.8]{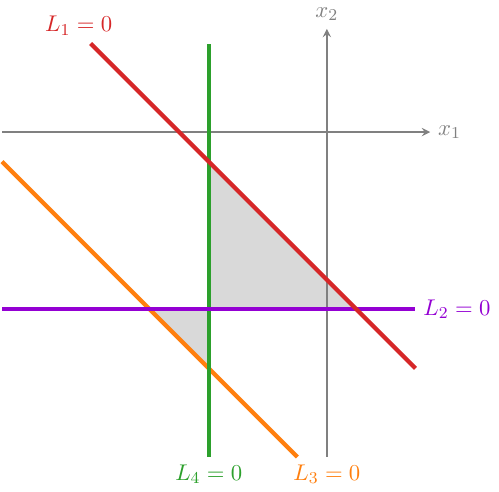}
	\caption{Visualization of the hyperplane arrangement defined by the singularities of the massive two-site integral in the $x_1$-$x_2$ plane.}
	\label{fig:hyperplanes}
\end{figure}

\subsubsection*{Hyperplane arrangements}

Describing the full features of the geometry underlying the universal massive integrand would take us too far afield from the main line of development. We take up this study in future work. Here, we content ourselves with describing the simplest nontrivial example of the two-site exchange graph.
Combining the terms appearing in~\eqref{eq:singleexchangemassive}, the function~$\uppsi^{(2)}$ can be written as\hs\footnote{It is worth noting that for real $\nu$, the two obviously special values of $A$ have natural interpretations. The value $A=1$ corresponds to the wavefunction, while $A=-1$ gives the corresponding in-in correlator~\cite{Benincasa:2024leu,Figueiredo:2025daa,Glew:2025arc}. In both cases there is a simplification of the numerator.}
\be
\uppsi^{(2)} = \frac{1}{2Y}\frac{(1-A)(X_1+X_2)^2+2(t_L X_1+t_R X_2)Y+(1+A)(t_L-t_R)^2Y^2}{(X_1+t_LY)(X_2+t_RY)(X_1+X_2+(t_L-t_R)Y)(X_1+X_2+(t_R-t_L)Y)}\,.
\label{eq:2sitemassiveint}
\ee
Writing all terms under a single denominator makes all singularities of the object manifest. The singularities are given by the vanishing loci of the following five lines
\beq
\begin{aligned}
\color{Red}{L_1} \,&{\color{Red}\equiv X_1+X_2+x_1+x_2+(t_R-t_L)Y}\,,\\
\color{purple3}{L_2}\, & {\color{purple3}\equiv X_2+x_2+t_R \hs Y}\,,\\
\color{orange3}{L_3}\, &{\color{orange3} \equiv X_1+X_2+ x_1+x_2+(t_L-t_R) Y}\,,\\
\color{Green}{L_4}\, & {\color{Green} \equiv X_1+x_1+t_L\hs Y}\,,\\
L_5\, &= Y\,,
\end{aligned}
\eeq
which defines a hyperplane arrangement in the (projective) space of integration variables $x_{1,2}$ and $t_{L,R}$. To visualize this arrangement, it is convenient to project either onto the two-dimensional spaces of $x$ or $t$-variables. In the $x$-space, the singular lines of~\eqref{eq:2sitemassiveint} correspond to the geometric picture shown in Fig.\,\ref{fig:hyperplanes}.
This  allows us to view the integrand~\eqref{eq:2sitemassiveint} as the canonical form of a geometric region in the space of $x$-variables. The simplest such association can be made by defining the dlog-form
\be
[L_1\cdots L_n] \equiv  \ud\log L_1\wedge \cdots \wedge \ud \log L_n\,,
\ee
so that the two-site integrand corresponds to the following sum of canonical forms associated to unbounded regions 
\be
\uppsi^{(2)}\, \ud x_1\wedge \ud x_2= [L_3L_4]-[L_1L_2]+A\hs [L_2L_4]\,.
\ee
Alternatively, we can view~\eqref{eq:2sitemassiveint} as the canonical form of a non-convex weighted geometry, corresponding to the gray-shaded region in Fig.\,\ref{fig:hyperplanes}~\cite{Benincasa:2024leu}.

\subsubsection*{Integral basis}
\begin{figure}[t!]
	\centering
\includegraphics[scale=.9]{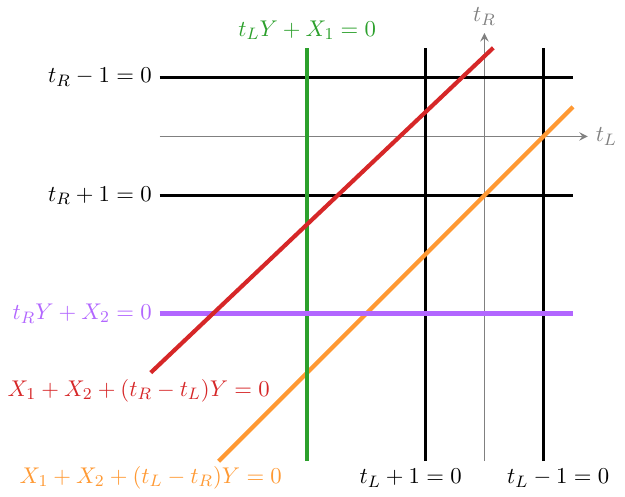}
	\caption{Visualization of the hyperplane arrangement defined by the singular lines of the massive two-site integral in the $t_L$-$t_R$ plane. Note that in addition to the singularities defined by the rational integrand, the twist factors define two pairs of singular lines $t_L \pm 1$ and $t_R\pm 1$.}
	\label{fig:hyperplanes2}
\end{figure}

It is interesting to investigate the integral basis used in the main text through this geometric lens (see  Section~\ref{sec:twistedint}). To do this, it is useful to visualize the hyperplane arrangement associated to~\eqref{eq:2sitemassiveint} in the $t$-space; see Fig.\,\ref{fig:hyperplanes2}. 
Comparing this arrangement with the analogous one in $x$-space (Fig.\,\ref{fig:hyperplanes}), a notable difference is that the twist factors involving $t_{L,R}$ correspond to  two pairs of singular lines $t_L \pm 1$ and $t_R\pm 1$. In~\eqref{eq:psi4 int rep}, both of these singularities are twisted with the same parameter. It is natural to allow these two lines to be twisted by different amounts and consider the integrals
\beq
\psi^{(\pmr\pmb)} \equiv \int \bigg(\prod_{a=1,2} \ud x_a \,x_a^\xi \prod_{b=L,R} \ud t_b \,(t_b^2-1)^{-\nu-\frac{1}{2}}  \bigg) \, (1\,\mpr\, t_L)(1\,\pmb\, t_R)\,\uppsi^{(2)} (X_a+x_a,t_b Y)\,,
\label{eq:Fintegrals}
\eeq
where the twists of the $(1\,\pm\, t_{L,R})$ singularities are shifted by one unit. 
One recovers the wavefunction by summing over all these basis elements as $\psi =\psi^{(++)}+\psi^{(--)}+\psi^{(+-)}+\psi^{(-+)}$.

\newpage
\section{One-Loop Bubble}
\label{app:loop}

The explicit examples in the main text all involved tree-level graphs, but it is worth emphasizing that the kinematic flow algorithm presented in Section~\ref{sec:Flow} also applies to {\it loop integrands} (i.e.~to loop-level wavefunctions before integrating over the three-momentum running in the loop, but after integrating over time.)
In fact, one nice feature of the massive case---as compared to the conformally coupled case---is that it is more uniform. The rules are precisely the same at both tree and loop level.
In the conformal case, 
 many of the basis functions identically vanish, and so
one must modify the flow rules to accommodate this~\cite{Baumann:2024mvm,Hang:2024xas}. In contrast, the massive case is fully generic, so all graphs follow the same rules.

\vskip4pt
In this appendix, we will illustrate the application of the procedure at loop level through the simplest example of the
one-loop bubble (with conformally coupled external legs):
\vspace{2pt}
\be
\raisebox{-35pt}{
\begin{tikzpicture}[line width=1. pt, scale=2]
\draw[lightgray, line width=2.pt] (-0.85,0.65) -- (0.85,0.65);
    \draw[line width=2.pt] (0,0) circle[radius=0.35];
    \draw[line width=1.pt,color=lightgray]  (-0.35,0)-- (-0.7,0.65);
    \draw[line width=1.pt,color=lightgray]  (-0.35,0) -- (-0.5,0.65);
    \draw[line width=1.pt,color=lightgray]  (0.35,0)-- (0.7,0.65);
    \draw[line width=1.pt,color=lightgray]  (0.35,0) -- (0.5,0.65);
        \draw[fill=Red, Red] (-0.35,0) circle (.03cm);
    \draw[fill=Blue, Blue] (0.35,0) circle (.03cm);
    \draw  (-0.47,0.0) node[below]  {\small $X_{1}$};  
        \draw  (0.47,0.0) node[below]  {\small $X_{2}$};  
            \draw  (0,0.47) node  {\small $Y_{u}$};  
             \draw  (0,-0.48) node  {\small $Y_{d}$};  
\end{tikzpicture}
}
\label{eq:bubblegraph}
\ee
Here, the ``up" ($u$) and ``down" ($d$) loop propagators have energies $Y_u$ and $Y_d$ and mass parameters $\nu_u=\half+\dev_u$ and $\nu_d=\half+\dev_d$, respectively.

\subsection{Letters and Functions}

We begin by presenting the letters and basis functions for the one-loop bubble~\eqref{eq:bubblegraph}. As in the previous examples, we truncate the (conformally coupled) external lines and mark the internal lines. The letters and functions then correspond to tubings of this marked graph, as in Section~\ref{ssec:tubings}.

\subsubsection*{Letters}

The following shows the letters appearing in the differential equations for the one-loop bubble:
\setlength{\extrarowheight}{10pt}
\begin{equation}\label{equ:letters}
    \begin{tabular}{ll}
 \includegraphics[valign=c]{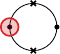} $~=~ \ud\log(X_1+Y_u+Y_d)\,,$
 \qquad\qquad&   \includegraphics[valign=c]{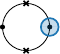} $~=~ \ud\log(X_2+Y_u+Y_d)\,,$
 \\
\includegraphics[valign=c]{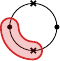} $~=~ \ud\log(X_1+Y_u-Y_d)\,,$
\qquad\qquad& \includegraphics[valign=c]{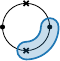} $~=~ \ud\log(X_2+Y_u-Y_d)\,,$
\\ 
\includegraphics[valign=c]{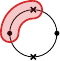} $~=~ \ud\log(X_1-Y_u+Y_d)\,,$
\qquad\qquad& \includegraphics[valign=c]{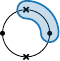} $~=~ \ud\log(X_2-Y_u+Y_d)\,,$
\\
\includegraphics[valign=c]{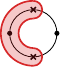} $~=~ \ud\log(X_1-Y_u-Y_d)\,,$
\qquad\qquad& \includegraphics[valign=c]{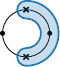} $~=~ \ud\log(X_2-Y_u-Y_d)\,,$
\\
\!\!\includegraphics[valign=c]{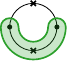} $~=~ \ud\log(X_1+X_2+2Y_u)\,,$
\qquad\qquad& 
\!\!\includegraphics[valign=c]{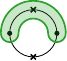} $~=~ \ud\log(X_1+X_2+2Y_d)\,,$
\\
\!\!\!\! \loopdlogY{1} \raisebox{-3pt}{~$~=~\ud\log(Y_u)\,,$}
\qquad\qquad& 
\!\!\!\!  \loopdlogY{-1} \raisebox{-3pt}{~$~=~\ud\log(Y_d)\,.$}
\end{tabular}
\end{equation}
Note that there are five equivalent representations of the total energy singularity:
\begin{equation}\label{equ:totalE}
    \begin{tabular}{ccccc}
\includegraphics[valign=c]{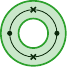} & 
\includegraphics[valign=c]{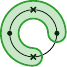} & 
\includegraphics[valign=c]{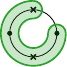} & 
\includegraphics[valign=c]{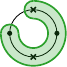} & 
\includegraphics[valign=c]{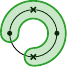} 
\end{tabular} = \ud\log(X_1+X_2)\,.
\end{equation}
We will see how each of these arises in the differential equations satisfied by the bubble.

\subsubsection*{Functions}

As in the main text, we restrict attention to fully connected diagrams. The functions associated with the one-loop bubble are then obtained by the same procedure as in Section~\ref{ssec:tubings}. 
It is straightforward to write down the tubing representation of each of these functions using the same rule as before. 

\vskip 4pt
We first consider tubings where each tube only contains one vertex.
We label these functions by $\psi^{(\pm\pm)}_{(\pm\pm)}$, where we
assign a minus sign when a tube encircles the corresponding cross and plus sign when it does not: 
\setlength{\extrarowheight}{10pt}
\begin{equation}\label{equ:tubes loop}
    \begin{tabular}{ccccc} 
$\psi^{(++)}_{(++)}$\Loopf{1}{1}{1}{1}\quad& \quad  $\psi^{(-+)}_{(++)}$ \Loopf{-1}{1}{1}{1}&  $\psi^{(+-)}_{(++)}$ \Loopf{1}{1}{-1}{1}\quad&\quad  $\psi^{(+-)}_{(-+)}$  \Loopf{1}{-1}{-1}{1}&  $\psi^{(+-)}_{(+-)}$  \Loopf{1}{1}{-1}{-1}\\
&\quad $\psi^{(++)}_{(-+)}$ \Loopf{1}{-1}{1}{1}&   $\psi^{(++)}_{(+-)}$ \Loopf{1}{1}{1}{-1} \quad& \quad  $\psi^{(-+)}_{(+-)}$  \Loopf{-1}{1}{1}{-1}& $\psi^{(-+)}_{(-+)}$  \Loopf{-1}{-1}{1}{1}\\ 
$\psi^{(--)}_{(--)}$ \Loopf{-1}{-1}{-1}{-1} \quad& \quad $\psi^{(+-)}_{(--)}$  \Loopf{1}{-1}{-1}{-1}& $\psi^{(-+)}_{(--)}$ \Loopf{-1}{-1}{1}{-1} \quad& \quad $\psi^{(--)}_{(++)}$   \Loopf{-1}{1}{-1}{1}& \\
 & \quad  $\psi^{(--)}_{(+-)}$\Loopf{-1}{1}{-1}{-1}& $\psi^{(--)}_{(-+)}$ \Loopf{-1}{-1}{-1}{1} \quad&  \quad $\psi^{(++)}_{(--)}$ \Loopf{1}{-1}{1}{-1}&
\end{tabular}
\end{equation}
Similar to what happens at tree level, the (connected) wavefunction coefficient $\psi$ is obtained by adding up these $16$ functions. 
There are 12 basis functions corresponding to the tubings with exposed crosses and overlapping tubes. As we will show in Section~\ref{ssec:conformal}, these vanish in the conformal limit. 

\vskip4pt
Additional functions appear when differentiating $\psi^{(\pm\pm)}_{(\pm\pm)}$. From the time-integral perspective, they correspond to collapsing internal propagators. Here, they are represented by tubes that contain two vertices:
\setlength{\extrarowheight}{10pt}
\be\label{eq:loop col list}
    \begin{tabular}{ccc|cccc|cc} 
 \LoopUu &  \LoopUd &&& \LoopCul & \LoopCur &&&\includegraphics[valign=c]{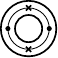}\\
  \LoopUdm & \LoopUum &&& \LoopCdl & \LoopCdr &&&
  \end{tabular}
\ee
As we will show below, the rightmost tubing doesn't actually appear in the kinematic flow equations. Of the remaining 8 tubings, only 6 are actually independent functions.

\subsection{Massive Kinematic Flow}

In Section~\ref{sec:kflowalgorithm}, we stated the three rules of the kinematic flow and illustrated them in selected tree-level examples. The same rules, without any modifications, apply to the differential equations for loop integrands. It is instructive to see this explicitly for the case of the one-loop bubble.

\vskip4pt
For concreteness, let us consider the function $\psi^{(-+)}_{(-+)}$. It satisfies the differential equation
\be
\begin{aligned}
\ud\raisebox{2pt}{\Loopf{-1}{-1}{1}{1}}~
=
~&\Bigg(\alpha_1 \, \includegraphics[valign=c]{Figures/Tubings/Bubble_Letters/DlogX1-Y1-Y2.pdf}  \,+ \alpha_2 \,\includegraphics[valign=c]{Figures/Tubings/Bubble_Letters/DlogX2+Y1+Y2.pdf}  \Bigg) \raisebox{2pt}{\Loopf{-1}{-1}{1}{1} }
~+~ \Bigg( \, \includegraphics[valign=c]{Figures/Tubings/Bubble_Letters/DlogX1-Y1-Y2.pdf}  \,-  \,\includegraphics[valign=c]{Figures/Tubings/Bubble_Letters/DlogX2+Y1+Y2.pdf}  \Bigg) \Bigg(\raisebox{2pt}{\LoopCdr} ~+\, \raisebox{2pt}{\LoopCur}\Bigg)\\[2pt]
&+\dev_u \, \Bigg(\, \includegraphics[valign=c]{Figures/Tubings/Bubble_Letters/DlogX1-Y1-Y2.pdf} \,-\!\raisebox{3pt}{\loopdlogY{1}}  \Bigg) \raisebox{1pt}{\Loopf{1}{-1}{1}{1}  }
 \, + \,  \dev_u \, \Bigg(\, \includegraphics[valign=c]{Figures/Tubings/Bubble_Letters/DlogX2+Y1+Y2.pdf} \,- \raisebox{3pt}{\loopdlogY{1}}  \Bigg)  \raisebox{3pt}{\Loopf{-1}{-1}{-1}{1}}   \\[2pt]
&+\, \dev_d\,\Bigg(\, \includegraphics[valign=c]{Figures/Tubings/Bubble_Letters/DlogX1-Y1-Y2.pdf} - \raisebox{3pt}{\loopdlogY{-1}}  \Bigg) \raisebox{4pt}{\Loopf{-1}{1}{1}{1} } \, +\,   \dev_d \,
\Bigg(\, \includegraphics[valign=c]{Figures/Tubings/Bubble_Letters/DlogX2+Y1+Y2.pdf} \,- \raisebox{3pt}{\loopdlogY{-1}}  \Bigg)  \raisebox{2pt}{\Loopf{-1}{-1}{1}{-1}  } \,.
\end{aligned}
\ee
The various terms appearing on the right-hand side of this equation can be understood from the same massive kinematic flow described in the main text. 
\begin{enumerate}
\item {\bf Activation}: Each tube of the graph tubing corresponding to the chosen parent function gets ``activated" and becomes a letter in the differential equation:
\be
\ud\raisebox{2pt}{\Loopf{-1}{-1}{1}{1}}~
\,\supset\,
~\Bigg(\alpha_1 \, \includegraphics[valign=c]{Figures/Tubings/Bubble_Letters/DlogX1-Y1-Y2.pdf}  \,+ \alpha_2 \,\includegraphics[valign=c]{Figures/Tubings/Bubble_Letters/DlogX2+Y1+Y2.pdf}  \Bigg) \raisebox{2pt}{\Loopf{-1}{-1}{1}{1} }\,.
\ee
Each letter is multiplied by the parameter $\alpha_{1,2}$ associated to the enclosed vertex.
\item {\bf Merger}: The two adjacent tubes of the parent function ``merge" to form a larger tube. There are two ways in which this can happen, so we get
\be
\ud\raisebox{2pt}{\Loopf{-1}{-1}{1}{1}}~\,\supset\,  
 \Bigg( \, \includegraphics[valign=c]{Figures/Tubings/Bubble_Letters/DlogX1-Y1-Y2.pdf}  \,-  \,\includegraphics[valign=c]{Figures/Tubings/Bubble_Letters/DlogX2+Y1+Y2.pdf}  \Bigg) \Bigg(\raisebox{2pt}{\LoopCdr} ~+\, \raisebox{2pt}{\LoopCur}\Bigg)\,.
\label{eq:merger loop}
\ee
The coefficient is the difference of the two letters corresponding to the two tubes involved in the merger.

\item  {\bf Mixing}: Both tubes of the parent function are pierced by the massive propagators and therefore can {\it grow} and {\it shrink} to produce new source functions.  There are four ways in which this can happen, so we get
\ba
\ud\raisebox{2pt}{\Loopf{-1}{-1}{1}{1}} \,\supset\,
~~&\dev_u \, \Bigg(\, \includegraphics[valign=c]{Figures/Tubings/Bubble_Letters/DlogX1-Y1-Y2.pdf} \,-\!\raisebox{3pt}{\loopdlogY{1}}  \Bigg) \raisebox{1pt}{\Loopf{1}{-1}{1}{1}  }
 \, + \,  \dev_u \, \Bigg(\, \includegraphics[valign=c]{Figures/Tubings/Bubble_Letters/DlogX2+Y1+Y2.pdf} \,- \raisebox{3pt}{\loopdlogY{1}}  \Bigg)  \raisebox{3pt}{\Loopf{-1}{-1}{-1}{1}}   \\ 
+\, &\dev_d\,\Bigg(\, \includegraphics[valign=c]{Figures/Tubings/Bubble_Letters/DlogX1-Y1-Y2.pdf} - \raisebox{3pt}{\loopdlogY{-1}}  \Bigg) \raisebox{4pt}{\Loopf{-1}{1}{1}{1} } \, +\,   \dev_d \,
\Bigg(\, \includegraphics[valign=c]{Figures/Tubings/Bubble_Letters/DlogX2+Y1+Y2.pdf} \,- \raisebox{3pt}{\loopdlogY{-1}}  \Bigg)  \raisebox{2pt}{\Loopf{-1}{-1}{1}{-1}  } \,.
\ea
The coefficients are the difference of the letter corresponding to the changed tube and the letter of the relevant line, times the parameter $\dev_{u,d}$ describing the massive field.
\end{enumerate}

Following the same procedure, one can derive the differential of all functions appearing in the bubble wavefunction. It is worth commenting on one additional feature of the differential involving collapsed functions, which appear as sources in~\eqref{eq:merger loop}. As a concrete example, consider the differential 
\be
\ud \, \raisebox{2pt}{\LoopCdr} \,\, = \raisebox{0pt}{ \includegraphics[valign=c,scale=0.9]{Figures/Tubings/Bubble_Letters/Zero1.pdf} }\, \raisebox{2pt}{\LoopCdr}\, +\, \dev_d \Bigg(  \raisebox{0pt}{ \includegraphics[valign=c,scale=0.9]{Figures/Tubings/Bubble_Letters/Zero1.pdf}} \,-   \raisebox{2pt}{\loopdlogY{-1}}  \Bigg) \Bigg(\raisebox{2pt}{\LoopUu}\,+\,\raisebox{2pt}{\LoopUdm}\Bigg)\,.
\ee
The terms on the right-hand side arise from activation and mixing. Note that there is no contribution from merger, despite the fact that the two sides of the tube on the left-hand side are adjacent to each other. A tube cannot merge with itself; instead we require two different tubes that are adjacent in order for merger to occur.\footnote{Alternatively, one could allow the self-merger of a tube, but the coefficient of the resulting source in the differential equation would be the difference of the same letter and itself, giving zero.} A consequence of this is that the rightmost tubing in~\reef{eq:loop col list} will never appear in the kinematic flow.

\vskip4pt
One can repeat this for all basis functions. In addition,
starting from the time-integral representation, it is straightforward to verify that this
kinematic flow procedure gives the correct differential equations for all basis functions of the one-loop bubble.

\subsection{Conformal Limit}
\label{ssec:conformal}

It is interesting to explore the conformally coupled limit of this system of differential equations, corresponding to $\dev_u,\dev_d\to 0$. In this limit, there is a drastic reduction in the size of the function basis. We can understand this from the 
explicit expressions for the basis functions in~\reef{equ:tubes loop}:
\beq
\begin{aligned}
\psi^{(\pmr \pmb)}_{(\pmr \pmb)} =  \int \frac{\ud \eta_1\, e^{iX_1\eta_1} }{(-\eta_1)^{1+\alpha_1+\dev_u+\dev_d}}   \frac{\ud \eta_2\,e^{iX_2 \eta_2}}{(-\eta_2)^{1+\alpha_2+\dev_u+\dev_d}} \, \bigg[\Big(\bar h_{Y_u}^{\pmr}(\eta_1)  \bar h_{Y_d}^{\pmr}(\eta_1) h_{Y_u}^{\pmb}(\eta_2) h_{Y_d}^{\pmb}(\eta_2)  \Big) &\,\theta_{12} \\
+ \Big(h_{Y_u}^{\pmr}(\eta_1)  h_{Y_d}^{\pmr}(\eta_1) \bar h_{Y_u}^{\pmb}(\eta_2) \bar h_{Y_d}^{\pmb}(\eta_2)  \Big) &\,\theta_{21}\bigg]\,.
\end{aligned}
\label{eq: loop f def}
\eeq
Recall that in the conformally coupled limit, $h_k^- = \bar h_k^+ = 0$. This means that the majority of the basis functions in~\eqref{equ:tubes loop} vanish. The only functions that survive the limit are $\psi^{(-+)}_{(-+)}$ and $\psi^{(+-)}_{(+-)}$.\footnote{Mechanically, the vanishing of these functions can be traced to two different mechanisms. The first is that the propagator associated to an internal line identically vanishes. This happens for the $++$ or $--$ assignments, so any $\psi$ with one of these assignments in either a superscript or a subscript vanishes. This is responsible for the vanishing of 12 of the functions in~\eqref{equ:tubes loop}. The remaining two functions vanish because the emergent time ordering of the propagators (discussed in Section~\ref{ssec:tubings}) is incompatible.} The tubings corresponding to these functions can naturally be given an orientation that agrees with the time ordering of the time integral. Explicitly, we have
\ba
\raisebox{2pt}{\Loopf{1}{1}{-1}{-1}}~&= \int  \frac{\ud \eta_1\, e^{iX_1\eta_1} }{(-\eta_1)^{1+\alpha_1+\dev_u+\dev_d}}   \frac{\ud \eta_2\,e^{iX_2 \eta_2}}{(-\eta_2)^{1+\alpha_2+\dev_u+\dev_d}} e^{i(\eta_1-\eta_2)(Y_u+Y_d)} \theta_{12} ~\,~\to~ 
\raisebox{3pt}{
\begin{tikzpicture}[scale=0.9,baseline=(current bounding box.center)]
    \lens \draw[thick] (0,0) circle[radius=\bigR];\filldraw[black] (-\bigR,0) circle (\Vsize);\filldraw[black] (\bigR,0) circle (\Vsize); \draw[-Latex] (0.1,\bigR-0.2) -- (0.1+0.03,\bigR-0.2);  \draw[-Latex] (0.1,-\bigR+0.2) -- (0.1+0.03,-\bigR+0.2); \end{tikzpicture}
    } \,, \\[1pt]
    \raisebox{2pt}{
\Loopf{-1}{-1}{1}{1}
} &= \int  \frac{\ud \eta_1\, e^{iX_1\eta_1} }{(-\eta_1)^{1+\alpha_1+\dev_u+\dev_d}}   \frac{\ud \eta_2\,e^{iX_2 \eta_2}}{(-\eta_2)^{1+\alpha_2+\dev_u+\dev_d}} e^{-i(\eta_1-\eta_2)(Y_u+Y_d)} \theta_{21}  ~\to\,
\raisebox{3pt}{
\begin{tikzpicture}[scale=0.9,baseline=(current bounding box.center)]
    \lens \draw[thick] (0,0) circle[radius=\bigR];\filldraw[black] (-\bigR,0) circle (\Vsize);\filldraw[black] (\bigR,0) circle (\Vsize); \draw[-Latex] (-0.1,\bigR-0.2) -- (-0.1-0.03,\bigR-0.2);  \draw[-Latex] (-0.1,-\bigR) -- (-0.1-0.03,-\bigR); \end{tikzpicture}
} \,,
\ea
where we have drawn the corresponding time-ordered graphs~\cite{Baumann:2025qjx}.  It is important to emphasize that the massive integrals do not have a definite time ordering because they have a mixture of theta functions. However, in the conformal limit, one of the theta function terms vanishes, so the propagators in the conformally coupled limit have a definite time order.

\vskip4pt
The simplification of the collapsed terms is even more pronounced. As enumerated in~\reef{eq:loop col list}, there are 9 possible tubings involving collapsed terms. However, as discussed above, only 8 of these appear in the kinematic flow equations, of which 6 are independent. 
In order to see this, it is convenient to display the explicit time-integral expressions for the collapsed functions:
\ba\label{eq:loop col def}
 \LoopUd &= \int  \ud \eta \,\zeta_u \,h^+_{Y_u}(\eta) \bar h^+_{Y_u}(\eta)\,, \qquad  \raisebox{3pt}{\LoopUu} = \int  \ud \eta \,\zeta_d \,h^+_{Y_d}(\eta) \bar h^+_{Y_d}(\eta)\,, \\[1pt]
  \raisebox{2pt}{\LoopUum} &= \int  \ud \eta \,\zeta_u \,h^-_{Y_u}(\eta) \bar h^-_{Y_u}(\eta) \,, \qquad   \raisebox{2pt}{\LoopUdm} = \int  \ud \eta \,\zeta_d \,h^-_{Y_d}(\eta) \bar h^-_{Y_d}(\eta) \,, \\[1pt]
\raisebox{2pt}{  \LoopCul} =\raisebox{2pt}{\LoopCur}&=\half \int \ud \eta \,\zeta_u \, \bigg(h^+_{Y_u}(\eta) \bar h^-_{Y_u}(\eta) + h^-_{Y_u}(\eta) \bar h^+_{Y_u}(\eta)  \bigg)\,,\\[1pt]
  \raisebox{2pt}{\LoopCdl} = \raisebox{2pt}{\LoopCdr}&=\half \int \ud \eta \,\zeta_d \, \bigg(h^+_{Y_d}(\eta) \bar h^-_{Y_d}(\eta) + h^-_{Y_d}(\eta) \bar h^+_{Y_d}(\eta)  \bigg)\,,\\[1pt]
\ea
where we have defined  
\beq
\zeta_{a}\equiv \frac{e^{i(X_1+X_2)\eta} }{(-\eta)^{1+\alpha_1+\alpha_2+2\dev_a}}\,, \quad {\rm for} \quad a=u,d\,.
\eeq
Here, we see explicitly which functions coincide in the massive case.
(Note that the equality of the pair of tubings in the last two lines of~\reef{eq:loop col def}  can also be verified from the kinematic flow, since they satisfy identical differential equations.)  
In the conformal limit, we have $h_k^- = \bar h_k^+  = 0$, along with $h_k^+ = e^{ik\eta}$ and $\bar h_k^- = e^{-ik\eta}$. It is easy to see that this makes the first four functions in~\reef{eq:loop col def} vanish, while the last four functions become identical and equal to one half of the function associated with the fully encircled tubing:
\be
\raisebox{2pt}{ \LoopCul} \,= \raisebox{2pt}{ \LoopCur} \,= \raisebox{2pt}{ \LoopCdl} \,= \raisebox{2pt}{ \LoopCdr} \, = \half \int \ud \eta \frac{ e^{i(X_1+X_2)\eta} }{(-\eta)^{1+\alpha_1+\alpha_2}}  ~\to~ \half \,\includegraphics[valign=c]{Figures/Tubings/Bubble_Letters/Z2form.pdf}\,.
\ee
It makes sense to denote all of these equivalent functions by the fully circled tubing. In this case, the merger step of the  kinematic flow in \reef{eq:merger loop} can also be reinterpreted as a double merging that results in a collapsed function with fully encircled tubing~\cite{Baumann:2024mvm}.

\vskip4pt
The same procedure can readily be applied to more complex processes involving more internal propagators or more loops.
The main takeaway is twofold: First, the massive kinematic flow behaves the same way for loop integrands as it does at tree level. In fact, this may be viewed as an advantage of this generalization. Second, in the conformally coupled limit, there are substantial simplifications due to the fact that many of the functions vanish. The massive flow rules then reduce to those of~\cite{Baumann:2024mvm,Hang:2024xas,Baumann:2025qjx}, suitably interpreted.

\newpage
\phantomsection
\addcontentsline{toc}{section}{References}
\bibliographystyle{utphys}
{\linespread{1.075}
	\bibliography{MassiveFlow-Refs}
}

\end{document}